\documentclass[9pt]{article}
\usepackage[utf8]{inputenc}
\usepackage{amsbsy}
\usepackage{graphicx}
\usepackage{enumerate}
\usepackage{makeidx}
\usepackage{geometry}
\usepackage{amssymb}
\usepackage{amsfonts}
\usepackage{amsmath}
\usepackage{amsthm}
\usepackage{amsbsy}
\usepackage{graphicx}
\usepackage{enumerate}
\usepackage{enumitem}
\usepackage{array}
\usepackage{lipsum}
\usepackage[nottoc]{tocbibind}
\usepackage{tikz}
\usepackage{mathtools}
\usepackage{alphabeta}
\usetikzlibrary{arrows,automata}
\newtheorem{defin}{Definition}
\newtheorem{examp}{Example}
\newtheorem{theo}{Theorem}
\newtheorem{prop}{Proposition}
\newtheorem*{prof*}{Proof}
\newtheorem{coro}{Corollary}
\newtheorem{lemma}{Lemma}
\newtheorem{rem}{Remark}

\usepackage{xfrac}

\usepackage{caption}
\usepackage{subcaption}

\usepackage{colortbl}
\usepackage{xcolor}
\usepackage{longtable}

\newcolumntype{?}{!{\vrule width 2pt}}

\usepackage{extarrows}

\usepackage{bbold}
\usepackage{array,multirow}
\usepackage{float}
\usepackage{booktabs}
\usepackage{array}
\newcommand\ChangeRT[1]{\noalign{\hrule height #1}}

\newcommand{\hide}[1]{}

\usepackage{xcolor}
\usepackage{centernot}
\usepackage{multirow}
\usepackage[colorlinks=true,linkcolor=black,anchorcolor=black,citecolor=black,filecolor=black,menucolor=black,runcolor=black,urlcolor=black]{hyperref}
\setlist[itemize,1]{label=$-$}
\setlist[itemize,2]{label=$\bullet$}
\usepackage{multicol}
\usepackage{pifont}

\allowdisplaybreaks
\providecommand{\keywords}[1]{\textbf{Keywords:} #1}

\newcommand{\skipitems}[1]{%
	\addtocounter{\@enumctr}{#1}%
}

\begin{document}
	
	\title{
		A formal algebraic approach for the quantitative \\ modeling of connectors in architectures}
	\author{Christina Chrysovalanti Fountoukidou$^a$  \ and Maria Pittou$^{b,}$\\Department of Mathematics\\Aristotle University of Thessaloniki\\54124 Thessaloniki, Greece\\$^{a}$christinafountoukidou1@gmail.com, $^{b}$mpittou@math.auth.gr}
	\date{}
	\maketitle

	\begin{abstract}
		In this paper we propose an algebraic formalization of connectors in the quantitative setting, in order to address their non-functional features in architectures of component-based systems. We firstly present a weighted Algebra of Interactions over a set of ports and a commutative and idempotent semiring, which is proved sufficient for modeling well-known coordination schemes in the weighted setup. In turn, we study a weighted Algebra of Connectors over a set of ports and a commutative and idempotent semiring, which extends the weighted Algebra of Interactions with types that encode Rendezvous and Broadcast synchronization. We show the expressiveness of the algebra by modeling the weighted connectors of several coordination schemes. Moreover, we derive two subalgebras, namely the weighted Algebra of Synchrons and the weighted Algebra of Triggers, and study their properties. Finally, we introduce a concept of congruence relation for connectors in the weighted setup and we provide conditions for proving such a congruence.

		\
		
		\noindent \keywords{Weighted Algebra of Interactions. Weighted Algebra of Connectors. Coordination schemes. Architectures. Component-based systems.
		}
	\end{abstract}

	\section{Introduction}\label{se1} 
	Software proliferation has induced the development of massively communicating systems with increasing growth in size and complexity. 
	Well-founded design of such systems can be achieved by component-based techniques which allow the separation of concerns between computation and coordination \cite{Br:St,Bl:Al,Si:Ri}. 
	Component-based systems are constructed by multiple individual components which interact according to a software architecture. Architectures have been proved fine-grained models for defining the communication of components, that is implemented by the concept of the so-called connectors \cite{Al:Fo,Be:On,Ma:Co,Si:Ri}. Connectors are architectural entities that regulate the synchronization mode among the permissible interactions of components, where the interactions are specified by the imposed coordination scheme \cite{Br:Ba,Ro:Fo}. For instance, in an architecture with a sender and two receiver components, a coordination 
	scheme may forbid any interaction between the receivers. In turn, a connector may impose Rendezvous
	synchronization mode requiring that all the components should interact simultaneously or Broadcast mode where a component, namely the sender,
	should initiate the interactions with some of the receivers.
	
	Connectors are distinguished in two main categories, namely the stateless and stateful connectors \cite{Br:Ba}. A connector
	is called stateless when the interaction constraints that it imposes on the ports
	stay the same at each round, and it is called stateful otherwise, i.e., when it supports dynamic interactions. 
	Applications of connectors occur in cloud and grid computing technologies with several functionalities
	including shared variable accesses, buffers, networking protocols, pipes etc. Rigorous formalization of connectors has been proved
	crucial for the efficient modeling and analysis of coordinated software systems \cite{Br:Ba,Su:Re}. 
	Connectors have been studied mainly in the qualitative setting with alternative frameworks and expressiveness, including
	process algebras \cite{Be:On,Br:Ba} and category theory \cite{Br:St,Br:Ba}, while they have been also supported by several architectural description languages in order to facilitate the specification of coordination among components \cite{Sa:Us,Oz:Ar,Su:Re}.
	
	However, well-founded design of systems communication should incorporate
	not only the required qualitative properties but also the related non-functional aspects
	(cf. \cite{Sa:Us,Si:Ri,Su:Re}). Such features include available
	resources, energy consumption, probabilities, etc., for implementing the interactions
	among the components. In this paper, we propose an algebraic formalization of the quantitative aspects of connectors in architectures. The quantitative modeling of connectors has only been addressed in the setting of architectural description languages and mainly deals with their probabilistic behavior (cf. \cite{Sa:Us,Su:Re}). On the other hand, there is lack of a formal algebraic framework for connectors in the weighted setup, which is the	main contribution of this work.

	In particular,  we extend the results of \cite{Bl:Al} for stateless connectors in the weighted setting. In \cite{Bl:Al}, the authors introduced two algebras for modeling the interactions and connectors in component based-systems, and studied their properties. In this paper, we extend these algebras in the weighted framework, and we show that the key results from \cite{Bl:Al} still hold. Our weighted algebras do not require knowledge on the behavior of components, and the only necessary information lies on the ports that serve for performing the communication. For the quantitative setup, we associate each port with a weight from a commutative and idempotent semiring $K$, expressing the ``cost'' of its
	participation in the interactions, which are described as sets of ports. In turn, we study two algebras that encode the weight of the interactions and of connectors, respectively. Also, we define an equivalence relation for the elements of the algebras and by their equivalence classes we derive several properties. In turn, we introduce a concept of congruence relation for weighted connectors and we extend the respective results from \cite{Bl:Al} in the weighted setup. Congruences are important for connectors since they allow to use them interchangeably without affecting the architecture \cite{Be:On,Br:Ba,Bl:Al}. Specifically, the contributions of the current paper are the following:
	
	(i) We introduce the \emph{weighted Algebra of Interactions} over a set of ports $P$ and a commutative and idempotent semiring $K$ ($wAI(P)$ for short). The syntax of the algebra is built over $P$, contains the symbols ``$0$'' and ``$1$'' such that $0,1\not \in P$, and allows two operators, namely the weighted union operator ``$\oplus$'' 
	and the weighted synchronization operator ``$\otimes$'', that encode the weight of independent and simultaneous interactions, respectively. We refer to the elements of the algebra simply as $wAI(P)$ elements, and we denote them by $z$. Given a set of interactions, we interpret the semantics of $z\in wAI(P)$ as polynomials over $P$ and $K$. Moreover, we denote by ``$\equiv$'' the equivalence relation of two $wAI(P)$ elements, i.e., elements that return the same weight on the same set of interactions over $P$. We define the quotient set $wAI(P)/\equiv$ of ``$\equiv$'' on $wAI(P)$ and for every $z\in wAI(P)$ we denote its equivalence class by $\overline{z}$. Then we show that the structure $(wAI(P)/\equiv,\oplus, \otimes,\bar{0},\bar{1})$ is a commutative and idempotent semiring with the binary operations $\oplus$ and $\otimes$ and constant elements $\bar{0} $ and $\bar{1} $, i.e., the equivalence class of $0$ and $1$, respectively. In turn, we apply this result for the computation of the semantics of our second algebra that encodes weighted connectors. Furthermore, we apply $wAI(P)$ in order to model several coordination schemes in the weighted setup, and specifically for Rendezvous, Broadcast, Atomic Broadcast and 
	Causality Chain \cite{Bl:Al}. 
	
	(ii) We introduce the \emph{weighted Algebra of Connectors} over a set of ports $P$ and a commutative and idempotent semiring $K$ ($wAC(P)$  for short) which extends $wAI(P)$ with two unary typing operators, that characterize the type of synchronization applied to ports, namely \emph{triggers} that can initiate an interaction and \emph{synchrons} which need synchronization with other ports in order to interact. Hence, the syntax of $wAC(P)$ allows two unary typing operators,
	trigger ``$\left[\cdot\right]'$'' and synchron ``$\left[\cdot\right]$'', and two binary operators ``$\oplus$'' and ``$\otimes$'', called weighted union and weighted
	fusion operator, respectively. Weighted union has the same meaning as in $wAI(P)$, while weighted fusion is a generalization of the weighted synchronization in $wAI(P)$. We define the semantics of $wAC(P)$ connectors as $wAI(P)$ elements, and then, applying the semantics of $wAI(P)$
	we derive the weight of the connectors over a concrete set of interactions in $P$. We  write $\left[ \zeta\right] ^{\alpha}$ for $\alpha\in \left\lbrace 0,1\right\rbrace $ to denote a typed $wAC(P)$ connector. When $\alpha=0$, the $wAC(P)$ connector is a synchron otherwise for $\alpha=1$ it is a trigger. Then we call $\zeta$ a \emph{fusion-$wAC(P)$ connector} when $\zeta=\left[ \zeta_{1} \right]^{\alpha_{1}} \otimes \ldots \otimes\left[ \zeta_{n}\right]^{\alpha_{n}}$, where $\zeta_{1}, \ldots, \zeta_{n}\in wAC(P)$ and $\alpha_1,\ldots, \alpha_n \in \lbrace 0,1\rbrace$. We obtain several nice properties for $wAC(P)$ and we show the expressiveness of $wAC(P)$
	by encoding the weight for the connectors of the coordination schemes Rendezvous, Broadcast, Atomic Broadcast
	and Causality Chain using fusion-$wAC(P)$ connectors.\hide{We also provide a $wAC(P)$ connector for encoding the coordination in a simple Request/Response architecture.} Furthermore, we consider two subalgebras of $wAC(P)$, the weighted Algebra of Synchrons ($wAS(P)$ for short) and of Triggers ($wAT(P)$ for short) over $P$ and $K$,
	where the former restricts to synchron elements and the latter to trigger elements, and study their properties.

	(iii) We show that due to the weighted fusion operator, 
	equivalent $wAC(P)$ connectors, i.e., connectors with the same $wAI(P)$ elements, are not in general interchangeable. For this, we are interested
	in a congruence relation for $wAC(P)$ connectors. 
	We
	explain why we cannot derive a congruence relation between $wAC(P)$ connectors in general, and in turn, we introduce a concept of congruence applied to fusion-$wAC(P)$ connectors. Finally,
	we  derive two theorems that provide conditions for proving such a congruence for fusion-$wAC(P)$ connectors, by extending the results of \cite{Bl:Al} 
	in our weighted framework. The first theorem
	shows that similarly typed equivalent fusion-$wAC(P)$ connectors are congruent. The second theorem shows
	that congruence relation between two fusion-$wAC(P)$ connectors is ensured when the following three conditions hold: 
	(i) they are equivalent, (ii) the equivalence is preserved under the weighted fusion with the trigger $[1]'$, and (iii) both connectors either contain only synchrons or some triggers.
	
	\section{Related work}\label{rel}

	The concept of connectors has been extensively studied in software engineering by versatile formal frameworks and architectural modeling languages. 
	Most of the existing work addresses the qualitative aspects of connectors, while there is a few work investigating the quantitative setting.\hide{According to our best knowledge there is lack of a formal framework that models connectors in the weighted setup in general.} In the present
	paper, we propose an algebraic formalization of the interactions and (stateless) connectors over a commutative and idempotent semiring.  In the sequel, we discuss some of the modeling theories and languages on connectors that are comparable with our methodology. 
	The work presented below differs with our framework in at least one of the following aspects: (i) they deal with stateful connectors and dynamic interactions,
	(ii) the modeling process incorporates the behavior of components and connectors, and (iii) they do not address the quantitative features of connectors, which is the main
	novelty of the current paper. 
	
	Representative work in the algebraic formalization of connectors includes process algebras, contracts, and category theory. 
	Process algebras are algebraic languages which support the compositional description of concurrent and distributed systems,  whose basic elements are their actions.
	In \cite{Be:On}, the authors formalized architectural types by a process algebra based on an architectural description language that supported dynamic interactions.
	Architectural types were described by components types and stateful connectors. Connectors were defined by a function of their behavior, specified as a family of process algebra terms,
	and their interactions, specified as a set of process algebra actions. The process algebra was build over a set of actions and contained a hiding, a relabeling,
	a composition and parallel composition operator, while the semantics was given by state transition graphs. 
	In turn, the authors presented a weak bisimulation equivalence based technique for verifying architectural compatibility and conformity. 
	
	In \cite{Pa:Co}, contracts were used to model components and connectors in UML language. Specifically,  
	p-calculus was combined with first-order modal reasoning in order to model connectors.  The extended calculus was
	interpreted in state-based algebraic structures (called objects) and
	captured dynamic interactions. Reasoning and refinement were also studied for component composition. 
	
	In \cite{Ro:Fo}, the authors proposed an architectural metamodel for describing stateful connectors focusing on the communication styles of
	message passing and remote procedure call mechanisms, which are common in distributed systems. The semantics of the styles were expressed by finite state machine models. 
	Then Alloy was chosen for formalizing those communication styles and for verifying conformance of the communication style at the
	model level.
	
	In contrast to \cite{Be:On,Pa:Co,Ro:Fo}, our framework studies stateless connectors, and hence does not deal with dynamic interactions.
	On the other hand, our algebra models the quantitative aspects of connectors that were not addressed in \cite{Be:On,Pa:Co,Ro:Fo}. 
	\hide{Moreover, in \cite{Be:On,Pa:Co,Ro:Fo}
		components and connectors were defined by their actions and their underlying behavior. On the contrary, in our setting
		the algebraic formalization in the weighted setup is achieved without considering the behavior of connectors. }The investigation of dynamic interactions and stateful connectors in the weighted setup is future work. 
	
	In the work of \cite{Br:St}, the authors developed an algebra for stateless connectors based on category theory. 
	Their algebra supported symmetry, synchronization, mutual exclusion, hiding and inaction connectors. The authors provided
	the operational, observational, and denotational semantics of connectors, and then,
	showed that the latter two coincide. Finally, a complete normal-form axiomatization was presented for the algebra and the proposed framework 
	was applied for modeling architectural and coordination connectors in CommUnity and Reo languages, respectively. 
	In contrast to our framework, the work of \cite{Br:St} studied the qualitative modeling of connectors.
	Also, our algebras encode sufficiently the weight of independent and synchronized actions, though do not involve hiding connectors. On the other hand, in the work of \cite{Br:St}, connectors were specified as entities with behavior and interactions, where concurrent actions, i.e., multiple actions that are executed in parallel, were permitted. 
	Studying the behavior of components and their connectors as well as the application of connectors in concurrent systems is future work. 
	
	In \cite{Sp:Co}, the authors introduced and studied connector wrappers in order to repair or augment communication-related properties of a system.
	A wrapper was defined as a new code interposed between component interfaces and infrastructure support. The intent of the code was
	to modify the behavior of the component with respect to the other components in the system, without actually modifying
	the component or the infrastructure of the system. Having applied a wrapper to a faulty connector, the authors studied
	whether the result was sound and if the wrapper was transparent to the caller role without changing the interface.
	The goal of wrappers was to avoid directly modifying the components in a system. In our approach,
	we introduce a concept of congruence relation for weighted connectors and we provide conditions that allow checking congruence
	between two fusion-$wAC(P)$ connectors. In contrast to \cite{Sp:Co}, this allows one to interchange directly fusion-$wAC(P)$ connectors without affecting the communication pattern of the architecture. 
	
	We point the reader to \cite{Br:Ba} for a nice survey on some well-known theories of connectors. In that paper, the authors studied the formal approaches of Reo, BIP, nets with boundaries, the algebra of stateless connectors, the tile model, and wire calculus, for the modeling, composing and analyzing connectors. The authors
	presented a comparison framework for those methodologies and discussed possible enhancements. In comparison to our work, the presented theories studied connectors
	in the qualitative setting. 
	
	Apart from the algebraic methods, the modeling of connectors has been supported by several architectural description languages. 
	In the recent work of \cite{Oz:Ar} can be found an extensive survey on several architectural description languages that support
	the modeling of several types of connectors. Among them, a well-known example is Reo, a channel-based coordination model, 
	that served as a language for coordination of concurrent processes or for compositional construction of connectors among component-based systems \cite{Am:Re,Ar:Re}. 
	In \cite{Am:Re,Ar:Re}, components performed input/output operations through connectors that do not have knowledge on the behavior of components. 
	Coordination of components was achieved through channels, which were
	considered as atomic connectors in Reo. Channels were used to transfer data and there were assumed two types of channel ends: sources and sinks. 
	Complex connectors were compositionally built out of simpler
	ones. Hence, a connector was a set of channel ends and their connecting channels organized in a graph.
	The topology of connectors in \cite{Am:Re,Ar:Re} was inherently dynamic and it allowed mobility in components connections. Some work has also investigated Reo connectors with probabilistic behavior (cf. \cite{Sa:Us,Su:Re}). In \cite{Su:Re}, the authors studied Reo connectors with probabilistic behavior as
	timed data distribution streams implemented in Coq. Moreover, in \cite{Sa:Us}, the authors formalized Reo connectors with random and probabilistic behavior in PVS. 
	In contrast to \cite{Sa:Us,Su:Re}, we propose a general algebraic modeling framework in the weighted setup. Another difference is that Reo treats input and output	ports separately, while in our framework, we use bidirectional ports.
	
	In \cite{Bl:Al}, the authors introduced the Algebra of Interactions over a set of ports $P$ ($AI(P)$ for short) for modeling the interactions of components. The syntax of the algebra was build over the set of ports $P$, two special symbols encoding empty interaction and the set of empty interactions, a union operator and a synchronization operator. 
	In turn, for their second algebra, the Algebra of Connectors ($AC(P)$ for short), each port was characterized by a type
	of synchronization, called trigger when initiating an interaction and synchron when
	synchronizing with other ports in order to interact. 
	The authors considered also two subalgebras, where
	all of the connectors had the same type (synchron or trigger), and discussed the relation of the presented algebras. In turn, they introduced and studied a congruence relation for connectors, expressed as the biggest equivalence relation that
	allowed using connectors interchangeably. Finally, the authors presented applications
	of their algebras for improving the language and the execution engine of BIP framework.
	We also point the reader to \cite{Si:De} and \cite{Si:Mo} for related methodologies on the characterization of connectors.
	
	Our paper is closely related to the work \cite{Bl:Al}, and in particular extends the presented results in the weighted setup. An important difference from \cite{Bl:Al} is that there was a clear distinction between syntactic equality and semantic equivalence. Moreover, most of the presented results in \cite{Bl:Al}
	were proved by syntactic equality and in particular by considering several axioms corresponding to important properties for the algebras. 
	On the contrary, in the weighted framework in general, we cannot state results by syntactic equality. For this, in our framework we explain
	that the notion of equivalence for the $wAI(P)$ elements and $wAC(P)$ connectors
	induces an equivalence relation, respectively. In turn, we derive the
	equivalence classes for $wAI(P)$ and $wAC(P)$. Then we
	formalize the required properties by equalities and we prove
	them using the respective equivalence classes. Due to this difference with the work of \cite{Bl:Al}, solving the congruence problem in the weighted setup is more difficult. Specifically, the investigation of a congruence relation for any $wAC(P)$ connector is an open problem (see also Section \ref{se7}).

	\section{Preliminaries}\label{se2}
	\subsection{Notations}
	For every natural number $n \geq 1$ we denote by $[n]$ the set $\{1, \ldots, n \}$. Hence, in the sequel, whenever we use the notation $[n]$ we always assume that $n \geq 1$. 
	\par Next, we recall the basic notions for semirings and series \cite{Dr:Ha,Sa:Ra}.

	\subsection{Semiring and series}
	A monoid $(K, +, \hat{0})$ is a non-empty set $K$ which is equipped with an associative operation $+$
	and a neutral element $\hat{0}$ such that $\hat{0} + k = k + \hat{0} = k$ for every $k \in K$. A monoid is called commutative
	if $+$ is commutative.
	
	\par A semiring $(K, +,\cdot,\hat{0},\hat{1})$ consists of a set $K$, two binary operations $+$ and $\cdot$, and two constant elements $\hat{0}$ and $\hat{1}$ in $K$, such that: 
	\begin{enumerate}[label=(\roman*)]
		\item $(K,+,\hat{0})$ is a commutative monoid,
		\item $(K,\cdot,\hat{1})$ is a monoid,
		\item  $\cdot$ distributes over $+$, i.e., $(k_{1} + k_{2})\cdot k_{3} = (k_{1} \cdot k_{3}) + (k_{2} \cdot k_{3})$ and $k_{1} \cdot (k_{2} + k_{3}) = (k_{1} \cdot k_{2}) + (k_{1} \cdot k_{3})$ for every $k_{1}, k_{2}, k_{3} \in K$, and
		\item  $\hat{0} \cdot k = k \cdot \hat{0} = \hat{0}$ for every $k \in K.$
	\end{enumerate}
	\noindent A semiring $K$ is called commutative if $(K,\cdot,\hat{1})$ is commutative. The semiring is denoted simply by $K$ if the
	operations and the constant elements are understood. Further, $K$ is called additively idempotent if $(K, + ,\hat{0})$ is an idempotent monoid, i.e., $k + k = k$ for every $k\in K$. By the distributivity law, this holds iff $\hat{1} + \hat{1} = \hat{1}$. In the sequel, we call an additively idempotent semiring simply an idempotent semiring. The following algebraic structures are well-known semirings:
	\begin{itemize}
		\item the semiring $(\mathbb{N},+,\cdot,0,1)$ of natural numbers, 
		\item the Boolean semiring $B=(\left\lbrace 0,1 \right\rbrace,+,\cdot,0,1 )$, 
		\item the tropical or $\mathrm{min}$-plus semiring $\mathbb{R}_{\mathrm{min}}=(\mathbb{R}_{+}\cup \left\lbrace \infty \right\rbrace,\mathrm{min},+,\infty,0 )$ where $\mathbb{R}_{+}=\left\lbrace r\in \mathbb{R}\mid r\geqslant 0 \right\rbrace $, 
		\item the arctical or $\mathrm{max}$-plus semiring $\mathbb{R}_{\mathrm{max}}=(\mathbb{R}_{+}\cup \left\lbrace -\infty \right\rbrace,\mathrm{max},+,-\infty,0 )$,
		\item the semiring $(\mathcal{P}(A),\cup, \cap, \emptyset, A)$ for every non-empty set $A$,
		\item the Viterbi semiring $(\left[ 0,1 \right],\mathrm{max},\cdot,0,1 )$ used in probability theory, and 
		\item the Fuzzy semiring $F=(\left[ 0,1\right] , \mathrm{max}, \mathrm{min}, 0, 1)$, and in general every bounded distributive lattice with the operations sup and inf.
	\end{itemize}
	All the above semirings are commutative, and all but the
	first one are idempotent.

	Let $K$ be a semiring and $P$ be a non-empty set. A formal series (or simply series) over $P$ and $K$ is a mapping $s: P\to K$. The support of $s$ is the set $\mathrm{supp}(s)=\left\lbrace p\in P\mid s(p)\neq \hat{0} \right\rbrace $. A series with finite support is called a polynomial. We denote by $K\left\langle P\right\rangle $ the class of all polynomials over $P$ and $K$. Let $s, r\in K\left\langle P\right\rangle $ and $k \in K$. The sum $s \oplus r$, the product with scalars $ks$ and $sk$, and the Hadamard product $s \otimes r$ are polynomials  in $K\left\langle P\right\rangle $, and are defined elementwise, by $s \oplus r(p)=s(p) + r(p), (ks)(p)=k \cdot s(p), (sk)(p)=s(p) \cdot k$ and $s \otimes r(p)=s(p) \cdot r(p)$ for every $p\in P$, respectively.
	
	\begin{quote}
		\noindent \textit{Throughout the paper, $(K, + ,\cdot, \hat{0},\hat{1})$ denotes a commutative and idempotent semiring.}
	\end{quote}

	\subsection{Interactions}\label{se3}
	In this work, we are interested in the coordination patterns of the systems, and specifically on the algebraic characterization of the concept of connectors in the quantitative setting. For this, we develop no theory about the semantics of  component-based systems. The investigation of the weighted behavior of a system, where communication is expressed
	with our weighted algebras, is part of future work. 
	
	Next, we use the basic notions and definitions of BIP for the communication of components in qualitative setting \cite{Bl:Al,Si:Ri}.
	Though, our results can be applied to every component-based framework where the interface of the system can be described by a set of ports. 
	
	In our setting, communication in architectures is performed by a set of ports and is defined by interactions. In turn, the permissible set of interactions is specified by the coordination scheme implemented in the architecture.

	\begin{defin}
		Let $P$ be a finite non-empty set of ports. Then an interaction $a$ is a set of ports over $P$, i.e., $a\in 2^{P}$. 
	\end{defin}

	\noindent Let $a=\left\lbrace p_1,\ldots ,p_n\right\rbrace  \subseteq P$. We often simplify the notation of $\textit{a}$ by writing $p_1 \ldots p_n$ instead of the set $\left\lbrace p_1, \ldots ,p_n\right\rbrace $. Then an \emph{interactions set} $\gamma$ is a set of interactions over $P$, i.e., $\gamma \in 2^{2^{P}}$. We let $I(P)=2^{P}$ denote the set of interactions over $P$ and $\Gamma(P)=2^{I(P)}$ denote the set of all subsets of $I(P)$. 
	
	It should be clear that we are not practically interested in the empty interaction $a=\emptyset$, and in turn the empty interactions set $\gamma=\lbrace \emptyset\rbrace$ since they correspond to the case that there is no communication among any components of a system. Still, we need to consider $a=\emptyset$ and $\gamma=\lbrace \emptyset\rbrace$ in the construction of our weighted algebras in order to be well-defined.
	
	In order to encode the weight of the interactions, we introduce a weighted Algebra of Interactions over a set of ports $P$ and a commutative and idempotent semiring $K$, presented in the next section. The proposed algebra is applied for modeling well-known coordination schemes in the weighted setup, and in turn serves as the basis for constructing  our second algebra, the weighted Algebra of Connectors, that formalizes of the concept of weighted connectors.

	\section{The Weighted Algebra of Interactions}\label{se4}
	Coordination schemes capture the permissible interactions among the components of an architecture. 
	In order to address the quantitative aspects of coordination schemes, we study the weighted Algebra of Interactions over a set of ports $P$ and a commutative and idempotent semiring $K$. The algebra is build over $P$, two further symbols ``$0$'' and ``$1$'' that do not belong in $P$, and permits two operators, the 
	weighted union ``$\oplus$'' and weighted synchronization ``$\otimes$'' operators. Weighted union serves to compute the weight of interactions executed independently by the involved terms,
	while weighted synchronization captures the weight of interactions applied simultaneously. The symbols	$0$ and $1$ serve as the neutral elements of the operators $\oplus$ and $\otimes$, respectively. We prove several nice properties for the algebra and we provide concrete examples of coordination schemes in the weighted setup, described by our algebra.  
	
	Let $P$ be a set of ports. Then we assign to each port $p \in P$ a unique weight from $K$, denoted by $k_{p}$.
	
	\begin{defin}
		Let $P$ be a set of ports such that $0,1\not \in P$. The syntax of the weighted Algebra of Interactions ($wAI(P)$ for short) over $P$ and $K$ is given by 
		$$z::=0\mid 1\mid p \mid z\oplus z \mid z \otimes z \mid \left( z \right)$$ 
		where $p\in P$, `` $\oplus$'' is the weighted union operator and `` $\otimes$'' is the weighted synchronization operator that binds stronger than `` $\oplus$''.
	\end{defin}
	\noindent It should be clear that ``$0$'' and ``$1$'' occurring in the syntax of the $wAI(P)$ are distinct from the respective elements $\hat{0}$ and $\hat{1}$ in semiring $K$. Indeed, the former serve to encode the weight on a specific $\gamma\in \Gamma(P)$, as presented in Definition \ref{sem-wAI(P)} below, while the latter correspond to the neutral elements of the operations ``$+$''  and ``$\cdot$'' in $K$, respectively. 
	\par We call $z$ a $wAI(P)$ element over $P$ and $K$. Whenever the latter are understood we simply refer to $z$ as a $wAI(P)$ element. The semantics of a $wAI(P)$ element $z$ over the set of ports $P$ and the semiring $K$ is defined by the function $\left\| \cdot \right\|: \Gamma(P) \to K$. Therefore, we represent the semantics of $z$ as polynomials $\left\|z\right\| \in K\left\langle \Gamma(P)\right\rangle $. Hence, given an interactions set $\gamma \in \Gamma(P)$, we can derive the weight of implementing $\gamma$ in a given architecture.
	\par Note that for different instantiations of the semiring $K$, the resulting weight is interpreted as a particular quantitative property.
	For example, if we are interested in ensuring a sufficient level of trustworthiness,
	then we opt for the interactions with the maximal probability to be executed.
	Hence, we may choose to work in the Viterbi semiring in order to
	address this issue. On the contrary, if the weights range over $\mathbb{R}_{\mathrm{min}} $,
	then we may be interested in the minimal cost, where the cost may refer to the energy
	consumption of implementing the communications. In such a setting,
	the derived weight returns the minimal cost of the interactions that could improve the efficiency of
	the architecture. 
	
	\begin{defin}\label{sem-wAI(P)}
		Let $z$ be a $wAI(P)$ element over $P$ and $K$. The semantics of $z$ is a polynomial $\left\|z\right\| \in K\left\langle \Gamma(P)\right\rangle $. For every interactions set $\gamma\in \Gamma(P)$, the value $\left\|z\right\|(\gamma)$ is defined inductively on $z$ as follows: 
		\begin{itemize}
			\renewcommand\labelitemi{--}
			\item $\left\|0\right\|(\gamma)=\hat{0}$,
			\item $\left\|1\right\|(\gamma)=\left\{	\begin{array}{l l}
				\hat{1} & \text{if} \quad  \emptyset\in\gamma  \\
				\hat{0} & \text{otherwise} 
			\end{array} \right.$,
			\item $\left\|p\right\|(\gamma)=\left\{	\begin{array}{l l}
				k_{p} & \text{if} \quad \exists \ a \in\gamma \ \text{such that} \  p\in a  \\
				\hat{0} & \text{otherwise} 
			\end{array} \right.$,
			\item $\left\|z_{1}\oplus z_{2}\right\|(\gamma)=
			\sum\limits_{a \in \gamma }    \big( \left\|z_{1}\right\|(\left\lbrace a\right\rbrace ) + \left\|z_{2}\right\|(\left\lbrace a\right\rbrace ) \big)$,
			\item $\left\|z_{1}\otimes z_{2}\right\|(\gamma)=
			\sum\limits_{a\in \gamma} \Big(  \sum\limits_{a=a_{1}\cup a_{2}} \big(  \left\|z_{1}\right\|(\left\lbrace a_{1}\right\rbrace) \cdot \left\|z_{2}\right\|(\left\lbrace a_{2}\right\rbrace) \big) \Big) $,
			\item $\left\| \left( z\right) \right\|(\gamma)=\left\| z \right\|(\gamma).$
		\end{itemize}
	\end{defin}

	\begin{rem}
		According to the above semantics, the $wAI(P)$ element `` $0$'' returns zero weight on any $\gamma$ while the $wAI(P)$ element `` $1$'' returns non-zero value, equal to $\hat{1}$, only if $\gamma$ contains $a=\emptyset$, i.e., the empty interaction. \hide{Moreover, the intuition behind the semantics of the element `` $p$'' is that when the port occurs in some interaction in $\gamma$, then it returns its assigned weight, whether it belongs in a permissible interaction or not, and independently of the presence of undesirable interactions in $\gamma$. This is a natural choice because it implies that the occurrence of a port activates the port with its associated weight, while the algebra sufficiently derives the expected overall weight of a coordination scheme, as it is shown by the examples presented later in this section.}Also, the $wAI(P)$ element $p$ returns its weight $k_p$ whenever it occurs in some interaction $a \in \gamma$, which implies that the port is ``activated'' with its assigned weight. Moreover, the semantics of weighted union and synchronization are justified by the fact that they encode the weight of independent and synchronous interactions, respectively. Hence, for the former operator we consider the weight on the same set $\lbrace a\rbrace$ for $z_1$ and $z_2$ for every $a\in \gamma$, while for the latter we analyze the interaction $a$ to the union of $a_1$ and $a_2$ and we compute the weight of $z_1$ and $z_{2}$ on $\lbrace a_{1} \rbrace$ and $\lbrace a_{2} \rbrace$, respectively, for every such analysis of $a$ and for every $a\in \gamma$.   
		Finally, the application of the parenthesis construct on a $wAI(P) $ element $z$ does not affect its weight, since parenthesis only serves for imposing the common order restrictions among the occurring operators. For this, $z$ and $(z)$ return the same semantics.
	\end{rem} 
	
	\begin{rem}
		Recall that an element of the Algebra of Interactions from \cite{Bl:Al} served to encode a specific interactions set $\gamma \in \Gamma(P)$. On the other hand, a $wAI(P)$ element can be interpreted for any $\gamma \in \Gamma(P)$, inducing a weight from $K$. This difference results from the semantics given to ports $p\in P$. In \cite{Bl:Al}, a term $p$ was associated only with $\gamma=\lbrace \lbrace p \rbrace \rbrace$, while in $wAI(P)$, a weighted term $p$ returns its weight $k_p$ whenever it occurs in some interaction $a \in \gamma$. This is a natural choice, and as a consequence, the algebra can sufficiently encode the expected overall weight of a coordination scheme, as it is shown by the examples presented later in this section.
	\end{rem}

	\begin{rem}\label{re1}
		Observe that if $\gamma=\emptyset$, then by the above definition we get that $\left\|z\right\|(\gamma)=\hat{0}$ for every $z\in wAI(P)$.
	\end{rem}
	
	\begin{rem}\label{singl}
		In case that $\gamma\in \Gamma(P)$ is a singleton, i.e., $\gamma=\left\lbrace a'\right\rbrace $, then for the semantics of the weighted union operator we have 
		\begin{align*}
			\left\|z_{1}\oplus z_{2}\right\|(\gamma)=&
			\sum\limits_{a \in \gamma }    \big( \left\|z_{1}\right\|(\left\lbrace a\right\rbrace ) + \left\|z_{2}\right\|(\left\lbrace a\right\rbrace ) \big)\\
			=& \sum\limits_{a \in \left\lbrace a'\right\rbrace  }    \big( \left\|z_{1}\right\|(\left\lbrace a\right\rbrace ) + \left\|z_{2}\right\|(\left\lbrace a\right\rbrace ) \big)\\
			=& \left\|z_{1}\right\|(\left\lbrace a'\right\rbrace ) + \left\|z_{2}\right\|(\left\lbrace a'\right\rbrace ).
		\end{align*} 
		Moreover, for the semantics of the weighted synchronization operator we have 
		\begin{align*}
			\left\|z_{1}\otimes z_{2}\right\|(\gamma)=& \sum\limits_{a \in \gamma }   \Bigg(
			\sum\limits_{a =a_{1}\cup a_{2}}    \big( \left\|z_{1}\right\|(\left\lbrace a_{1}\right\rbrace )\cdot\left\|z_{2}\right\|(\left\lbrace a_{2}\right\rbrace ) \big)\Bigg)\\
			=& \sum\limits_{a \in \left\lbrace a'\right\rbrace  }   \Bigg(
			\sum\limits_{a =a_{1}\cup a_{2}}    \big( \left\|z_{1}\right\|(\left\lbrace a_{1}\right\rbrace )\cdot\left\|z_{2}\right\|(\left\lbrace a_{2}\right\rbrace ) \big)\Bigg)\\
			=& \sum\limits_{a' =a_{1}\cup a_{2}}    \big( \left\|z_{1}\right\|(\left\lbrace a_{1}\right\rbrace ) \cdot \left\|z_{2}\right\|(\left\lbrace a_{2}\right\rbrace ) \big).
		\end{align*} 
		Hence, in what follows, when $\gamma$ is a singleton, i.e., $\gamma=\left\lbrace a'\right\rbrace \in \Gamma(P)$ we often write
		\begin{itemize}
			\item $\left\|z_{1}\oplus z_{2}\right\|(\gamma)=\left\|z_{1}\right\|(\left\lbrace a'\right\rbrace ) + \left\|z_{2}\right\|(\left\lbrace a'\right\rbrace )$, and
			\item $\left\|z_{1}\otimes z_{2}\right\|(\gamma)=\sum\limits_{a' =a_{1}\cup a_{2}}    \big( \left\|z_{1}\right\|(\left\lbrace a_{1}\right\rbrace ) \cdot \left\|z_{2}\right\|(\left\lbrace a_{2}\right\rbrace ) \big)$.
		\end{itemize}
		
	\end{rem}

	\noindent In \cite{Bl:Al}, the authors derived several properties in their Algebra of Interactions, as axioms obtained directly by syntactic equality. On the contrary, in the weighted setup, we derive an equivalence relation induced by the semantics of $wAI(P)$. In turn, we use the equivalence classes of $wAI(P)$ for proving the properties that hold for the weighted algebra. We say that $z_{1}, z_{2}\in wAI(P)$ are equivalent and we write $z_{1}\equiv z_{2}$, when they return the same weight on the same set of interactions, i.e., when $ \left\|z_{1}\right\|(\gamma)=\left\|z_{2}\right\|(\gamma)$ for every $\gamma\in \Gamma(P)$. 
	\par Obviously the relation ``$\equiv$'' is an equivalence relation. We define the quotient set $wAI(P)/\equiv$ of ``$\equiv$'' on $wAI(P)$. For every $z\in wAI(P)$, we simply denote by $\overline{z}$ its equivalence class. We define on $wAI(P)/\equiv$ two operations as follows: 
	$\overline{z_{1}}\oplus \overline{z_{2}}=\overline{z_{1}\oplus z_{2}}$
	and $\overline{z_{1}}\otimes \overline{z_{2}}=\overline{z_{1}\otimes z_{2}}$
	for every $z_{1}, z_{2}\in wAI(P)$. It can be easily proved that these operations depend only on the classes and not on their representatives, i.e., they are well-defined. Thus, if $\overline{z_{1}}=\overline{z_{3}}$ and $\overline{z_{2}}=\overline{z_{4}}$, then $\overline{z_{1}}\oplus\overline{z_{2}}=\overline{z_{3}}\oplus\overline{z_{4}}$ and $\overline{z_{1}}\otimes \overline{z_{2}}=\overline{z_{3}}\otimes \overline{z_{4}}$ for every $z_{1}, z_{2}, z_{3}, z_{4} \in wAI(P)$.

	\par In the next proposition, we prove several properties satisfied by the $wAI(P)$ elements over $P$ and $K$. As a consequence we obtain that $(wAI(P)/\equiv,\oplus,\otimes, \bar{0},\bar{1})$ is a commutative and idempotent semiring with two binary operations $\oplus$, i.e., weighted union and $\otimes$, i.e., weighted synchronization, and two constant elements $\bar{0}$ and $\bar{1}$, respectively. In turn, we apply this result to our weighted algebra for connectors, presented in the following section. Firstly, we consider the subsequent lemma which is needed for proving the equalities iii), iv) and vii) of Proposition \ref{propeties_walg_in}. The lemma actually formalizes the intuitive result that the weight of a $wAI(P)$ element $z$ on a given set of interactions $\gamma\in \Gamma(P)$ equals the sum of the
	weight of $z$ on $\lbrace a \rbrace $ for every interaction $a \in \gamma$.

	\begin{lemma}\label{l1}
		Let $z\in wAI(P)$. Then for every interactions set $\gamma\in \Gamma(P)$ it holds that
		$$\sum\limits_{a \in \gamma }  \Big( \left\|z\right\|(\left\lbrace a\right\rbrace) \Big) = \left\|z\right\|(\gamma).$$
	\end{lemma}
	
	\begin{prof*}
		We prove the lemma by induction on the structure of $z\in wAI(P)$. By Remark \ref{re1}, we assume that $\gamma\in \Gamma(P)\setminus \left\lbrace \emptyset\right\rbrace $.
		\begin{itemize}
			\item If $z=0$, then $\sum\limits_{a \in \gamma }  \Big( \left\|0\right\|(\left\lbrace a\right\rbrace)  \Big) =	
			\sum\limits_{a \in \gamma }    \big( \hat{0}  \big) 
			=\hat{0}$ and $\left\|0\right\|(\gamma)=\hat{0}.$
			
			\item If $z=1$ and $\emptyset \in \gamma$, then	$\sum\limits_{a\in \gamma} \Big( \left\|1\right\|(\left\lbrace a\right\rbrace )\Big)= \left\|1\right\|(\left\lbrace \emptyset\right\rbrace )+ \overset{}{\underset{\substack{a\in \gamma,\\a\neq \emptyset }}{\sum}}\big( \left\|1\right\|(\left\lbrace a\right\rbrace ) \big) = \hat{1} + \overset{}{\underset{\substack{a\in \gamma,\\a\neq \emptyset }}{\sum}}\big( \hat{0} \big) 
			= \hat{1} + \hat{0} = \hat{1}$ and $\left\|1\right\|(\gamma)=\hat{1}.$ On the other hand, if $z=1$ and $\emptyset \notin \gamma$, then
			$\sum\limits_{a\in \gamma} \Big( \left\|1\right\|(\left\lbrace a\right\rbrace )\Big)=\overset{}{\underset{a\in \gamma}{\sum}}\big( \hat{0} \big)=\hat{0}$	and $\left\|1\right\|(\gamma)=\hat{0}.$	Hence, in each case we obtain that $\sum\limits_{a\in \gamma} \Big( \left\|1\right\|(\left\lbrace a\right\rbrace )\Big)=\left\|1\right\|(\gamma).$	
			\item If $z=p$ and $\exists \ a\in \gamma$ such that $p \in a$, then
			$\sum\limits_{a\in \gamma} \Big( \left\|p\right\|(\left\lbrace a\right\rbrace )\Big)=\overset{}{\underset{\substack{a\in \gamma,\\ p\in a}}{\sum}}\big(\left\|p\right\|(\left\lbrace a\right\rbrace ) \big) + \overset{}{\underset{\substack{a\in \gamma,\\ p\notin a}}{\sum}}\big(\left\|p\right\|(\left\lbrace a\right\rbrace ) \big)
			= \overset{}{\underset{\substack{a\in \gamma,\\ p\in a}}{\sum}}\big(k_{p} \big) + \overset{}{\underset{\substack{a\in \gamma,\\ p\notin a}}{\sum}}\big(\hat{0} \big) = k_{p} + \hat{0} = k_{p}$ and $\left\|p\right\|(\gamma)=k_{p}.$ On the other hand, if $z=p$ and $\nexists \ a\in \gamma$ such that $p \in a$, then $\sum\limits_{a\in \gamma} \Big( \left\|p\right\|(\left\lbrace a\right\rbrace )\Big)=\overset{}{\underset{\substack{a\in \gamma,\\p\notin a }}{\sum}}\big( \hat{0} \big)=\hat{0} $ and	$\left\|p\right\|(\gamma)=\hat{0}.$	Hence, in each case we obtain that $\sum\limits_{a\in \gamma} \Big( \left\|p\right\|(\left\lbrace a\right\rbrace )\Big)=\left\|p\right\|(\gamma).$    
			\item If $z=z_{1}\oplus z_{2}$, then 
			\begin{align*}
				\sum\limits_{a \in \gamma }    \Big( \left\|z_{1}\oplus z_{2}\right\|(\left\lbrace a\right\rbrace)  \Big) 
				=&
				\sum\limits_{a \in \gamma } \Bigg(  \sum\limits_{a' \in \left\lbrace a\right\rbrace  }  \big( \left\|z_{1}\right\|(\left\lbrace a'\right\rbrace )+ \left\|z_{2}\right\|(\left\lbrace a'\right\rbrace )  \big) \Bigg) \\
				=&\sum\limits_{a \in \gamma }    \big( \left\|z_{1}\right\|(\left\lbrace a\right\rbrace )+\left\|z_{2}\right\|(\left\lbrace a\right\rbrace ) \big)\\
				=& \left\|z_{1}\oplus z_{2}\right\|(\gamma),
			\end{align*}
			where the second equality holds since the set $\left\lbrace a\right\rbrace $ contains only the interaction $a$, thus we get that $a'=a$.
			\item If $z=z_{1}\otimes z_{2}$, then 
			\begin{align*}
				\sum\limits_{a \in \gamma }    \Big( \left\|z_{1}\otimes z_{2}\right\|(\left\lbrace a\right\rbrace)  \Big)
				=&\sum\limits_{a \in \gamma } \Bigg(  \sum\limits_{a' \in \left\lbrace a\right\rbrace  }  \Bigg( \sum\limits_{a'=a_{1}\cup a_{2}} \big( \left\|z_{1}\right\|(\left\lbrace a_{1}\right\rbrace )\cdot \left\|z_{2}\right\|(\left\lbrace a_{2}\right\rbrace ) \big) \Bigg) \Bigg)\\
				=&\sum\limits_{a \in \gamma }    \Bigg(   \sum\limits_{a=a_{1}\cup a_{2}} \big( \left\|z_{1}\right\|(\left\lbrace a_{1}\right\rbrace )\cdot\left\|z_{2}\right\|(\left\lbrace a_{2}\right\rbrace ) \big) \Bigg)\\
				=&\left\|z_{1}\otimes z_{2}\right\|(\gamma),
			\end{align*}
			where the second equality holds since the set $\left\lbrace a\right\rbrace $ is singleton, and hence $a'=a$.
			\item If $z=(z_{1})$, then 
			\begin{align*}
				\sum\limits_{a \in \gamma }    \Big( \left\|(z_{1})\right\| (\left\lbrace a\right\rbrace) \Big)=&\sum\limits_{a \in \gamma } \Big( \left\|z_{1}\right\|(\left\lbrace a\right\rbrace)  \Big)\\
				=& \left\|z_{1}\right\|(\gamma) \\
				=& \left\|(z_{1})\right\|(\gamma),
			\end{align*}
			where the second equality holds by induction, and our proof is completed.
		\end{itemize}  
		Therefore, for any $z\in wAI(P)$, we proved that $\sum\limits_{a \in \gamma } \Big( \left\|z\right\|(\left\lbrace a\right\rbrace)  \Big)=\left\|z\right\|(\gamma)$.\qed
	\end{prof*}
	
	\begin{prop}\label{propeties_walg_in}
		Let $\overline{z_{1}}, \overline{z_{2}}, \overline{z_{3}}\in wAI(P)/\equiv$. Then
		\begin{multicols}{2}
			\begin{enumerate}[label=\roman*)]
				\item $(\overline{z_{1}}\oplus \overline{z_{2}})\oplus \overline{z_{3}} = \overline{z_{1}}\oplus (\overline{z_{2}}\oplus \overline{z_{3}})$
				\item $\overline{z_{1}}\oplus \overline{z_{2}} = \overline{z_{2}}\oplus \overline{z_{1}}$
				\item $\overline{z_{1}}\oplus \overline{z_{1}} = \overline{z_{1}}$
				\item $\overline{z_{1}}\oplus \overline{0} = \overline{z_{1}}$
				\item $(\overline{z_{1}}\otimes \overline{z_{2}})\otimes \overline{z_{3}} = \overline{z_{1}}\otimes (\overline{z_{2}}\otimes \overline{z_{3}})$
				\item $\overline{z_{1}}\otimes \overline{z_{2}} = \overline{z_{2}}\otimes \overline{z_{1}}$
				\item $\overline{z_{1}}\otimes \overline{1} = \overline{z_{1}}$
				\item $\overline{z_{1}}\otimes \overline{0} = \overline{0}$
				\item $\overline{z_{1}}\otimes (\overline{z_{2}}\oplus \overline{z_{3}}) = (\overline{z_{1}}\otimes \overline{z_{2}})\oplus (\overline{z_{1}}\otimes \overline{z_{3}})$
				\item $(\overline{z_{1}}\oplus \overline{z_{2}})\otimes \overline{z_{3}} = (\overline{z_{1}}\otimes \overline{z_{3}})\oplus (\overline{z_{2}}\otimes \overline{z_{3}}).$
			\end{enumerate}
		\end{multicols}
	\end{prop}
	\begin{prof*}
		Let $\gamma\in \Gamma(P)$. By Remark \ref{re1}, we prove the above equalities for $\gamma\in \Gamma(P)\setminus\left\lbrace  \emptyset\right\rbrace $.
		\begin{enumerate}[label=\roman*)]
			\item \begin{align*}
				&(\overline{z_{1}}\oplus \overline{z_{2}})\oplus \overline{z_{3}}=(\overline{z_{1}\oplus z_{2}}) \oplus \overline{z_{3}}=\overline{(z_{1}\oplus z_{2})} \oplus \overline{z_{3}}=\overline{(z_{1}\oplus z_{2})\oplus z_{3}}\\
				=& \left\lbrace z\in wAI(P) \mid z\equiv (z_{1}\oplus z_{2})\oplus z_{3} \right\rbrace \\
				=& \left\lbrace z\in wAI(P) \mid \left\|z\right\|(\gamma)=\left\| (z_{1}\oplus z_{2})\oplus z_{3}\right\|(\gamma) \ \text{for every} \ \gamma\in \Gamma(P) \right\rbrace \\
				=& \bigg\{ z\in wAI(P) \mid \left\|z\right\|(\gamma)=\sum\limits_{a \in \gamma }    \big(\left\|z_{1}\oplus z_{2}\right\|(\left\lbrace a\right\rbrace )+ \left\|z_{3}\right\|(\left\lbrace a\right\rbrace ) \big) \ \text{for every} \\
				&\gamma\in \Gamma(P) \bigg\} \\
				=& \bigg\{ z\in wAI(P) \mid \left\|z\right\|(\gamma)=\sum\limits_{a \in \gamma }    \Big( \sum\limits_{a' \in \left\lbrace a\right\rbrace   } \big( \left\|z_{1}\right\|(\left\lbrace a'\right\rbrace )+ \left\|z_{2}\right\|(\left\lbrace a'\right\rbrace ) \big)+ \left\|z_{3}\right\|(\left\lbrace a\right\rbrace ) \Big) \\ 
				& \text{for every} \ \gamma\in \Gamma(P) \bigg\} \\
				=& \bigg\{ z\in wAI(P) \mid \left\|z\right\|(\gamma)=\sum\limits_{a \in \gamma }   \Big(  \big( \left\|z_{1}\right\|(\left\lbrace a\right\rbrace )+ \left\|z_{2}\right\|(\left\lbrace a\right\rbrace ) \big)+ \left\|z_{3}\right\|(\left\lbrace a\right\rbrace ) \Big) \\ 
				&\ \text{for every} \ \gamma\in \Gamma(P) \bigg\} \\
				=& \bigg\{ z\in wAI(P) \mid \left\|z\right\|(\gamma)=\sum\limits_{a \in \gamma }   \Big(  \left\|z_{1}\right\|(\left\lbrace a\right\rbrace )+ \big( \left\|z_{2}\right\|(\left\lbrace a\right\rbrace ) + \left\|z_{3}\right\|(\left\lbrace a\right\rbrace ) \big) \Big) \\ 
				&\ \text{for every} \ \gamma\in \Gamma(P) \bigg\} \\
				=& \bigg\{ z\in wAI(P) \mid \left\|z\right\|(\gamma)=\sum\limits_{a \in \gamma }    \big(\left\|z_{1}\right\|(\left\lbrace a\right\rbrace ) + \left\|z_{2}\oplus z_{3}\right\|(\left\lbrace a\right\rbrace ) \big) \ \text{for every} \\
				& \gamma\in \Gamma(P) \bigg\} \\
				=& \left\lbrace z\in wAI(P) \mid \left\|z\right\|(\gamma)=\left\| z_{1}\oplus (z_{2}\oplus z_{3})\right\|(\gamma) \ \text{for every} \ \gamma\in \Gamma(P) \right\rbrace \\
				=& \left\lbrace z\in wAI(P) \mid z\equiv z_{1}\oplus (z_{2}\oplus z_{3}) \right\rbrace \\
				=&\overline{z_{1}\oplus (z_{2}\oplus z_{3})}\\
				=&\overline{z_{1}}\oplus \overline{(z_{2} \oplus z_{3})}\\
				=&\overline{z_{1}}\oplus (\overline{z_{2} \oplus z_{3}})\\
				=&\overline{z_{1}}\oplus (\overline{z_{2}}\oplus \overline{z_{3}})
			\end{align*}
			where the eighth equality holds since $\left\lbrace a\right\rbrace$ is a singleton and hence $a'\in \left\lbrace a\right\rbrace $ implies that $a'=a$. The ninth equality holds by the associativity property of `` $+$'' in semiring $K$.
			\item \begin{align*}	
				&\overline{z_{1}}\oplus \overline{z_{2}}=\overline{z_{1}\oplus z_{2}} = \left\lbrace z\in wAI(P) \mid z\equiv z_{1}\oplus z_{2} \right\rbrace \\
				=& \left\lbrace z\in wAI(P) \mid \left\|z\right\|(\gamma)=\left\| z_{1}\oplus z_{2}\right\|(\gamma) \ \text{for every} \ \gamma\in \Gamma(P) \right\rbrace \\
				=& \bigg\{ z\in wAI(P) \mid \left\|z\right\|(\gamma)=\sum\limits_{a \in \gamma }    \big(\left\|z_{1}\right\|(\left\lbrace a\right\rbrace )+ \left\|z_{2}\right\|(\left\lbrace a\right\rbrace ) \big) \ \text{for every} \ \gamma\in \Gamma(P) \bigg\} \\
				=& \bigg\{ z\in wAI(P) \mid \left\|z\right\|(\gamma)=\sum\limits_{a \in \gamma }    \big(\left\|z_{2}\right\|(\left\lbrace a\right\rbrace )+ \left\|z_{1}\right\|(\left\lbrace a\right\rbrace ) \big) \ \text{for every} \ \gamma\in \Gamma(P) \bigg\} \\
				=& \left\lbrace z\in wAI(P) \mid \left\|z\right\|(\gamma)=\left\| z_{2}\oplus z_{1}\right\|(\gamma) \ \text{for every} \ \gamma\in \Gamma(P) \right\rbrace \\
				=& \left\lbrace z\in wAI(P) \mid z\equiv z_{2}\oplus z_{1} \right\rbrace \\
				=&\overline{z_{2}\oplus z_{1}}\\
				=&\overline{z_{2}}\oplus \overline{z_{1}}	
			\end{align*} 
			where the fifth equality holds by the commutativity property of `` $+$'' in semiring $K$.
			\item \begin{align*}
				&\overline{z_{1}}\oplus \overline{z_{1}}=\overline{z_{1}\oplus z_{1}} = \left\lbrace z\in wAI(P) \mid z\equiv z_{1}\oplus z_{1} \right\rbrace \\
				=& \left\lbrace z\in wAI(P) \mid \left\|z\right\|(\gamma)=\left\| z_{1}\oplus z_{1}\right\|(\gamma) \ \text{for every} \ \gamma\in \Gamma(P) \right\rbrace \\
				=& \bigg\{ z\in wAI(P) \mid \left\|z\right\|(\gamma)=\sum\limits_{a \in \gamma }    \big(\left\|z_{1}\right\|(\left\lbrace a\right\rbrace )+ \left\|z_{1}\right\|(\left\lbrace a\right\rbrace ) \big) \ \text{for every} \  \gamma\in \Gamma(P) \bigg\} \\
				=& \bigg\{ z\in wAI(P) \mid \left\|z\right\|(\gamma)=\sum\limits_{a \in \gamma }    \big(\left\|z_{1}\right\|(\left\lbrace a\right\rbrace ) \big) \ \text{for every} \ \gamma\in \Gamma(P) \bigg\} \\
				=& \left\lbrace z\in wAI(P) \mid \left\|z\right\|(\gamma)=\left\| z_{1}\right\|(\gamma) \ \text{for every} \ \gamma\in \Gamma(P) \right\rbrace \\
				=& \left\lbrace z\in wAI(P) \mid z\equiv  z_{1} \right\rbrace\\
				=&\overline{z_{1}}
			\end{align*}	
			where the fifth and sixth equality hold since the semiring $K$ is idempotent and by a direct application of Lemma \ref{l1}, respectively.
			\item \begin{align*}
				&\overline{z_{1}}\oplus \overline{0}=\overline{z_{1}\oplus 0} = \left\lbrace z\in wAI(P) \mid z\equiv z_{1}\oplus 0 \right\rbrace \\
				=& \left\lbrace z\in wAI(P) \mid \left\|z\right\|(\gamma)=\left\| z_{1}\oplus 0\right\|(\gamma) \ \text{for every} \ \gamma\in \Gamma(P) \right\rbrace \\
				=& \bigg\{ z\in wAI(P) \mid \left\|z\right\|(\gamma)=\sum\limits_{a \in \gamma }    \big(\left\|z_{1}\right\|(\left\lbrace a\right\rbrace )+ \left\|0\right\|(\left\lbrace a\right\rbrace ) \big) \ \text{for every} \ \gamma\in \Gamma(P) \bigg\} \\
				=& \bigg\{ z\in wAI(P) \mid \left\|z\right\|(\gamma)=\sum\limits_{a \in \gamma }    \big(\left\|z_{1}\right\|(\left\lbrace a\right\rbrace ) + \hat{0} \big) \ \text{for every} \ \gamma\in \Gamma(P) \bigg\} \\
				=& \bigg\{ z\in wAI(P) \mid \left\|z\right\|(\gamma)=\sum\limits_{a \in \gamma }    \big(\left\|z_{1}\right\|(\left\lbrace a\right\rbrace )  \big) \ \text{for every} \ \gamma\in \Gamma(P) \bigg\} \\
				=& \left\lbrace z\in wAI(P) \mid \left\|z\right\|(\gamma)=\left\| z_{1}\right\|(\gamma) \ \text{for every} \ \gamma\in \Gamma(P) \right\rbrace \\
				=& \left\lbrace z\in wAI(P) \mid z\equiv  z_{1} \right\rbrace\\
				=&\overline{z_{1}}
			\end{align*}
			where the sixth and the seventh equality hold since `` $\hat{0}$'' is the neutral element of `` $+$'' in semiring $K$ and by applying Lemma \ref{l1}, respectively.
			\item \begin{align*}
				&(\overline{z_{1}}\otimes \overline{z_{2}})\otimes \overline{z_{3}}=(\overline{z_{1}\otimes z_{2}}) \otimes \overline{z_{3}}=\overline{(z_{1}\otimes z_{2})} \otimes \overline{z_{3}}=\overline{(z_{1}\otimes z_{2})\otimes z_{3}}\\
				=& \left\lbrace z\in wAI(P) \mid z\equiv (z_{1}\otimes z_{2})\otimes z_{3} \right\rbrace \\
				=& \left\lbrace z\in wAI(P) \mid \left\|z\right\|(\gamma)=\left\| (z_{1}\otimes z_{2})\otimes z_{3}\right\|(\gamma) \ \text{for every} \ \gamma\in \Gamma(P) \right\rbrace \\
				=& \bigg\{ z\in wAI(P) \mid \left\|z\right\|(\gamma)=\sum\limits_{a \in \gamma }    \Big(  \sum\limits_{a=a'\cup a_{3}} \big( \left\|z_{1}\otimes z_{2}\right\|(\left\lbrace a'\right\rbrace )\cdot \left\|z_{3}\right\|(\left\lbrace a_{3}\right\rbrace ) \big) \Big) \\
				& \text{for every} \ \gamma\in \Gamma(P) \bigg\} \\
				=& \Bigg\{ z\in wAI(P) \mid \left\|z\right\|(\gamma)=\sum\limits_{a \in \gamma }    \Bigg(  \sum\limits_{a=a'\cup a_{3}} \bigg( \sum\limits_{a'' \in \left\lbrace a'\right\rbrace } \Big( \sum\limits_{a''=a_{1}\cup a_{2}} \big( \left\|z_{1}\right\|(\left\lbrace a_{1}\right\rbrace )\cdot \\
				& \left\|z_{2}\right\|(\left\lbrace a_{2}\right\rbrace ) \big) \Big)\cdot \left\|z_{3}\right\|(\left\lbrace a_{3}\right\rbrace ) \bigg) \Bigg) \ \text{for every} \ \gamma\in \Gamma(P) \Bigg\} \\
				=& \bigg\{ z\in wAI(P) \mid \left\|z\right\|(\gamma)=\sum\limits_{a \in \gamma }    \bigg(  \sum\limits_{a=a'\cup a_{3}} \Big(   \sum\limits_{a'=a_{1}\cup a_{2}} \big( \left\|z_{1}\right\|(\left\lbrace a_{1}\right\rbrace) \cdot \left\|z_{2}\right\|(\left\lbrace a_{2}\right\rbrace ) \big)  \cdot \\
				& \left\|z_{3}\right\|(\left\lbrace a_{3}\right\rbrace ) \Big) \bigg) \ \text{for every} \ \gamma\in \Gamma(P) \bigg\} \\
				=& \bigg\{ z\in wAI(P) \mid \left\|z\right\|(\gamma)=\sum\limits_{a \in \gamma }    \bigg(  \sum\limits_{a=a_{1}\cup a_{2}\cup a_{3}}     \Big( \big(  \left\|z_{1}\right\|(\left\lbrace a_{1}\right\rbrace) \cdot \left\|z_{2}\right\|(\left\lbrace a_{2}\right\rbrace ) \big)  \cdot \\
				& \left\|z_{3}\right\|(\left\lbrace a_{3}\right\rbrace ) \Big)  \bigg) \ \text{for every} \ \gamma\in \Gamma(P) \bigg\} \\
				=& \bigg\{ z\in wAI(P) \mid \left\|z\right\|(\gamma)=\sum\limits_{a \in \gamma }    \bigg(  \sum\limits_{a=a_{1}\cup a_{2}\cup a_{3}}     \Big(  \left\|z_{1}\right\|(\left\lbrace a_{1}\right\rbrace) \cdot \big( \left\|z_{2}\right\|(\left\lbrace a_{2}\right\rbrace )  \cdot \\
				& \left\|z_{3}\right\|(\left\lbrace a_{3}\right\rbrace ) \big) \Big)  \bigg) \ \text{for every} \ \gamma\in \Gamma(P) \bigg\} \\
				=& \Bigg\{ z\in wAI(P) \mid \left\|z\right\|(\gamma)=\sum\limits_{a \in \gamma }    \Bigg(  \sum\limits_{a=a_{1}\cup a'''}   \bigg(  \left\|z_{1}\right\|(\left\lbrace a_{1}\right\rbrace) \cdot \Big( \sum\limits_{a'''=a_{2}\cup a_{3}} \big( \left\|z_{2}\right\|(\left\lbrace a_{2}\right\rbrace )  \cdot \\
				& \left\|z_{3}\right\|(\left\lbrace a_{3}\right\rbrace ) \big) \Big) \bigg)  \Bigg) \ \text{for every} \ \gamma\in \Gamma(P) \Bigg\} \\
				=& \bigg\{ z\in wAI(P) \mid \left\|z\right\|(\gamma)=\sum\limits_{a \in \gamma }    \Bigg(  \sum\limits_{a=a_{1}\cup a'''}   \bigg(  \left\|z_{1}\right\|(\left\lbrace a_{1}\right\rbrace) \cdot  \left\|z_{2}\otimes z_{3}\right\|(\left\lbrace a'''\right\rbrace )     \bigg)  \Bigg) \\
				& \text{for every} \ \gamma\in \Gamma(P) \bigg\} \\
				=& \left\lbrace z\in wAI(P) \mid \left\|z\right\|(\gamma)=\left\| z_{1}\otimes (z_{2}\otimes z_{3})\right\|(\gamma) \ \text{for every} \ \gamma\in \Gamma(P) \right\rbrace \\
				=& \left\lbrace z\in wAI(P) \mid z\equiv z_{1}\otimes (z_{2}\otimes z_{3}) \right\rbrace \\
				=&\overline{z_{1}\otimes (z_{2}\otimes z_{3})}\\
				=&\overline{z_{1}}\otimes \overline{(z_{2} \otimes z_{3})}\\
				=& \overline{z_{1}}\otimes (\overline{z_{2} \otimes z_{3}})\\
				=&\overline{z_{1}}\otimes (\overline{z_{2}}\otimes \overline{z_{3}})
			\end{align*}
			where the eighth equality holds since $a''\in \left\lbrace a'\right\rbrace $ implies that $a''=a'$. The tenth equality holds by the associativity property of `` $\cdot$'' in semiring $K$, and the eleventh equality holds by the distributivity property of `` $\cdot$'' over `` $+$'' in semiring $K$.
			\item \begin{align*}
				&\overline{z_{1}}\otimes \overline{z_{2}}=\overline{z_{1}\otimes z_{2}} = \left\lbrace z\in wAI(P) \mid z\equiv z_{1}\otimes z_{2} \right\rbrace \\
				=& \left\lbrace z\in wAI(P) \mid \left\|z\right\|(\gamma)=\left\| z_{1}\otimes z_{2}\right\|(\gamma) \ \text{for every} \ \gamma\in \Gamma(P) \right\rbrace \\
				=& \bigg\{ z\in wAI(P) \mid \left\|z\right\|(\gamma)=\sum\limits_{a \in \gamma } \Big(  \sum\limits_{a=a_{1}\cup a_{2}}  \big(\left\|z_{1}\right\|(\left\lbrace a_{1}\right\rbrace )\cdot \left\|z_{2}\right\|(\left\lbrace a_{2}\right\rbrace ) \big) \Big) \ \text{for every} \\
				& \gamma\in \Gamma(P) \bigg\} \\
				=& \bigg\{ z\in wAI(P) \mid \left\|z\right\|(\gamma)=\sum\limits_{a \in \gamma } \Big(  \sum\limits_{a=a_{2}\cup a_{1}}  \big(\left\|z_{2}\right\|(\left\lbrace a_{2}\right\rbrace )\cdot \left\|z_{1}\right\|(\left\lbrace a_{1}\right\rbrace ) \big) \Big) \ \text{for every} \\
				& \gamma\in \Gamma(P) \bigg\} \\
				=& \left\lbrace z\in wAI(P) \mid \left\|z\right\|(\gamma)=\left\| z_{2}\otimes z_{1}\right\|(\gamma) \ \text{for every} \ \gamma\in \Gamma(P) \right\rbrace \\
				=& \left\lbrace z\in wAI(P) \mid z\equiv z_{2}\otimes z_{1} \right\rbrace\\
				=&\overline{z_{2}\otimes z_{1}}\\
				=&\overline{z_{2}}\otimes \overline{z_{1}}	
			\end{align*}
			where the fifth equality holds since `` $\cdot$'' is commutative in semiring $K$.
			\item \begin{align*}
				&\overline{z_{1}}\otimes \overline{1}=\overline{z_{1}\otimes 1} = \left\lbrace z\in wAI(P) \mid z\equiv z_{1}\otimes 1 \right\rbrace \\
				=& \left\lbrace z\in wAI(P) \mid \left\|z\right\|(\gamma)=\left\| z_{1}\otimes 1\right\|(\gamma) \ \text{for every} \ \gamma\in \Gamma(P) \right\rbrace \\
				=& \bigg\{ z\in wAI(P) \mid \left\|z\right\|(\gamma)=\sum\limits_{a \in \gamma } \Big(  \sum\limits_{a=a_{1}\cup a_{2}}  \big(\left\|z_{1}\right\|(\left\lbrace a_{1}\right\rbrace )\cdot \left\|1\right\|(\left\lbrace a_{2}\right\rbrace ) \big) \Big) \ \text{for every} \\
				& \gamma\in \Gamma(P) \bigg\} \\
				=& \bigg\{ z\in wAI(P) \mid \left\|z\right\|(\gamma)=\sum\limits_{a \in \gamma } \big(  \left\|z_{1}\right\|(\left\lbrace a\right\rbrace )\cdot \left\|1\right\|(\left\lbrace \emptyset \right\rbrace )  \big) \ \text{for every} \ \gamma\in \Gamma(P) \bigg\} \\
				=& \bigg\{ z\in wAI(P) \mid \left\|z\right\|(\gamma)=\sum\limits_{a \in \gamma } \big(  \left\|z_{1}\right\|(\left\lbrace a\right\rbrace )\cdot \hat{1}  \big) \ \text{for every} \ \gamma\in \Gamma(P) \bigg\} \\
				=& \bigg\{ z\in wAI(P) \mid \left\|z\right\|(\gamma)=\sum\limits_{a \in \gamma } \big(  \left\|z_{1}\right\|(\left\lbrace a\right\rbrace ) \big) \ \text{for every} \ \gamma\in \Gamma(P) \bigg\} \\
				=& \left\lbrace z\in wAI(P) \mid \left\|z\right\|(\gamma)=\left\|  z_{1}\right\|(\gamma) \ \text{for every} \ \gamma\in \Gamma(P) \right\rbrace \\
				=& \left\lbrace z\in wAI(P) \mid z\equiv  z_{1} \right\rbrace\\
				=&\overline{z_{1}}	
			\end{align*}
			where the fifth equality holds since the only interactions set with non-zero weight for `` $1$'' is $\left\lbrace a_{2}\right\rbrace = \left\lbrace \emptyset\right\rbrace $. Hence, we obtain that $a_{1}=a$ and $a_{2}=\emptyset$. Also, the seventh equality holds since `` $\hat{1}$'' is the neutral element for `` $\cdot$'' in semiring $K$, while the eighth equality results by an application of Lemma \ref{l1}. 
			\item \begin{align*}
				&\overline{z_{1}}\otimes \overline{0}=\overline{z_{1}\otimes 0} = \left\lbrace z\in wAI(P) \mid z\equiv z_{1}\otimes 0 \right\rbrace \\
				=& \left\lbrace z\in wAI(P) \mid \left\|z\right\|(\gamma)=\left\| z_{1}\otimes 0\right\|(\gamma) \ \text{for every} \ \gamma\in \Gamma(P) \right\rbrace \\
				=& \bigg\{ z\in wAI(P) \mid \left\|z\right\|(\gamma)=\sum\limits_{a \in \gamma } \Big(  \sum\limits_{a=a_{1}\cup a_{2}}  \big(\left\|z_{1}\right\|(\left\lbrace a_{1}\right\rbrace )\cdot \left\|0\right\|(\left\lbrace a_{2}\right\rbrace ) \big) \Big) \ \text{for every} \\
				& \gamma\in \Gamma(P) \bigg\} \\
				=& \bigg\{ z\in wAI(P) \mid \left\|z\right\|(\gamma)=\sum\limits_{a \in \gamma } \Big(  \sum\limits_{a=a_{1}\cup a_{2}}  \big(\left\|z_{1}\right\|(\left\lbrace a_{1}\right\rbrace )\cdot \hat{0} \big) \Big) \ \text{for every} \ \gamma\in \Gamma(P) \bigg\} \\
				=& \bigg\{ z\in wAI(P) \mid \left\|z\right\|(\gamma)=\sum\limits_{a \in \gamma } \Big(  \sum\limits_{a=a_{1}\cup a_{2}}  \big( \hat{0} \big) \Big) \ \text{for every} \ \gamma\in \Gamma(P) \bigg\} \\
				=& \bigg\{ z\in wAI(P) \mid \left\|z\right\|(\gamma)=\hat{0} \ \text{for every} \ \gamma\in \Gamma(P) \bigg\} \\
				=& \left\lbrace z\in wAI(P) \mid \left\|z\right\|(\gamma)=\left\|  0\right\|(\gamma) \ \text{for every} \ \gamma\in \Gamma(P) \right\rbrace \\
				=& \left\lbrace z\in wAI(P) \mid z\equiv  0 \right\rbrace\\
				=&\overline{0}	
			\end{align*}
			where the sixth equality holds since `` $\hat{0}$'' is the absorbing element for `` $\cdot$'' in semiring $K$.
			\item \label{w-dist} \begin{align*}
				&\overline{z_{1}}\otimes (\overline{z_{2}}\oplus \overline{z_{3}})=\overline{z_{1}}\otimes (\overline{z_{2} \oplus z_{3}})=\overline{z_{1}}\otimes \overline{(z_{2} \oplus z_{3})}=\overline{z_{1}\otimes (z_{2}\oplus z_{3})}\\
				=& \left\lbrace z\in wAI(P) \mid z\equiv z_{1}\otimes (z_{2} \oplus z_{3}) \right\rbrace \\
				=& \left\lbrace z\in wAI(P) \mid \left\|z\right\|(\gamma)=\left\| z_{1}\otimes (z_{2} \oplus z_{3})\right\|(\gamma) \ \text{for every} \ \gamma\in \Gamma(P) \right\rbrace \\
				=& \bigg\{ z\in wAI(P) \mid \left\|z\right\|(\gamma)=\sum\limits_{a \in \gamma }    \Big(  \sum\limits_{a=a_{1}\cup a'} \big( \left\|z_{1}\right\|(\left\lbrace a_{1}\right\rbrace )\cdot \left\|z_{2}\oplus z_{3}\right\|(\left\lbrace a'\right\rbrace ) \big) \Big) \\
				& \text{for every} \ \gamma\in \Gamma(P) \bigg\} \\
				=& \bigg\{ z\in wAI(P) \mid \left\|z\right\|(\gamma)=\sum\limits_{a \in \gamma }    \bigg(  \sum\limits_{a=a_{1}\cup a'} \Big( \left\|z_{1}\right\|(\left\lbrace a_{1}\right\rbrace )\cdot \Big( \sum\limits_{a'' \in \left\lbrace a'\right\rbrace  } \big( \left\|z_{2}\right\|(\left\lbrace a''\right\rbrace )+ \\
				& \left\|z_{3}\right\|(\left\lbrace a''\right\rbrace ) \big) \Big) \Big) \bigg)\ \text{for every} \ \gamma\in \Gamma(P) \bigg\} \\
				=& \bigg\{ z\in wAI(P) \mid \left\|z\right\|(\gamma)=\sum\limits_{a \in \gamma } \bigg(  \sum\limits_{a=a_{1}\cup a'} \Big( \left\|z_{1}\right\|(\left\lbrace a_{1}\right\rbrace )\cdot  \Big( \left\|z_{2}\right\|(\left\lbrace a'\right\rbrace )+ \\
				& \left\|z_{3}\right\|(\left\lbrace a'\right\rbrace )  \Big) \Big) \bigg) \ \text{for every} \ \gamma\in \Gamma(P) \bigg\} \\
				=& \bigg\{ z\in wAI(P) \mid \left\|z\right\|(\gamma)=\sum\limits_{a \in \gamma }    \bigg(  \sum\limits_{a=a_{1}\cup a'} \Big( \big( \left\|z_{1}\right\|(\left\lbrace a_{1}\right\rbrace )\cdot  \left\|z_{2}\right\|(\left\lbrace a'\right\rbrace ) \big) + \\
				& \big( \left\|z_{1}\right\|(\left\lbrace a_{1}\right\rbrace )\cdot  \left\|z_{3}\right\|(\left\lbrace a'\right\rbrace )  \big) \Big) \bigg) \ \text{for every} \ \gamma\in \Gamma(P) \bigg\} \\
				=& \bigg\{ z\in wAI(P) \mid \left\|z\right\|(\gamma)=\sum\limits_{a \in \gamma }  \bigg(  \sum\limits_{a=a_{1}\cup a'} \Big(  \left\|z_{1}\right\|(\left\lbrace a_{1}\right\rbrace )\cdot  \left\|z_{2}\right\|(\left\lbrace a'\right\rbrace )  \Big)  + \\
				& \sum\limits_{a=a_{1}\cup a'} \Big( \left\|z_{1}\right\|(\left\lbrace a_{1}\right\rbrace )\cdot  \left\|z_{3}\right\|(\left\lbrace a'\right\rbrace )   \Big) \bigg)  \ \text{for every} \ \gamma\in \Gamma(P) \bigg\} \\
				=& \bigg\{ z\in wAI(P) \mid \left\|z\right\|(\gamma)=\sum\limits_{a \in \gamma }  \bigg( \left\|z_{1}\otimes  z_{2}\right\|(\left\lbrace a\right\rbrace ) + \left\|z_{1}\otimes  z_{3}\right\|(\left\lbrace a\right\rbrace ) \bigg)  \ \text{for} \\
				& \text{every} \ \gamma\in \Gamma(P) \bigg\} \\
				=& \left\lbrace z\in wAI(P) \mid \left\|z\right\|(\gamma)=\left\| (z_{1}\otimes z_{2})\oplus (z_{1}\otimes z_{3}) \right\|(\gamma) \ \text{for every} \ \gamma\in \Gamma(P)\right\rbrace \\
				=& \left\lbrace z\in wAI(P) \mid z\equiv (z_{1}\otimes z_{2})\oplus (z_{1}\otimes z_{3}) \right\rbrace \\
				=&\overline{(z_{1}\otimes z_{2})\oplus (z_{1}\otimes z_{3})}\\
				=&\overline{(z_{1}\otimes z_{2})}\oplus \overline{(z_{1} \otimes z_{3})}\\
				=&(\overline{z_{1}\otimes z_{2}})\oplus (\overline{z_{1} \otimes z_{3}})\\
				=&(\overline{z_{1}}\otimes \overline{z_{2}}) \oplus (\overline{z_{1}}\otimes \overline{z_{3}})
			\end{align*}
			where the ninth equality holds by distributivity of `` $\cdot$'' over `` $+$'' in semiring $K$, and the tenth equality results by associativity property of `` $+$'' in semiring $K$.
			\item The equality is proved as in the previous case.
		\end{enumerate} 
		Thus, our proof is completed.\qed
	\end{prof*}

	\begin{coro}\label{walg_semi}
		The structure $(wAI(P)/\equiv,\oplus,\otimes, \bar{0},\bar{1})$ is a commutative and idempotent semiring.
	\end{coro}
	
	\subsection{Examples of coordination schemes encoded in $wAI(P)$}
	
	\noindent Next we present several coordination schemes in the weighted setup and we encode the cost of their implementation by the weighted Algebra of Interactions. 
	\hide{\par Let $z$ be the $wAI(P)$ element describing a scheme in the weighted setup and consider an interactions set $\gamma \in \Gamma(P)$. In order to derive the semantics of the weighted terms of $z$ that are connected by the ``$\otimes$'' operator, we need to obtain the analyses $a_1\cup a_2$ of every interaction $a$ of the given $\gamma$, and in turn the analyses of the obtained interactions for the rest ``$\oplus$'' and ``$\otimes$'' operators. We apply these computation steps as many times as needed to the operators comprising the corresponding $wAI(P)$ element $z$.} 
	\par Let $z$ be the $wAI(P)$ element describing a scheme in the weighted setup and consider an interactions set $\gamma \in \Gamma(P)$. In order to derive the semantics of the $z$, we use for every $a\in \gamma$ two tables, namely the primary and the auxiliary tables. In the primary tables, we derive the overall weight of element $z$ using the corresponding  auxiliary tables. Primary and auxiliary tables contain in the first row the $wAI(P)$ element $z$ and the $wAI(P)$ terms of $z$, respectively, whose weight is calculated on a given interaction $a\in \gamma$ as indicated by the label of the tables. The primary tables, referred to as tables, have four columns. The first one includes all the possible analyses $a_1\cup a_2$ of the given $a\in \gamma$, while the second and third column contain the semantics of the respective $wAI(P)$ terms comprising $z$. In the fourth column we compute the weight of the weighted synchronization applied on the derived $wAI(P)$ terms. Then in the last row of the tables we obtain the overall weight of $z$ on the particular $\lbrace a\rbrace$. The auxiliary tables, are structured as the primary ones with the difference that they return the weight on the $wAI(P)$ terms of $z$ appearing on the third column of the primary tables. The auxiliary tables are also referred to as tables and are identified by their labels.
	
	\par For representation reasons, all the tables used in our examples, are found in the Appendix \ref{appen}. There we present firstly the primary tables of the $wAI(P)$ elements for every interaction $a\in \gamma$, and then, the auxiliary tables which are necessary for our calculations. Also, when a $wAI(P)$ element contains the ``$\oplus$'' operator and its set of interactions is singleton, we compute directly its corresponding weight within its table, without considering further tables.
	\par It should be clear that the analysis of $z$ in the respective $wAI(P)$ elements is not unique, and hence it is chosen arbitrarily. In what follows, we usually split $z$ in the terms occurring at the left and right of the first weighted synchronization operator. Obviously, for any other analyses of $z$ we derive the same overall weight on the given interactions set $\gamma$. Also, we choose to compute the weight of each $wAI(P)$ element on the interactions set $\gamma \in \Gamma(P)$ that contains only the interactions of the scheme. Obviously, by Definition \ref{sem-wAI(P)}, for any other $\gamma \in \Gamma(P)$ our $wAI(P)$ algebra returns the expected weight.

	\begin{examp}\label{eq:4}
		Consider an architecture with a sender and two receivers, each having a single port $s$ and $r_{1}, r_{2}$, for sending and receiving messages, respectively. Hence,  $P=\left\lbrace  s,r_{1}, r_{2}\right\rbrace $, and let $k_{s}, k_{r_{1}}, k_{r_{2}}$, denote the weights of the ports $s, r_{1}, r_{2}$, respectively. We consider the coordination schemes of Rendezvous, Broadcast, Atomic Broadcast and Causality Chain for defining the communication among the three components of the architecture. Next, we formalize each of the aforementioned schemes in the weighted setup by a $wAI(P)$ element $z$.   \\ 
		
		\noindent	\underline{\textbf{Weighted Rendezvous}}: This coordination scheme requires strong synchronization between the sender and each of the two receivers. Hence, it consists of a single interaction involving all ports, namely $sr_1r_2$. The $wAI(P)$ element describing the weighted Rendezvous is
		$$z=s\otimes r_{1}\otimes r_{2}.$$
		We let $\gamma=\left\lbrace \left\lbrace s,r_{1},r_{2}\right\rbrace \right\rbrace \in\Gamma(P)$ and we compute the weight of $z$ on $\gamma$ by applying the semantics of $wAI(P)$.
		In Table \ref{tab-re1}, we list in the first column all the possible analyses for $a=\left\lbrace s,r_{1},r_{2}\right\rbrace \in \gamma$ such that $a=a_{1}\cup a_{2}$, in the second and third column we compute the semantics of the $wAI(P)$ elements $s$ and $r_{1}\otimes r_{2}$, respectively, and in the last column we derive the cost of the respective weighted synchronization. The semantics for $r_{1}\otimes r_{2}$ are obtained by the auxiliary Tables \ref{tab-re2}-\ref{tab-re9}, for each set $a_{2}$ where $a_{2}=a_{2,1}\cup a_{2,2}$. Then the last row in each table serves for computing the sum of the weights occurring in the last column. Specifically, in Table \ref{tab-re1}, the resulting value $k_{s}\cdot k_{r_{1}}\cdot k_{r_{2}}$, which is highlighted, corresponds to the overall weight of Rendezvous scheme on the given $\gamma$. The weight of $z\in wAI(P)$ on $\gamma $ is computed as follows:
		\begin{align*}
			&\left\|s\otimes r_{1}\otimes r_{2} \right\|(\gamma)\\ 
			=&\sum\limits_{a\in \gamma} \Big(  \sum\limits_{a=a_{1}\cup a_{2}} \big(  \left\|s\right\|(\left\lbrace a_{1}\right\rbrace) \cdot \left\|r_{1}\otimes r_{2}\right\|(\left\lbrace a_{2}\right\rbrace) \big) \Big)\\
			=&\sum\limits_{a\in \gamma} \Bigg(  \sum\limits_{a=a_{1}\cup a_{2}} \bigg(  \left\|s\right\|(\left\lbrace a_{1}\right\rbrace) \cdot \bigg( \sum\limits_{a'\in \left\lbrace a_{2}\right\rbrace } \Big( \sum\limits_{a'=a_{2,1}\cup a_{2,2}} \big( \left\|r_{1}\right\|(\left\lbrace a_{2,1}\right\rbrace)\cdot \left\|r_{2}\right\|(\left\lbrace a_{2,2}\right\rbrace) \big) \Big) \bigg) \bigg) \Bigg)\\
			=&\sum\limits_{a\in \gamma} \Bigg(  \sum\limits_{a=a_{1}\cup a_{2}} \bigg(  \left\|s\right\|(\left\lbrace a_{1}\right\rbrace) \cdot \Big( \sum\limits_{a_{2}=a_{2,1}\cup a_{2,2}} \big( \left\|r_{1}\right\|(\left\lbrace a_{2,1}\right\rbrace)\cdot \left\|r_{2}\right\|(\left\lbrace a_{2,2}\right\rbrace) \big)  \Big) \bigg) \Bigg)\\
			=&k_{s}\cdot k_{r_{1}}\cdot k_{r_{2}}.
		\end{align*}
		
		\noindent For instance, let us consider the Fuzzy semiring $F=(\left[ 0,1\right] , \mathrm{max}, \mathrm{min}, 0, 1)$. Then the value 
		\begin{align*}
			&\left\|s\otimes r_{1}\otimes r_{2}\right\|(\gamma) \\
			=&\underset{a \in \gamma }{\mathrm{max}} \Bigg( \underset{a=a_{1}\cup a_{2}}{\mathrm{max}} \bigg( \mathrm{min}\Big(\left\|s\right\|(\left\lbrace a_{1}\right\rbrace ), \underset{a_{2}=a_{2,1}\cup a_{2,2} }{\mathrm{max}} \big( \mathrm{min}( \left\|r_{1}\right\|(\left\lbrace a_{2,1}\right\rbrace ) , \left\|r_{2}\right\|(\left\lbrace a_{2,2}\right\rbrace ) ) \big) \Big) \bigg) \Bigg) \\ 
			=& \underset{a \in \gamma }{\mathrm{max}} \Bigg( \underset{a=a_{1}\cup a_{2}}{\mathrm{max}} \bigg( \underset{a_{2}=a_{2,1}\cup a_{2,2}}{\mathrm{max}} \Big( \mathrm{min}\big(\left\|s\right\|(\left\lbrace a_{1}\right\rbrace ),\mathrm{min}( \left\|r_{1}\right\|(\left\lbrace a_{2,1}\right\rbrace ) , \left\|r_{2}\right\|(\left\lbrace a_{2,2}\right\rbrace ) ) \big) \Big) \bigg) \Bigg) \\
			=& \underset{a \in \gamma }{\mathrm{max}} \bigg( \underset{a=a_{1}\cup a_{2,1}\cup a_{2,2}}{\mathrm{max}} \Big(  \mathrm{min}\big(\left\|s\right\|(\left\lbrace a_{1}\right\rbrace ), \left\|r_{1}\right\|(\left\lbrace a_{2,1}\right\rbrace ) , \left\|r_{2}\right\|(\left\lbrace a_{2,2}\right\rbrace )  \big) \Big) \bigg)
		\end{align*}
		returns for the Rendezvous scheme the maximum of the minimum weights associated with each port. 
		\par Let now $\gamma=\left\lbrace \left\lbrace s,r_{1},r_{2}\right\rbrace, \left\lbrace s, r_{2}\right\rbrace \right\rbrace \in\Gamma(P)$. We use the primary Tables \ref{tab-re1} and \ref{tab-re10} as well as the auxiliary Tables \ref{tab-re2}-\ref{tab-re9}. Then the weight of $z\in wAI(P)$ on $\gamma $ occurs as follows:
		\begin{align*}
			&\left\|s\otimes r_{1}\otimes r_{2} \right\|(\gamma)\\ 
			=&\sum\limits_{a\in \gamma} \Big(  \sum\limits_{a=a_{1}\cup a_{2}} \big(  \left\|s\right\|(\left\lbrace a_{1}\right\rbrace) \cdot \left\|r_{1}\otimes r_{2}\right\|(\left\lbrace a_{2}\right\rbrace) \big) \Big)\\
			=&\sum\limits_{a\in \gamma} \Bigg(  \sum\limits_{a=a_{1}\cup a_{2}} \bigg(  \left\|s\right\|(\left\lbrace a_{1}\right\rbrace) \cdot \bigg( \sum\limits_{a'\in \left\lbrace a_{2}\right\rbrace } \Big( \sum\limits_{a'=a_{2,1}\cup a_{2,2}} \big( \left\|r_{1}\right\|(\left\lbrace a_{2,1}\right\rbrace)\cdot \left\|r_{2}\right\|(\left\lbrace a_{2,2}\right\rbrace) \big) \Big) \bigg) \bigg) \Bigg)\\
			=&\sum\limits_{a\in \gamma} \Bigg(  \sum\limits_{a=a_{1}\cup a_{2}} \bigg(  \left\|s\right\|(\left\lbrace a_{1}\right\rbrace) \cdot  \Big( \sum\limits_{a_{2}=a_{2,1}\cup a_{2,2}} \big( \left\|r_{1}\right\|(\left\lbrace a_{2,1}\right\rbrace)\cdot \left\|r_{2}\right\|(\left\lbrace a_{2,2}\right\rbrace) \big)  \Big) \bigg) \Bigg)\\
			=&(k_{s}\cdot k_{r_{1}}\cdot k_{r_{2}}) + \hat{0} \\
			=&k_{s}\cdot k_{r_{1}}\cdot k_{r_{2}}. 
		\end{align*} 
		Hence, in the case that $\gamma$ includes further interactions than the expected ones, then the given $wAI(P)$ element returns the expected weight for the scheme. Moreover, let us consider the case that $\gamma$ consists only of interactions that do not result from the given Rendezvous scheme, and let $\gamma=\left\lbrace \left\lbrace s,r_{2}\right\rbrace \right\rbrace \in\Gamma(P)$. Then by Table \ref{tab-re10} the weight of $z\in wAI(P)$ on $\gamma $ equals to $\hat{0}$. The cases for $\gamma=\left\lbrace \left\lbrace s,r_{1},r_{2}\right\rbrace, \left\lbrace s, r_{2}\right\rbrace \right\rbrace$ and $\gamma=\left\lbrace \left\lbrace s,r_{2}\right\rbrace \right\rbrace $ verify the robustness of our $wAI(P)$ and justify the interpretation of the presented semantics.
		\hide{\begin{align*}
				&\left\|s\otimes r_{1}\otimes r_{2} \right\|(\gamma)\\ 
				=&\sum\limits_{a\in \gamma} \Big(  \sum\limits_{a=a_{1}\cup a_{2}} \big(  \left\|s\right\|(\left\lbrace a_{1}\right\rbrace) \cdot \left\|r_{1}\otimes r_{2}\right\|(\left\lbrace a_{2}\right\rbrace) \big) \Big)\\
				=&\sum\limits_{a\in \gamma} \Bigg(  \sum\limits_{a=a_{1}\cup a_{2}} \bigg(  \left\|s\right\|(\left\lbrace a_{1}\right\rbrace) \cdot \bigg( \sum\limits_{a'\in \left\lbrace a_{2}\right\rbrace } \Big( \sum\limits_{a'=a_{2,1}\cup a_{2,2}} \big( \left\|r_{1}\right\|(\left\lbrace a_{2,1}\right\rbrace)\cdot \left\|r_{2}\right\|(\left\lbrace a_{2,2}\right\rbrace) \big) \Big) \bigg) \bigg) \Bigg)\\
				=&\sum\limits_{a\in \gamma} \Bigg(  \sum\limits_{a=a_{1}\cup a_{2}} \bigg(  \left\|s\right\|(\left\lbrace a_{1}\right\rbrace) \cdot \Big( \sum\limits_{a_{2}=a_{2,1}\cup a_{2,2}} \big( \left\|r_{1}\right\|(\left\lbrace a_{2,1}\right\rbrace)\cdot \left\|r_{2}\right\|(\left\lbrace a_{2,2}\right\rbrace) \big)  \Big) \bigg) \Bigg)\\
				=&\hat{0}. \tag*{\qed}
		\end{align*} }
		\\
		
		\noindent \underline{\textbf{Weighted Broadcast}}: This coordination scheme allows executing all interactions involving the sender and any subset of receivers, possibly the empty one. Hence, each permissible interaction should contain the port $s$. Thus, all the possible interactions are $s, sr_{1}, sr_{2}, sr_{1}r_{2}$. The $wAI(P)$ element $z$ describing weighted Broadcast is
		$$z=s\otimes (1\oplus r_{1})\otimes (1\oplus r_{2}) .$$
		We set $\gamma=\left\lbrace  \left\lbrace  s\right\rbrace , \left\lbrace  s,r_{1}\right\rbrace , \left\lbrace  s,r_{2}\right\rbrace , \left\lbrace  s,r_{1},r_{2}\right\rbrace  \right\rbrace\in\Gamma(P)$ and we compute the weight of $z$ on $\gamma$. For this, we need to obtain the analyses and the respective weight for each of the four interactions occurring in $\gamma$. We firstly consider the interaction $a=\left\lbrace s\right\rbrace \in \gamma$ for which we obtain Table \ref{tab-bra}. Then we derive the weight of $z$  for $a=\left\lbrace s,r_{1}\right\rbrace \in \gamma$ and $a=\left\lbrace s,r_{2}\right\rbrace \in \gamma$, presented in Tables \ref{tab-brb} and \ref{tab-brc}, respectively. Finally, we get the weight of $z$ for $a=\left\lbrace s,r_{1}, r_{2}\right\rbrace \in \gamma$, as shown in Table \ref{tab-brd}. The above weights are derived using the auxiliary Tables \ref{tab-brd1}-\ref{tab-brd8}. The overall weight of $z$ on $\gamma$ is computed as follows:
		\begin{align*}
			&\big\|s\otimes (1\oplus r_{1})\otimes (1\oplus r_{2}) \big\|(\gamma)\\
			=& \sum\limits_{a \in \gamma } \Big( \sum\limits_{a=a_{1}\cup a_{2}} \big( \left\|s\right\|(\left\lbrace a_{1}\right\rbrace )\cdot \left\| (1\oplus r_{1})\otimes (1\oplus r_{2})\right\|(\left\lbrace a_{2}\right\rbrace ) \big) \Big)\\
			=& \sum\limits_{a \in \gamma } \Bigg( \sum\limits_{a=a_{1}\cup a_{2}} \bigg( \left\|s\right\|(\left\lbrace a_{1}\right\rbrace )\cdot \bigg( \sum\limits_{a' \in \left\lbrace a_{2}\right\rbrace  } \Big( \sum\limits_{a'=a_{2,1}\cup a_{2,2}} \big( \left\| 1\oplus r_{1}\right\|(\left\lbrace a_{2,1}\right\rbrace )\cdot \\
			& \left\|1\oplus r_{2}\right\|(\left\lbrace a_{2,2}\right\rbrace ) \big) \Big) \bigg) \bigg) \Bigg)\\
			=& \sum\limits_{a \in \gamma } \Bigg( \sum\limits_{a=a_{1}\cup a_{2}} \bigg( \left\|s\right\|(\left\lbrace a_{1}\right\rbrace )\cdot  \Big( \sum\limits_{a_{2}=a_{2,1}\cup a_{2,2}} \big( \left\| 1\oplus r_{1}\right\|(\left\lbrace a_{2,1}\right\rbrace )\cdot \left\|1\oplus r_{2}\right\|(\left\lbrace a_{2,2}\right\rbrace ) \big) \Big) \bigg)\Bigg)\\
			=& \sum\limits_{a \in \gamma } \Bigg( \sum\limits_{a=a_{1}\cup a_{2}} \bigg( \left\|s\right\|(\left\lbrace a_{1}\right\rbrace )\cdot  \bigg( \sum\limits_{a_{2}=a_{2,1}\cup a_{2,2}} \Big( \sum\limits_{a'' \in \left\lbrace a_{2,1}\right\rbrace  } \big( \left\| 1\right\|(\left\lbrace a''\right\rbrace )+ \left\|r_{1}\right\|(\left\lbrace a''\right\rbrace ) \big) \cdot \\
			& \Big(\sum\limits_{a''' \in \left\lbrace a_{2,2}\right\rbrace  } \big( \left\| 1\right\|(\left\lbrace a'''\right\rbrace )+ \left\|r_{2}\right\|(\left\lbrace a'''\right\rbrace ) \big) \Big) \Big) \bigg) \bigg) \Bigg)\\
			=& \sum\limits_{a \in \gamma } \Bigg( \sum\limits_{a=a_{1}\cup a_{2}} \bigg( \left\|s\right\|(\left\lbrace a_{1}\right\rbrace )\cdot   \bigg( \sum\limits_{a_{2}=a_{2,1}\cup a_{2,2}} \Big( \big( \left\| 1\right\|(\left\lbrace a_{2,1}\right\rbrace )+ \left\|r_{1}\right\|(\left\lbrace a_{2,1}\right\rbrace ) \big) \cdot \\
			&\big( \left\| 1\right\|(\left\lbrace a_{2,2}\right\rbrace ) +  \left\|r_{2}\right\|(\left\lbrace a_{2,2}\right\rbrace ) \big) \Big) \bigg) \bigg) \Bigg)\\
			=& k_{s} + \big(k_{s} + (k_{s}\cdot k_{r_{1}})\big) + \big(k_{s} + (k_{s}\cdot k_{r_{2}})\big) +  \big(k_{s} + (k_{s} \cdot k_{r_{1}}) + (k_{s}\cdot k_{r_{2}}) + \\
			& (k_{s} \cdot k_{r_{1}}\cdot k_{r_{2}})\big)\\
			=& k_{s} + (k_{s}\cdot k_{r_{1}})+ (k_{s}\cdot k_{r_{2}}) + (k_{s}\cdot k_{r_{1}}\cdot k_{r_{2}}).
		\end{align*}
		\noindent Consider for instance the $\mathbb{R}_{\mathrm{max}}$ semiring. Then the resulting value represents for weighted Broadcast scheme the maximum sum of the weights associated with the ports occurring in the interactions of $\gamma$.

		\
		
		\noindent\underline{\textbf{Weighted Atomic Broadcast}}: In the Atomic Broadcast scheme a message is either received by all receivers or by none of them, which implies that we have only two interactions, $s$ and $sr_{1}r_{2}$. The $wAI(P)$ element $z$ describing the weighted Atomic Broadcast is
		$$z=s\otimes (1 \oplus r_{1}\otimes r_{2}).$$
		We let $\gamma=\lbrace \lbrace s\rbrace, \lbrace s, r_{1},r_{2}\rbrace\rbrace \in \Gamma(P)$. We firstly compute the weight of the above $wAI(P)$ element for $a=\left\lbrace s\right\rbrace\in \gamma$, and hence we derive Table \ref{tab-ata}. Afterwards, we obtain the weight of $z$ for $a=\left\lbrace s,r_{1}, r_{2}\right\rbrace \in \gamma$, as shown in Table \ref{tab-atb}. For this, we also need the auxiliary Tables \ref{tab-re2}-\ref{tab-re9} that we obtained for the weighted Rendezvous scheme. Then the weight of the Atomic Broadcast on $\gamma$ is obtained as follows:
		\begin{align*}
			&\big\|s\otimes (1 \oplus r_{1}\otimes r_{2} )\big\|(\gamma)\\
			=& \sum\limits_{a \in \gamma } \Big( \sum\limits_{a=a_{1}\cup a_{2}} \big( \left\|s\right\|(\left\lbrace a_{1}\right\rbrace )\cdot \left\| 1\oplus r_{1}\otimes r_{2}\right\|(\left\lbrace a_{2}\right\rbrace ) \big) \Big)\\
			=& \sum\limits_{a \in \gamma } \Bigg( \sum\limits_{a=a_{1}\cup a_{2}} \bigg( \left\|s\right\|(\left\lbrace a_{1}\right\rbrace )\cdot \Big( \sum\limits_{a' \in \left\lbrace a_{2}\right\rbrace  } \big( \left\| 1\right\|(\left\lbrace a'\right\rbrace )+ \left\|r_{1}\otimes r_{2}\right\|(\left\lbrace a'\right\rbrace ) \big) \Big) \bigg) \Bigg) \\
			=& \sum\limits_{a \in \gamma } \bigg( \sum\limits_{a=a_{1}\cup a_{2}} \Big( \left\|s\right\|(\left\lbrace a_{1}\right\rbrace )\cdot \big( \left\| 1\right\|(\left\lbrace a_{2}\right\rbrace )+\left\|r_{1}\otimes r_{2}\right\|(\left\lbrace a_{2}\right\rbrace ) \big) \Big) \bigg) \\
			=& \sum\limits_{a \in \gamma } \Bigg( \sum\limits_{a=a_{1}\cup a_{2}} \bigg( \left\|s\right\|(\left\lbrace a_{1}\right\rbrace )\cdot \Big( \left\| 1\right\|(\left\lbrace a_{2}\right\rbrace )+\sum\limits_{a''\in \left\lbrace a_{2}\right\rbrace } \big( \sum\limits_{a''=a_{2,1}\cup a_{2,2}} ( \left\|r_{1}\right\|(\left\lbrace a_{2,1}\right\rbrace )\cdot \\
			& \left\|r_{2}\right\|(\left\lbrace a_{2,2}\right\rbrace ) ) \big) \Big) \bigg) \Bigg) \\
			=& \sum\limits_{a \in \gamma } \Bigg( \sum\limits_{a=a_{1}\cup a_{2}} \bigg( \left\|s\right\|(\left\lbrace a_{1}\right\rbrace )\cdot \Big( \left\| 1\right\|(\left\lbrace a_{2}\right\rbrace )+ \sum\limits_{a_{2}=a_{2,1}\cup a_{2,2}} \big( \left\|r_{1}\right\|(\left\lbrace a_{2,1}\right\rbrace )\cdot \left\|r_{2}\right\|(\left\lbrace a_{2,2}\right\rbrace ) \big) \Big) \bigg) \Bigg) \\
			=& k_{s} + \big(k_{s}+ (k_{s}\cdot k_{r_{1}}\cdot k_{r_{2}})\big)\\
			=& k_{s}+ (k_{s}\cdot k_{r_{1}}\cdot k_{r_{2}}).
		\end{align*}
		
		\noindent In the Viterbi semiring for instance, the above value represents the maximum weight between $k_{s}$ and $k_{s}\cdot k_{{r_1}}\cdot k_{r_{2}}$ for the Atomic Broadcast scheme. 
		
		\
		
		\noindent\underline{\textbf{Weighted Causality Chain}}: In a Causality Chain scheme, a message is either not received by none of the two receivers, or it is received by the first one of them, or by both of them. Thus, there are three possible interactions, namely $s, sr_{1}, sr_{1}r_{2}$. The $wAI(P)$ element describing the weighted Causality Chain is
		$$z=s\otimes (1\oplus r_{1} \otimes (1\oplus r_{2})).$$
		We let $\gamma=\left\lbrace \left\lbrace s \right\rbrace ,\left\lbrace s,r_{1}\right\rbrace , \left\lbrace s, r_{1}, r_{2}\right\rbrace \right\rbrace\in\Gamma(P) $. In order to compute the weight of $z$ on $\gamma$ we firstly obtain the weight of $z$ for $a=\left\lbrace s\right\rbrace \in \gamma $ derived in Table \ref{tab-caua}. Then we compute the weight of $z$ for $a=\left\lbrace s,r_{1}\right\rbrace \in \gamma$, shown in Table \ref{tab-caub}, and finally, in Table \ref{tab-cauc} is derived the resulting weight for $a=\left\lbrace s,r_{1},r_{2}\right\rbrace  \in \gamma$. For these computations we use the auxiliary Tables \ref{tab-cauc1}-\ref{tab-cauc8}.
		\noindent Therefore, the weight for implementing the Causality Chain scheme in $\gamma$ is computed as follows:
		\begin{align*}
			&\big\|s\otimes \big(1\oplus r_{1} \otimes (1\oplus r_{2})\big)\big\|(\gamma)\\
			=& \sum\limits_{a \in \gamma } \Big( \sum\limits_{a=a_{1}\cup a_{2}} \big( \left\|s\right\|(\left\lbrace a_{1}\right\rbrace )\cdot \left\| 1\oplus r_{1} \otimes (1\oplus r_{2})\right\|(\left\lbrace a_{2}\right\rbrace ) \big) \Big)\\
			=& \sum\limits_{a \in \gamma } \Bigg( \sum\limits_{a=a_{1}\cup a_{2}} \bigg( \left\|s\right\|(\left\lbrace a_{1}\right\rbrace )\cdot \Big( \sum\limits_{a' \in \left\lbrace a_{2}\right\rbrace  } \big( \left\| 1\right\|(\left\lbrace a'\right\rbrace )+\left\| r_{1} \otimes (1\oplus r_{2})\right\|(\left\lbrace a'\right\rbrace ) \big) \Big) \bigg) \Bigg)\\
			=& \sum\limits_{a \in \gamma } \bigg( \sum\limits_{a=a_{1}\cup a_{2}} \Big( \left\|s\right\|(\left\lbrace a_{1}\right\rbrace )\cdot \big( \left\| 1\right\|(\left\lbrace a_{2}\right\rbrace )+ \left\|r_{1}\otimes (1\oplus r_{2})\right\|(\left\lbrace a_{2}\right\rbrace ) \big) \Big) \bigg) \\
			=& \sum\limits_{a \in \gamma } \Bigg( \sum\limits_{a=a_{1}\cup a_{2}} \bigg( \left\|s\right\|(\left\lbrace a_{1}\right\rbrace )\cdot \Big( \left\| 1\right\|(\left\lbrace a_{2}\right\rbrace )+ \sum\limits_{a'' \in \left\lbrace a_{2}\right\rbrace  } \big( \sum\limits_{a''=a_{2,1}\cup a_{2,2}} ( \left\|r_{1}\right\|(\left\lbrace a_{2,1}\right\rbrace )\cdot \\
			& \left\| 1\oplus r_{2}\right\|(\left\lbrace a_{2,2}\right\rbrace ) ) \big) \Big) \bigg) \Bigg) \\
			=& \sum\limits_{a \in \gamma } \Bigg( \sum\limits_{a=a_{1}\cup a_{2}} \bigg( \left\|s\right\|(\left\lbrace a_{1}\right\rbrace )\cdot \Big( \left\| 1\right\|(\left\lbrace a_{2}\right\rbrace )+  \sum\limits_{a_{2}=a_{2,1}\cup a_{2,2}} \big( \left\|r_{1}\right\|(\left\lbrace a_{2,1}\right\rbrace )\cdot \\
			& \left\| 1\oplus r_{2}\right\|(\left\lbrace a_{2,2}\right\rbrace ) \big) \Big) \bigg) \Bigg) \\
			=& \sum\limits_{a \in \gamma } \Bigg( \sum\limits_{a=a_{1}\cup a_{2}} \bigg( \left\|s\right\|(\left\lbrace a_{1}\right\rbrace )\cdot \bigg( \left\| 1\right\|(\left\lbrace a_{2}\right\rbrace )+ \sum\limits_{a_{2}=a_{2,1}\cup a_{2,2}} \Big( \left\|r_{1}\right\|(\left\lbrace a_{2,1}\right\rbrace )\cdot \\
			& \Big(\sum\limits_{a''' \in \left\lbrace a_{2,2}\right\rbrace  } ( \left\| 1\right\|(\left\lbrace a'''\right\rbrace )+ \left\|r_{2}\right\|(\left\lbrace a'''\right\rbrace ) ) \Big) \Big)  \bigg) \bigg)  \Bigg)\\
			=& \sum\limits_{a \in \gamma } \Bigg( \sum\limits_{a=a_{1}\cup a_{2}} \bigg( \left\|s\right\|(\left\lbrace a_{1}\right\rbrace )\cdot \Big( \left\| 1\right\|(\left\lbrace a_{2}\right\rbrace )+ \sum\limits_{a_{2}=a_{2,1}\cup a_{2,2}} \big( \left\|r_{1}\right\|(\left\lbrace a_{2,1}\right\rbrace )\cdot \\
			& ( \left\| 1\right\|(\left\lbrace a_{2,2}\right\rbrace )+ \left\|r_{2}\right\|(\left\lbrace a_{2,2}\right\rbrace ) ) \big) \Big) \bigg)  \Bigg)\\
			=&k_{s} + \big( k_{s} + (k_{s} \cdot k_{r_{1}})\big) + \big(k_{s} + (k_{s} \cdot k_{r_{1}}) + (k_{s} \cdot k_{r_{1}} \cdot k_{r_{2}})\big)\\
			=& k_{s} + (k_{s} \cdot k_{r_{1}}) +  (k_{s} \cdot k_{r_{1}} \cdot k_{r_{2}}).
		\end{align*}
		Consider for instance the $\mathbb{R}_{\mathrm{min}}$ semiring. Then the above value represents for weighted Causality Chain the minimum sum of the weights associated with the ports occurring in the interactions of $\gamma$.
	\end{examp}
	
	\hide{
		\begin{remark}
			Note that, the choice of $\gamma$ in the presented examples was such that it contained exactly the permissible interactions. Though, it should be clear that for any other choice of $\gamma$ the $wAI(P)$ returns the expected weight. Specifically, consider the Rendezvous scheme and let $\gamma=\left\lbrace \left\lbrace s,r_{1},r_{2}\right\rbrace, \left\lbrace s, r_{2}\right\rbrace \right\rbrace \in\Gamma(P)$. Then $\left\|s\otimes r_{1}\otimes r_{2} \right\|(\gamma)={s}\otimes k_{r_{1}}\otimes k_{r_{2}}$. Hence, if $\gamma$ includes both permissible and erroneous interactions, then the $wAI(P)$ element returns the expected weight for the scheme. Now let $\gamma$ consist only of not allowed interactions for Rendezvous scheme, for instance $\gamma=\lbrace\lbrace s, r_{2} \rbrace\rbrace$. Then $\left\|s\otimes r_{1}\otimes r_{2} \right\|(\gamma)=\hat{0}$. These results verify the robustness of $wAI(P)$ and justify the interpretation of the presented semantics.
	\end{remark}	}
	
	\hide{\begin{examp}
			Request/Response
	\end{examp}}
	
	\section{The Weighted Algebra of Connectors}\label{se5}
	In architectures, the components communicate through their ports and their allowed interactions are defined by the imposed coordination scheme. In turn, connectors 	specify the synchronization constraints among these interactions by relating a set of typed ports. Types extend ports with synchronization modes, and specifically in this work, with Rendezvous and Broadcast mode \cite{Bl:Al}. Rendezvous requires that all the components should interact simultaneously, while in Broadcast, a component initiates the interactions with some of the rest components. 
	
	In this section, we are interested in encoding the weight of connectors in architectures. For this, we study the weighted Algebra of Connectors over $P$ and $K$, that extends $wAI(P)$ with two typing operators, namely triggers and synchrons that correspond to Rendezvous and Broadcast mode, respectively. We prove several properties for $wAC(P)$ and we show that it can encode sufficiently several connectors in the weighted setup.
	\par Let $P$ be a set of ports. Similarly to $wAI(P)$, we assign to each port $p \in P$ a unique weight from $K$, denoted by $k_{p}$.

	\begin{defin}  Let $P$ be a set of ports, such that $0,1 \notin P$. The syntax of the weighted  Algebra of Connectors ($wAC(P)$ for short)  over $P$ and $K$ is given by
		$$	\sigma ::=[0] \mid [1] \mid [p] \mid [\zeta] \ \ \ (synchron)$$
		$$	\tau ::=[0]' \mid [1]' \mid [p]' \mid [\zeta]' \ \ \ (trigger)$$ 
		\quad\quad\quad\quad\quad\quad\quad\quad\quad\quad\quad\quad\quad	$\ \zeta ::=\sigma \mid \tau \mid \zeta \oplus \zeta \mid \zeta \otimes \zeta $ \\
		\\
		\noindent where $p\in P$, `` $\oplus$'' denotes the weighted union operator, `` $\otimes$'' denotes the weighted fusion operator, and `` $\left[ \cdot \right]$ '',`` $\left[ \cdot \right]'$ '' are the unary synchron and trigger typing operators, respectively.
	\end{defin}
	
	Similarly to the Algebra of Connectors introduced in \cite{Bl:Al}, the new operators in $wAC(P)$, specifically, `` $\left[ \cdot \right]$ '' and `` $\left[ \cdot \right]'$ '',  are assigned the characterization ``typing operators'' since they encode the type of synchronization mode applied to the respective ports. Particularly, a trigger is responsible for initiating an interaction, while a synchron requires to interact simultaneously with other ports. It should be clear that the typing operators in $wAC(P)$ coincide with the ones from the work of \cite{Bl:Al}, since the synchronization mode that they encode is a qualitative	feature. The difference is that in this work, the typing operators are applied among connectors in the weighted setup.
	\par Weighted union has the same meaning both in $wAC(P)$ and $wAI(P)$, while weighted fusion is a generalization of weighted synchronization in $wAI(P)$. This is clarified by the semantics of $wAC(P)$ presented in Definition \ref{connecttointer}.
	\par We call $\zeta$ a $wAC(P)$ connector over $P$ and $K$. Whenever the latter are understood, we simply refer to $\zeta$ as a $wAC(P)$ connector. Also, we write $\left[ \zeta\right] ^{\alpha}$ for $\alpha\in \left\lbrace 0,1\right\rbrace $ to denote a typed $wAC(P)$ connector. When $\alpha=0$, the $wAC(P)$ connector is a synchron, otherwise for $\alpha=1$ it is a trigger. When the type is irrelevant we write $\left[ \cdot\right] ^{\ast}$.  Moreover, we call $\zeta$ a  \emph{fusion-$wAC(P)$ connector} when $\zeta=\left[ \zeta_{1} \right]^{\alpha_{1}} \otimes \ldots \otimes\left[ \zeta_{n}\right]^{\alpha_{n}}$, where $\zeta_{1}, \ldots, \zeta_{n}\in wAC(P)$ and $\alpha_1,\ldots, \alpha_n \in \lbrace 0,1\rbrace$.
	\par Next we introduce the notations relating to the degree of a $wAC(P)$ connector $\zeta$. In particular, for a fusion-$wAC(P)$ connector $\zeta=\left[ \zeta_{1} \right]^{\alpha_{1}} \otimes \ldots \otimes\left[ \zeta_{n}\right]^{\alpha_{n}}$, where $\zeta_{1}, \ldots, \zeta_{n}\in wAC(P)$, we denote by $\#_{T}\zeta$ the number of its trigger elements, which we call the degree of $\zeta$. Then for $\zeta = \bigoplus_{i \in [n] } \zeta_{i}$, where all $\zeta_{i}$ are fusion-$wAC(P)$ connectors, we let $\#_{T} \zeta = \mathrm{max} \left\lbrace \#_{T}\zeta_{i} \mid i \in [n]\right\rbrace $. We say that $\zeta$ has a strictly positive degree iff $\mathrm{min} \left\lbrace \#_{T}\zeta_{i} \mid i \in [n]\right\rbrace > 0$. We use the notion of the degree in the semantics of $wAC(P)$ and in Section \ref{se7}, for proving a concept of congruence relation between fusion-$wAC(P)$ connectors.
	
	\par  The intuition behind the semantics of $wAC(P)$, presented in Definition \ref{connecttointer}, is to encode the weight of a $wAC(P)$ connector according to the coordination scheme imposed on an architecture. This implies that given an interactions set $\gamma$, we would like to get for each $wAC(P)$ connector a value from 
	$K$. For this, we firstly relate each $wAC(P)$ connector with a $wAI(P)$ element. In turn, we obtain
	the weight of the $wAC(P)$ connector through the semantics of $wAI(P)$. Formally, the semantics of a $wAC(P)$ connector $\zeta$ over a set of ports $P$ and the semiring $K$ is defined by the function $\left| \cdot \right|: wAC(P) \to wAI(P)$, which is presented below in Definition \ref{connecttointer}.\hide{Therefore, the semantics of $\zeta$ are $wAI(P)$ elements. In turn we can  obtain the weight of $wAC(P)$ connectors through the semantics of $wAI(P)$.} For every $\zeta\in wAC(P)$ we firstly compute its $wAI(P)$ element $\left| \zeta\right|$, and then, applying the semantics of $wAI(P)$, we compute its weight over an interactions set $\gamma \in \Gamma(P)$ by the polynomial $\big\| \left| \zeta \right| \big\|\in K \left\langle \Gamma(P)\right\rangle $.

	\begin{defin} \label{connecttointer}
		Let $\zeta$ be a $wAC(P)$ connector over $P$ and $K$. The semantics of $\zeta$ is a $wAI(P)$ element defined by the function  $\left|\cdot \right|: wAC(P) \to wAΙ(P)$ as follows:
		\begin{itemize}
			\renewcommand\labelitemi{--}
			\item $\big|\left[ p\right]\big|=p$, for $p\in P\cup \left\lbrace 0,1\right\rbrace  $,
			\item $\big|\left[ p\right]'\big|=p$, for $p\in P\cup \left\lbrace 0,1\right\rbrace $,
			\item  $\left| \left[ \zeta \right]  \right|=\left| \zeta \right|$,
			\item $\big| \left[ \zeta \right] ' \big|=\left| \zeta \right|$,
			\item $\left|  \zeta_{1} \oplus \zeta_{2}\right|  =\left|  \zeta_{1}\right|  \oplus \left|  \zeta_{2}\right|  $,
			\item $\big| \left[ \zeta_{1}\right] \otimes \left[ \zeta_{2}\right] \big| = \left|\zeta_{1} \right| \otimes \left|\zeta_{2} \right|$,
			\hide{\item $\big|\left[ \zeta_{1}\right]' \otimes \left[ \zeta_{2}\right]'\big|=\big(\left| \zeta_{1}\right| \otimes \left( 1\oplus \left|\zeta_{2} \right| \right)\big)\oplus \big(\left| \zeta_{2}\right| \otimes \left( 1\oplus \left|\zeta_{1} \right| \right)\big)$,}
			\item $\big| \left[ \zeta_{1} \right]^{\alpha_{1}} \otimes \ldots \otimes\left[ \zeta_{n}\right]^{\alpha_{n}} \big|  =\overset{}{\underset{\substack{i\in[n],\\\alpha_{i}=1}}{\bigoplus}}\Big(\left| \zeta_{i}\right| \otimes \overset{}{\underset{\substack{k\neq i,\\\alpha_{k}\in \lbrace 0,1 \rbrace}}{\bigotimes}}  \left(  1\oplus \left| \zeta_{k}\right| \right) \Big)$, where $\#_{T}(\left[ \zeta_{1} \right]^{\alpha_{1}} \otimes \ldots \otimes\left[ \zeta_{n}\right]^{\alpha_{n}})> 0$ and $\alpha_{1}, \ldots, \alpha_{n} \in \lbrace 0, 1\rbrace$.
		\end{itemize}
	\end{defin}	
	\noindent Observe that in the last case of the $wAC(P)$ semantics, the index $k$ refers both
	to connectors which act as triggers and to connectors which act as synchrons.
	
	\par Obviously, we can extend the application of the weighted fusion operator between synchrons as well as the union operator for more than two  $wAC(P)$ connectors. Specifically, for $\zeta_1,\ldots, \zeta_n \in wAC(P)$ it holds that $\big| \otimes_{i\in [n]}\left[ \zeta_{i}\right] \big| =\otimes_{ i\in [n]} \left|\zeta_{i} \right| $ and $\big| \oplus_{i\in [n]} \zeta_{i}\big|=\oplus_{i\in [n]} \left|\zeta_{i} \right|$.
	
	\begin{rem}
		Note that the weighted union operator between $wAC(P)$ connectors does not require any further typing operator. On the other hand, the weighted fusion operator serves to encode a $wAI(P)$ element for 
		a Rendezvous or Broadcast type of synchronization,
		which necessitates the typing of the involved $wAC(P)$ connectors. Specifically, we need to distinguish the following cases, relating to the last two semantics of $wAC(P)$, respectively. 
		\begin{enumerate}
			\item No component initiates the communication
			and all of them interact simultaneously. This is a case of synchronization
			where each one of the respective $wAC(P)$ connectors is typed with a synchron. In turn, by applying the weighted fusion operator we derive the $wAI(P)$ elements for synchronizing all of the involved $wAC(P)$ connectors.
			\hide{item Each of the involved elements eager to initiate the communication. This is a case of synchronization,
				where each of the occurring $wAC(P)$ connectors are typed with a trigger. 
				Then the intuition behind these fusion semantics is to compute the $wAI(P)$ element for the case that each of the $wAC(P)$ connectors initiates the interactions, while the rest ones are treated as synchrons.  }
			\item At least one of involved components eagers to initiate the communication, i.e., at least one of the $wAC(P)$ connectors is typed with a trigger. Then the idea behind these fusion semantics is to fix at each time the $wAC(P)$ connector that will act as a trigger and to compute the $wAI(P)$ element by treating the rest of the triggers as synchrons, along with the actual synchrons.
		\end{enumerate}
		Applying the semantics of $wAI(P)$ we can derive in turn, the weight of the $wAC(P)$ connectors for the above modes of synchronization. To simplify the semantics of $wAC(P)$, we restricted the computation on two $wAC(P)$ connectors for the above cases but the last one, where we obviously need to
		consider a finite but arbitrary number of triggers and synchrons. 
	\end{rem}
	
	\noindent Next, for simplicity we write $0, 1, p$, for $[0], [1], [p]$, respectively, and $0', 1', p'$, for $[0]', [1]', [p]'$, respectively. For instance, $\big[[p]\big]'$ is written $[p]'$, $[p] \oplus [q]'$ is written $p \oplus q'$, $\big[[p] \oplus [q]' \big]$ is written $[p \oplus q']$, and $\big[[p]\big]' \otimes \big[[q]\big]$ is written $[p]' \otimes [q]$.
	\par It should be clear that given a $\zeta\in wAC(P)$, in order to proceed the computations on $\left| \zeta \right| \in wAI(P)$, we need to consider the corresponding equivalence class and apply Corollary \ref{walg_semi}, i.e., that $(wAI(P)/\equiv, \oplus, \otimes, \bar{0}, \bar{1})$ is a commutative and idempotent semiring. Though, for simplicity, in the rest of the paper, we identify $\overline{\left|\zeta\right|}$ with the representative $\left|\zeta\right| $.
	\par In the next example, we clarify the above conventions in our notations.
	\begin{examp} Consider the ports $p,q,r\in P$. We apply the weighted fusion operator to $wAC(P)$ connectors $p$ and $q\oplus r$. For this, we need to specify a typing operator on them. We choose to synchronize the connectors $p$ and $q\oplus r$. Therefore, the resulting connector is $ \left[p \right]\otimes \left[ q\oplus  r \right]$ and its $wAI(P)$ element is computed as follows:
		\begin{align*}
			\big| \left[p \right]\otimes \left[ q\oplus  r \right] \big|=& \left| p\right| \otimes \left| q\oplus r \right|\\
			=& \left| p\right| \otimes (\left| q\right| \oplus \left| r \right|)\\
			=&p\otimes (q\oplus r)\\
			=& (p\otimes q) \oplus (p\otimes r).
		\end{align*}
		On the other hand, if we let $p$ serve as a trigger, then we obtain the connector $ \left[p \right]'\otimes \left[ q\oplus  r \right]$. In turn, its $wAI(P)$ element is computed as follows:
		\begin{align*}
			\big| \left[p \right]'\otimes \left[ q\oplus  r \right] \big|=& \left| p\right| \otimes (1\oplus \left| q\oplus r \right|)\\
			=& \left| p\right| \otimes \big( 1\oplus (\left| q\right| \oplus \left| r \right|)\big)\\
			=&p\otimes \big(1 \oplus (q\oplus r) \big)\\
			=&p \otimes (1\oplus q\oplus r) \\
			=& p \oplus (p \otimes q) \oplus (p\otimes r).
		\end{align*}
	\end{examp}
	
	Next we introduce a notion of equivalence for $wAC(P)$ connectors and by their equivalence classes we derive several nice properties for $wAC(P)$. 
	\par Two connectors $\zeta_{1}, \zeta_{2}\in wAC(P)$ are equivalent, and we write $\zeta_{1}\equiv \zeta_{2}$ when $\left|\zeta_{1}\right| =\left|\zeta_{2}\right|$, i.e., when they return the same $wAI(P)$ elements.
	Obviously, this in turn implies that $\big\|\left|\zeta_{1}\right|\big\|(\gamma)=\big\|\left|\zeta_{2} \right|\big\|(\gamma)$ for every $\gamma \in \Gamma(P)$, i.e., equivalent $wAC(P)$ connectors return the same weight on the same interactions set $\gamma$. Clearly ``$\equiv$'' is an equivalence relation.
	We define the quotient set $wAC(P) / \equiv$ of ``$\equiv$'' on $wAC(P)$. For every $\zeta\in wAC(P)$ we simply denote by $\overline{\zeta} $ its equivalence class. We define on $wAC(P)/\equiv$ the operations 
	\begin{equation*}\label{o5}
		\overline{\zeta_{1}} \oplus \overline{\zeta_{2}}=\overline{\zeta_{1}\oplus \zeta_{2}}  \tag{$O5$}
	\end{equation*}
	and 
	\begin{equation*}\label{o6}
		\overline{[\zeta_{1}]^{\alpha}} \otimes \overline{[\zeta_{2}]^{\beta}}=\overline{[\zeta_{1}]^{\alpha}\otimes [\zeta_{2}]^{\beta}}  \tag{$O6$}
	\end{equation*}
	for every $\zeta_{1}, \zeta_{2}\in wAC(P)$ and $\alpha, \beta \in \left\lbrace 0,1\right\rbrace $. The above two operations are well-defined, since, if $\overline{\zeta_{1}}=\overline{\zeta_{3}}$ and $\overline{\zeta_{2}}=\overline{\zeta_{4}}$, then we get that $\overline{\zeta_{1}} \oplus \overline{\zeta_{2}}=\overline{\zeta_{3}} \oplus \overline{\zeta_{4}}$ and $\overline{[\zeta_{1}]^{\alpha}} \otimes \overline{[\zeta_{2}]^{\beta}}=\overline{[\zeta_{3}]^{\alpha}} \otimes \overline{[\zeta_{4}]^{\beta}}$ for every $\zeta_1,\zeta_2,\zeta_3,\zeta_4 \in wAC(P)$ and $\alpha, \beta \in \left\lbrace 0,1\right\rbrace $.
	\par In \cite{Bl:Al}, the authors proved several properties for the Algebra of Connectors using axioms. In the sequel, we show that  $wAC(P)$ acknowledges the respective properties in the weighted setup. For this, we use the equivalence classes of $wAC(P)$.
	\par Next proposition states that connector $[1]$ is the neutral element of weighted fusion operator over the quotient set $wAC(P)/ \equiv$. 
	\begin{prop}
		Let $\zeta \in wAC(P)$ and $\alpha\in \left\lbrace 0,1\right\rbrace $. Then
		$$\overline{\left[\zeta\right]^{\alpha}}  \otimes \overline{[1]} = \overline{\left[\zeta\right]^{\alpha}}=\overline{[1]}\otimes \overline{\left[\zeta\right]^{\alpha}}.$$
	\end{prop}
	\begin{prof*}
		We use the equivalence classes of $wAC(P)$ and we apply Definition	\ref{connecttointer} and Corollary \ref{walg_semi}. \qed
	\end{prof*}
	\noindent In the following proposition, we show that $wAC(P)$ satisfies the associativity property of weighted fusion operator when a specific type, synchron or trigger, is applied for simple grouping. Associativity property is useful since it 
	indicates that independently of the order in which we apply the typing operator, the resulting synchronization remains the same.
	
	\begin{prop}\label{assoc}
		Let $\zeta_{1}, \zeta_{2}, \zeta_{3}\in wAC(P)$. Then
		\begin{enumerate}[label=\roman*)]
			\item $\overline{\big[\left[ \zeta_{1} \right]\otimes \left[ \zeta_{2} \right]\big]}\otimes \overline{\left[ \zeta_{3} \right]} =\overline{\left[ \zeta_{1} \right] }\otimes \overline{\big[\left[ \zeta_{2} \right]\otimes \left[ \zeta_{3} \right]\big]}  \label{synch-assoc}$
			\item $\overline{\big[\left[ \zeta_{1} \right]'\otimes \left[ \zeta_{2} \right]'\big]'}\otimes \overline{\left[ \zeta_{3} \right]'} =\overline{\left[ \zeta_{1} \right]' }\otimes \overline{\big[\left[ \zeta_{2} \right]'\otimes \left[ \zeta_{3} \right]'\big]'}.\label{trig-assoc}$
		\end{enumerate}
	\end{prop}
	\begin{prof*}
		In order to prove the above equalities we apply Corollary \ref{walg_semi}.
		\begin{enumerate}[label=\roman*)]
			\item We compute the left part of the first equality as follows:
			\begin{align*}
				&\overline{\big[\left[ \zeta_{1} \right]\otimes \left[ \zeta_{2} \right]\big]}\otimes \overline{\left[ \zeta_{3} \right]} =\overline{\big[\left[\zeta_{1}\right]\otimes\left[\zeta_{2}\right]\big]\otimes \left[\zeta_{3}\right]}\\
				=&\left\lbrace \zeta\in wAC(P) \mid \zeta \equiv \big[\left[\zeta_{1}\right]\otimes \left[\zeta_{2}\right]\big]\otimes \left[\zeta_{3}\right]\right\rbrace \\
				=&\left\lbrace \zeta\in wAC(P) \mid \left|\zeta\right|= \Big|\big[\left[\zeta_{1}\right]\otimes \left[\zeta_{2}\right]\big]\otimes \left[\zeta_{3}\right] \Big| \right\rbrace \\
				=&\left\lbrace \zeta\in wAC(P) \mid \left|\zeta\right|= \big| \left[ \zeta_{1}\right] \otimes \left[ \zeta_{2}\right]\big| \otimes  \big| \zeta_{3} \big| \right\rbrace \\
				=&\left\lbrace \zeta\in wAC(P) \mid \left|\zeta\right|=\big( \left| \zeta_{1}\right| \otimes \left| \zeta_{2}\right|\big) \otimes  \left| \zeta_{3} \right| \right\rbrace \\
				=&\left\lbrace \zeta\in wAC(P) \mid \left|\zeta\right|= \left| \zeta_{1}\right| \otimes \left| \zeta_{2}\right| \otimes  \left| \zeta_{3} \right| \right\rbrace
			\end{align*} 
			
			\noindent Now, we compute the right part of the first equality as follows:
			\begin{align*}
				&\overline{\left[ \zeta_{1} \right] }\otimes \overline{\big[\left[ \zeta_{2} \right]\otimes \left[ \zeta_{3} \right]\big]} =\overline{\left[\zeta_{1}\right]\otimes\big[\left[\zeta_{2}\right]\otimes \left[\zeta_{3}\right]\big]}\\
				=&\left\lbrace \zeta\in wAC(P) \mid \zeta \equiv \left[\zeta_{1}\right]\otimes \big[\left[\zeta_{2}\right]\otimes \left[\zeta_{3}\right]\big]\right\rbrace \\
				=&\left\lbrace \zeta\in wAC(P) \mid \left|\zeta\right|= \Big|\left[\zeta_{1}\right]\otimes\big[ \left[\zeta_{2}\right]\otimes \left[\zeta_{3}\right]\big] \Big| \right\rbrace \\
				=&\left\lbrace \zeta\in wAC(P) \mid \left|\zeta\right|=  \left| \zeta_{1}\right| \otimes \big|\left[ \zeta_{2}\right] \otimes  \left[ \zeta_{3} \right] \big| \right\rbrace \\
				=&\left\lbrace \zeta\in wAC(P) \mid \left|\zeta\right|= \left| \zeta_{1}\right| \otimes \big(\left| \zeta_{2}\right| \otimes  \left| \zeta_{3} \right|\big) \right\rbrace \\
				=&\left\lbrace \zeta\in wAC(P) \mid \left|\zeta\right|= \left| \zeta_{1}\right| \otimes \left| \zeta_{2}\right| \otimes  \left| \zeta_{3} \right| \right\rbrace.
			\end{align*}
			
			\noindent Consequently,  $\overline{\big[\left[ \zeta_{1} \right]\otimes \left[ \zeta_{2} \right]\big]}\otimes \overline{\left[ \zeta_{3} \right]} $ and $\overline{\left[ \zeta_{1} \right] }\otimes \overline{\big[\left[ \zeta_{2} \right]\otimes \left[ \zeta_{3} \right]\big]} $ are equal.
			\item  For the left part of the second equality we have: 
			\begin{align*}
				&\overline{\big[\left[ \zeta_{1} \right]'\otimes \left[ \zeta_{2} \right]'\big]'}\otimes \overline{\left[ \zeta_{3} \right]'} =\overline{\big[\left[\zeta_{1}\right]'\otimes\left[\zeta_{2}\right]'\big]'\otimes \left[\zeta_{3}\right]'}\\
				=&\left\lbrace \zeta\in wAC(P) \mid \zeta \equiv \big[\left[\zeta_{1}\right]'\otimes \left[\zeta_{2}\right]'\big]'\otimes \left[\zeta_{3}\right]'\right\rbrace \\
				=&\left\lbrace \zeta\in wAC(P) \mid \left|\zeta\right|=\Big| \big[ \left[ \zeta_{1}\right]'\otimes \left[ \zeta_{2}\right]'\big]'  \otimes \left[ \zeta_{3}\right]' \Big|\right\rbrace \\
				=&\left\lbrace \zeta\in wAC(P) \mid \left|\zeta\right|= \Big(\big| \left[ \zeta_{1}\right]'\otimes \left[ \zeta_{2}\right]' \big|\otimes (1\oplus \left| \zeta_{3}\right|) \Big) \oplus \Big( \left| \zeta_{3}\right| \otimes \big( 1\oplus  \big| \left[ \zeta_{1}\right]'\otimes \left[ \zeta_{2}\right]' \big| \big)\Big)\right\rbrace \\
				=&\left\lbrace \zeta\in wAC(P) \mid \left|\zeta\right|= \bigg(\Big(\big( \left| \zeta_{1}\right| \otimes (1 \oplus \left| \zeta_{2}\right| )\big)\oplus \big(\left| \zeta_{2}\right| \otimes (1 \oplus \left| \zeta_{1}\right| ) \big) \Big)\otimes (1\oplus \left| \zeta_{3}\right|)\bigg) \oplus\right.\\
				&\left.\bigg( \left|\zeta_{3} \right|\otimes \Big( 1\oplus  \big(\left| \zeta_{1}\right| \otimes (1 \oplus \left| \zeta_{2}\right| )\big)\oplus \big(\left| \zeta_{2}\right| \otimes (1 \oplus \left| \zeta_{1}\right| )\big) \Big)\bigg)\right\rbrace \\
				=&\left\lbrace \zeta\in wAC(P) \mid \left|\zeta\right|= \bigg(\Big( \left|\zeta_{1}\right| \oplus (\left| \zeta_{1}\right| \otimes \left|\zeta_{2}\right|) \oplus \left|\zeta_{2}\right|\oplus (\left| \zeta_{2}\right|\otimes \left|\zeta_{1}\right|) \Big)\otimes (1\oplus \left| \zeta_{3}\right|) \bigg)\oplus \right. \\
				&\left. \bigg(\left| \zeta_{3}\right| \otimes \Big( 1\oplus \left|\zeta_{1}\right| \oplus  (\left| \zeta_{1}\right| \otimes \left|\zeta_{2}\right|) \oplus  \left|\zeta_{2}\right|\oplus (\left| \zeta_{2}\right|\otimes \left|\zeta_{1}\right|) \Big)\bigg)\right\rbrace \\
				=& \left\lbrace \zeta\in wAC(P) \mid \left|\zeta\right|=\bigg(\Big( \left|\zeta_{1}\right| \oplus \left| \zeta_{2}\right|\oplus (\left| \zeta_{1}\right| \otimes \left|\zeta_{2}\right|) \Big)\otimes (1\oplus \left| \zeta_{3}\right|)\bigg) \oplus \right.\\
				&\left. \bigg(\left| \zeta_{3}\right| \otimes \Big( 1 \oplus \left|\zeta_{1}\right| \oplus \left| \zeta_{2}\right|\oplus (\left| \zeta_{1}\right| \otimes \left|\zeta_{2}\right|) \Big)\bigg)\right\rbrace \\
				=& \Big\{ \zeta\in wAC(P) \mid \left|\zeta\right|=\Big(\left|\zeta_{1}\right| \oplus \left| \zeta_{2}\right|\oplus (\left| \zeta_{1}\right| \otimes \left|\zeta_{2}\right|) \oplus (\left|\zeta_{1}\right|\otimes \left|\zeta_{3}\right|) \oplus (\left| \zeta_{2}\right|\otimes \left|\zeta_{3}\right|)\oplus  \\
				&  (\left| \zeta_{1}\right| \otimes \left|\zeta_{2}\right|\otimes \left|\zeta_{3}\right|) \Big)\oplus \Big( \left|\zeta_{3}\right|\oplus (\left|\zeta_{3}\right|\otimes \left|\zeta_{1}\right|) \oplus (\left|\zeta_{3}\right|\otimes \left|\zeta_{2}\right|) \oplus (\left|\zeta_{3}\right|\otimes \left| \zeta_{1}\right| \otimes \left|\zeta_{2}\right|)\Big)\Big\} \\
				=&\big\{ \zeta\in wAC(P) \mid \left|\zeta\right|=\left|\zeta_{1}\right| \oplus \left| \zeta_{2}\right|\oplus \left| \zeta_{3}\right| \oplus (\left| \zeta_{1}\right| \otimes \left|\zeta_{2}\right|) \oplus (\left|\zeta_{1}\right|\otimes \left|\zeta_{3}\right|) \oplus \\
				& (\left|\zeta_{2}\right| \otimes \left| \zeta_{3}\right|) \oplus  (\left| \zeta_{1}\right| \otimes \left|\zeta_{2}\right|\otimes \left|\zeta_{3}\right|) \big\}.
			\end{align*}
			\noindent	Finally, for the right part in the second equality, we have:
			\begin{align*}
				&\overline{\left[ \zeta_{1} \right]'} \otimes \overline{\big[\left[ \zeta_{2} \right]'\otimes \left[ \zeta_{3} \right]'\big]'} =\overline{\left[\zeta_{1}\right]'\otimes\big[\left[\zeta_{2}\right]'\otimes \left[\zeta_{3}\right]'\big]'}\\
				=&\left\lbrace \zeta\in wAC(P) \mid \zeta \equiv \left[\zeta_{1}\right]'\otimes\big[\left[\zeta_{2}\right]'\otimes \left[\zeta_{3}\right]'\big]'\right\rbrace \\
				=&\left\lbrace \zeta\in wAC(P) \mid \left|\zeta\right|=\Big|  \left[ \zeta_{1}\right]'\otimes \big[\left[ \zeta_{2}\right]'  \otimes \left[ \zeta_{3}\right]'\big]' \Big|\right\rbrace \\
				=&\Big\{ \zeta\in wAC(P) \mid \left|\zeta\right|=\Big(\left| \zeta_{1}\right|\otimes \big( 1\oplus \big|\left[ \zeta_{2}\right]'  \otimes \left[ \zeta_{3}\right]'\big|\big)\Big) \oplus \Big( \big| \left[ \zeta_{2}\right]'  \otimes \left[ \zeta_{3}\right]'\big| \otimes \big( 1\oplus \left|\zeta_{1}\right| \big)\Big)\Big\} \\
				=&  \left\lbrace \zeta\in wAC(P) \mid \left|\zeta\right|=\bigg(\left| \zeta_{1}\right| \otimes \Big(1 \oplus \big(\left| \zeta_{2}\right|\otimes (1\oplus \left|\zeta_{3}\right|)\big) \oplus  \big(\left| \zeta_{3}\right|\otimes (1\oplus \left|\zeta_{2}\right|)\big)\Big) \bigg) \oplus\right.\\
				&\left.\bigg( \Big(\big(\left| \zeta_{2}\right|\otimes (1\oplus \left|\zeta_{3}\right|)\big) \oplus \big( \left| \zeta_{3}\right|\otimes (1\oplus \left|\zeta_{2}\right|)\big)\Big) \otimes  \big( 1\oplus \left|\zeta_{1}\right| \big)\bigg)\right\rbrace \\
				=&\left\lbrace \zeta\in wAC(P) \mid \left|\zeta\right|=\bigg( \left|\zeta_{1}\right| \otimes \Big( 1\oplus \left|\zeta_{2}\right| \oplus (\left|\zeta_{2}\right|\otimes \left|\zeta_{3}\right|) \oplus  \left|\zeta_{3}\right| \oplus (\left|\zeta_{3}\right|\otimes \left|\zeta_{2}\right|) \Big)\bigg) \oplus\right.\\
				&\left. \bigg(\Big( \left|\zeta_{2}\right| \oplus (\left|\zeta_{2}\right|\otimes \left|\zeta_{3}\right|) \oplus \left|\zeta_{3}\right| \oplus  (\left|\zeta_{3}\right|\otimes \left|\zeta_{2}\right|) \Big) \otimes \big( 1\oplus \left|\zeta_{1}\right| \big)\bigg)\right\rbrace \\
				=& \left\lbrace \zeta\in wAC(P) \mid \left|\zeta\right|=\bigg(\left|\zeta_{1}\right| \otimes \Big( 1\oplus \left|\zeta_{2}\right| \oplus  \left|\zeta_{3}\right| \oplus (\left|\zeta_{2}\right|\otimes \left|\zeta_{3}\right|) \Big)\bigg) \oplus \right.\\
				& \left. \bigg(\Big( \left|\zeta_{2}\right| \oplus \left|\zeta_{3}\right| \oplus  (\left|\zeta_{2}\right|\otimes \left|\zeta_{3}\right|) \Big) \otimes \big( 1\oplus \left|\zeta_{1}\right| \big)\bigg)\right\rbrace \\
				=& \Big\{ \zeta\in wAC(P) \mid \left|\zeta\right|=\Big(\left|\zeta_{1}\right| \oplus (\left|\zeta_{1}\right| \otimes \left|\zeta_{2}\right|) \oplus (\left|\zeta_{1}\right| \otimes \left|\zeta_{3}\right|) \oplus (\left|\zeta_{1}\right| \otimes \left|\zeta_{2}\right|\otimes \left|\zeta_{3}\right|)\Big) \oplus \\
				&\Big( \left|\zeta_{2}\right| \oplus \left|\zeta_{3}\right| \oplus (\left|\zeta_{2}\right|\otimes \left|\zeta_{3}\right|) \oplus (\left|\zeta_{2}\right|\otimes \left|\zeta_{1}\right|) \oplus (\left|\zeta_{3}\right|\otimes \left|\zeta_{1}\right|) \oplus (\left|\zeta_{2}\right|\otimes \left|\zeta_{3}\right|\otimes \left|\zeta_{1}\right|)\Big)\Big\} \\
				=& \big\{ \zeta\in wAC(P) \mid \left|\zeta\right|= \left|\zeta_{1}\right| \oplus \left| \zeta_{2}\right|\oplus \left| \zeta_{3}\right| \oplus (\left| \zeta_{1}\right| \otimes \left|\zeta_{2}\right|) \oplus (\left|\zeta_{1}\right|\otimes \left|\zeta_{3}\right|) \oplus \\
				& (\left|\zeta_{2}\right| \otimes \left| \zeta_{3}\right|) \oplus  (\left| \zeta_{1}\right| \otimes \left|\zeta_{2}\right|\otimes \left|\zeta_{3}\right|)\big\}.
			\end{align*}
		\end{enumerate}
		\noindent Hence,  $\overline{\big[\left[ \zeta_{1} \right]'\otimes \left[ \zeta_{2} \right]'\big]'}\otimes \overline{\left[ \zeta_{3} \right]'}$ and $ \overline{\left[ \zeta_{1} \right]' }\otimes \overline{\big[\left[ \zeta_{2} \right]'\otimes \left[ \zeta_{3} \right]'\big]'}$ are equal.\qed
	\end{prof*}

	\vspace{5mm} \noindent Next proposition serves for simplifying the computation of the semantics for some of the $wAC(P)/\equiv$ elements. Specifically, part a) indicates that in case we apply two typing operators to a $wAC(P)$ connector, then the inner typing can be omitted, part b)
	expresses that the typing of weighted union of two $wAC(P)$ connectors coincides with the weighted union of their respective typing, 
	while part c) and d) verify the commutativity property of the weighted union and the weighted fusion operator with respect to the typing operator when a specific type, synchron or trigger, is applied for simple grouping. \hide{In the sequel, we use the results of these propositions on Theorem \ref{eq4} and Theorem \ref{th1} that follows, in order to simplify our proofs.} 
	
	\begin{prop}\label{propos3}
		Let $\zeta_{1}, \zeta_{2} \in wAC(P)$ and $\alpha, \beta\in \left\lbrace 0,1\right\rbrace $. Then
		\begin{enumerate}[label=\alph*)]
			\item $\overline{\big[ \left[ \zeta_{1}\right] ^{\alpha}\big] ^{\beta}} = \overline{\left[ \zeta_{1}\right] ^{\beta}}\label{eq-two-types}$ \label{itm:propos3a}
			\item $\overline{\left[ \zeta_{1}\oplus \zeta_{2}\right] ^{\alpha}} = \overline{\left[ \zeta_{1}\right] ^{\alpha}}\oplus \overline{\left[ \zeta_{2} \right] ^{\alpha}}$\label{itm:propos3b}
			\item \label{com-union}$\overline{\left[ \zeta_{1}\right] ^{\alpha}} \oplus \overline{\left[ \zeta_{2}\right] ^{\beta}} = \overline{\left[ \zeta_{2}\right] ^{\beta}} \oplus \overline{\left[ \zeta_{1}\right] ^{\alpha}}$
			\item \label{com-fusion} $\overline{\left[ \zeta_{1}\right] ^{\alpha}} \otimes \overline{\left[ \zeta_{2}\right] ^{\beta}} = \overline{\left[ \zeta_{2}\right] ^{\beta}} \otimes \overline{\left[ \zeta_{1}\right] ^{\alpha}}.$ 
			
		\end{enumerate}
	\end{prop}
	
	\begin{prof*}
		We apply Corollary \ref{walg_semi} and we compute the equivalence classes of the involved $wAC(P)$ connectors. \qed
	\end{prof*}

	\noindent In the sequel, we apply $wAC(P)$ for modeling several connectors in the weighted setup. Then we apply the semantics of $wAC(P)$ and in turn by the semantics of the respective $wAI(P)$ elements, we are able to derive the weight of implementing each $wAC(P)$ connector for a given interactions set. 
	\par For the $wAC(P)$ connectors presented in the sequel, we follow the representation considered in \cite{Bl:Al}. In particular, we use triangles and circles in order to represent triggers and synchrons, respectively. Then we connect the involved $wAC(P)$ connectors with lines which are labeled by the respective weight from $K$. In turn, we draw the resulting $wAC(P)$ connector incrementally, as it described in the following examples.
	
	\hide{\par In the previous section we investigated four coordination schemes in the weighted setup. Now we present the $wAC(P)$ connectors that return the $wAI(P)$ elements of those schemes.}
	\begin{examp}\label{eq2}

		\begin{figure}[H]
			\begin{minipage}[b]{0.45\linewidth}
				\centering
				\includegraphics[scale=1.7]{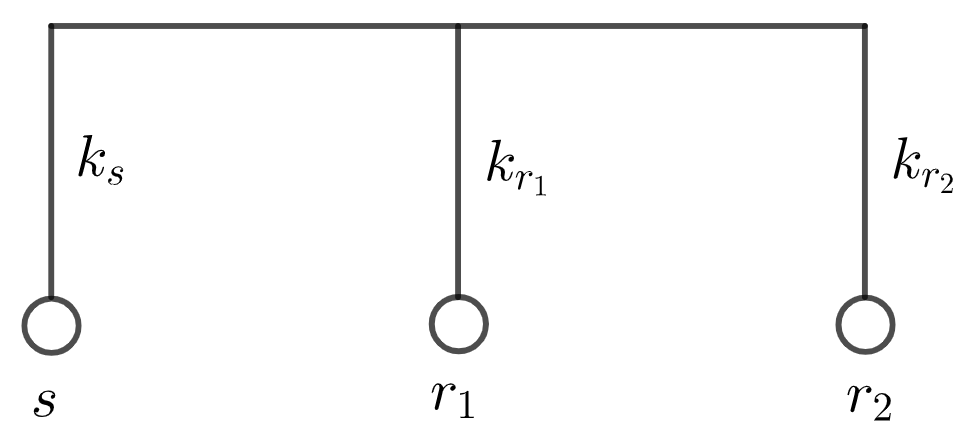}
				\subcaption{Weighted Rendezvous}
				\label{fig:f3a}
			\end{minipage}
			\begin{minipage}[b]{0.45\linewidth}
				\centering
				\includegraphics[scale=1.7]{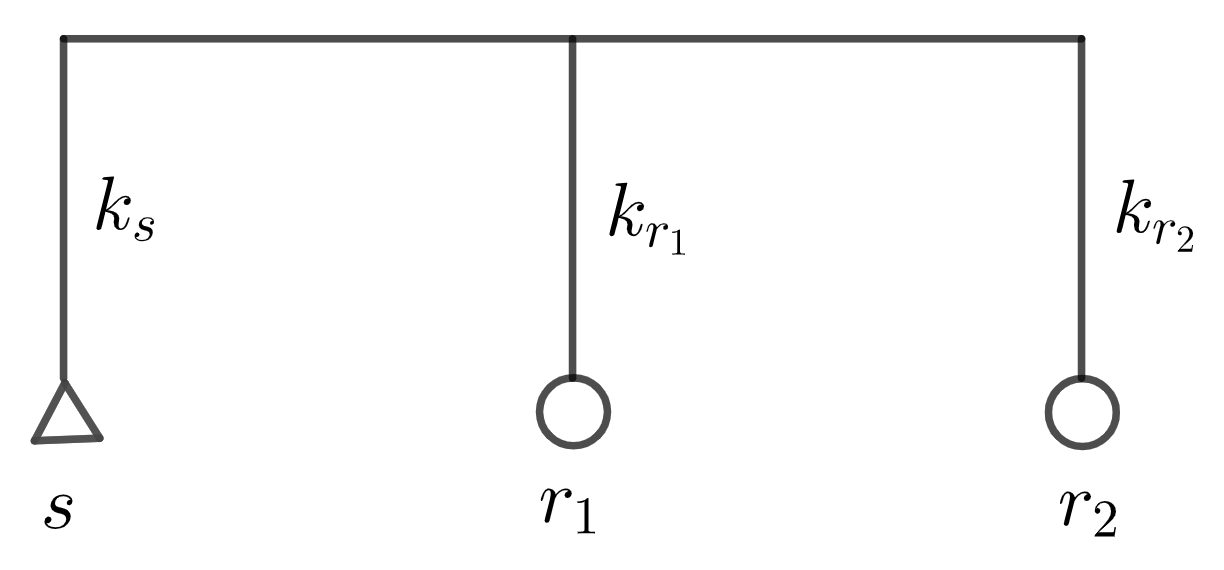}
				\subcaption{Weighted Broadcast}
				\label{fig:f3b}
			\end{minipage}
			\begin{minipage}[b]{0.45\linewidth}
				\centering
				\includegraphics[scale=1.7]{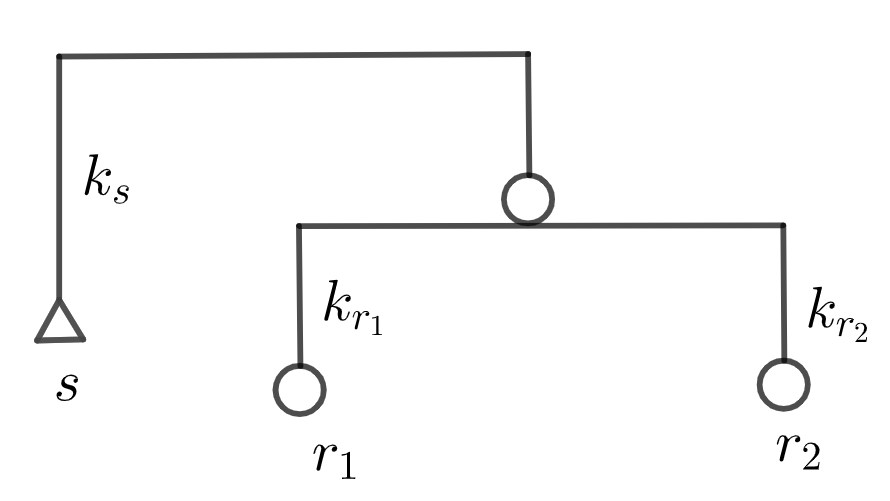}
				\subcaption{Weighted Atomic Broadcast}
				\label{fig:f3c}
			\end{minipage}
			\hspace{11mm}
			\begin{minipage}[b]{0.45\linewidth}
				\centering
				\includegraphics[scale=1.7]{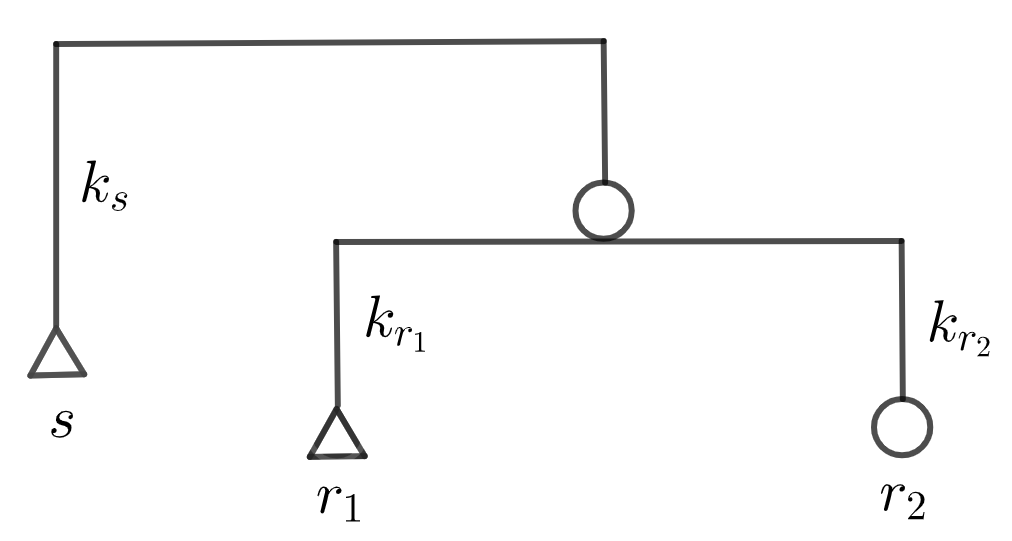}
				\subcaption{Weighted Causality Chain}
				\label{fig:f3d}
			\end{minipage}
			\caption{The $wAC(P)$ connectors of four coordination schemes.}
			\label{fig4}
		\end{figure}

		We present the $wAC(P)$ connectors of the coordination schemes Rendezvous, Broadcast, Atomic Broadcast and Causality Chain, in the weighted setup, described in Example \ref{eq:4}. Recall that we considered a sender and two receivers with ports $s, r_1,r_2$, respectively, and hence, $P=\left\lbrace s,r_1,r_2\right\rbrace $. Also, we denoted by  $k_{s}, k_{r_{1}}, k_{r_{2}}$, the weight associated with the ports $s,r_1,r_2$, respectively. \\ \\
		\noindent\underline{\textbf{Weighted Rendezvous}}: In Rendezvous coordination scheme, the involved components should be strongly synchronized. Hence, the respective connector should not
		contain any trigger operator, but it would rather apply the synchron typing in each of the terms. Then in the weighted setting the connector should encode the ``cost'' for this synchronization. Specifically, the $wAC(P)$ connector for weighted Rendezvous in our example is given by
		$$\zeta = [s]\otimes [r_{1}]\otimes [r_{2}].$$ The $wAC(P)$ connector is shown in Figure \ref{fig:f3a}. In particular, the $wAC(P)$ connectors $s, r_{1}, r_{2}$, are represented with circles, since they are all typed with synchrons. Moreover,	the connection is performed simultaneously, and hence the three $wAC(P)$ connectors occur at the same level. Then we obtain the $wAI(P)$ element of the connector as follows:
		\begin{align*}
			\left| \zeta \right| =&\big| [s]\otimes [r_{1}]\otimes [r_{2}]\big|\\
			=&\left|s\right|\otimes\left|r_{1}\right|\otimes\left|r_{2}\right|\\
			=&s\otimes r_{1}\otimes r_{2}.
		\end{align*}
		The weight of the above $wAI(P)$ element was computed in the corresponding part of Example \ref{eq:4} for $\gamma=\lbrace \lbrace s,r_1,r_2\rbrace\rbrace$, and we showed that it equals to $k_{s}\cdot k_{r_{1}}\cdot k_{r_{2}}$. \hide{In particular, we derived the cost of implementing the connector over the fuzzy semiring $F$.} \\ \\
		\noindent\underline{\textbf{Weighted Broadcast}}: In Broadcast communication, a component should trigger the interaction with the rest components. As a result, we should allow a
		trigger typing along with some synchrons in the respective connector. For our example, the sender initiates the interactions
		with some of the two receivers, and hence we consider the trigger typing for $s$, while $r_1$ and $r_2$ are typed with synchrons. In turn, the $wAC(P)$ connector for Broadcast scheme is
		$$\zeta = \left[s\right]'\otimes \left[r_{1}\right]\otimes [r_{2}].$$
		The connector is shown in Figure \ref{fig:f3b}, where the trigger for $s$ and the synchrons for $r_1$ and $r_2$ are indicated by a triangle and two circles, respectively, all occurring at the same level. Then the $wAI(P)$ element of the above connector is computed as follows:
		\begin{align*}
			\left|\zeta\right| =& \left|\left[s\right]'\otimes \left[r_{1}\right]\otimes [r_{2}]\right|\\
			=&\left|s\right|\otimes (1\oplus \left|r_{1}\right|)\otimes (1\oplus \left|r_{2}\right|)\\
			=&s\otimes (1\oplus r_{1})\otimes (1\oplus r_{2}).
		\end{align*}
		The weight of the latter was obtained in the weighted Broadcast scheme of Example \ref{eq:4} for $\gamma=\lbrace \lbrace s\rbrace, \lbrace s,r_1\rbrace, \lbrace s,r_2\rbrace, \lbrace s,r_1,r_2\rbrace\rbrace$, and it was computed equal to $k_{s}+ (k_{s}\cdot k_{r_{1}})+ (k_{s}\cdot k_{r_{2}})+ (k_{s}\cdot k_{r_{1}}\cdot k_{r_{2}})$. Then in $\mathbb{R}_{\mathrm{min}}$ semiring the resulting value corresponds to the minimum sum of the weights associated with the ports occurring in the interactions of $\gamma$. 
		In turn, if we would interpret the weight as the energy consumption associated with each of the allowed interactions, then the architecture could opt for the most efficient one. \hide{Moreover, we derived the weight of this component in $\mathbb{R}_{\mathrm{max}}$ semiring.}\\ \\
		\noindent\underline{\textbf{Weighted Atomic Broadcast}}: Similarly to  Broadcast, in the Atomic Broadcast scheme there is a component that initiates the interactions with the other components. 
		The difference is that it requires the communication with all the other components or with none of them, and hence we should
		obtain the overall weight for the two cases. For our example, the unique sender is assigned a trigger typing,
		while we compute the weight of the simultaneous presence or absence of both receivers, by the application of the synchron typing to the 
		weighted fusion of $[r_1] $ and $[r_2]$. The $wAC(P)$ connector for the weighted Atomic Broadcast is given by 
		$$\zeta = [s]'\otimes \big[ [r_{1}]\otimes [r_{2}]\big]$$
		and is shown in Figure \ref{fig:f3c}. The synchrons for $r_{1}$ and $r_{2}$ are denoted with circles and occur at the same level, the trigger for	$s$ is represented with a triangle, and a circle is used for the synchronization of the latter with $ [r_{1}]\otimes [r_{2}]$. The corresponding $wAI(P)$ element of $\zeta$ is obtained as follows:
		\begin{align*}
			\left|\zeta\right|=& \big| [s]'\otimes \big[ [r_{1}]\otimes [r_{2}]\big] \big|\\
			=& \big|s\big| \otimes \Big(1 \oplus \big| [r_{1}]\otimes [r_{2}] \big|\Big)\\
			=& \left|s\right|\otimes \big(1\oplus  \left|r_{1}\right|\otimes \left|r_{2}\right|\big)\\
			=& s\otimes (1\oplus r_{1}\otimes r_{2}).
		\end{align*}
		The weight of the above $wAI(P)$ element was computed in the corresponding scheme in Example \ref{eq:4} for $\gamma=\lbrace \lbrace s\rbrace, \lbrace s,r_1,r_2\rbrace\rbrace$, and equals to $k_{s}+ (k_{s}\cdot k_{{r_1}}\cdot k_{r_{2}})$. Then in Viterbi semiring the resulting value corresponds to executing the interaction with the maximum probability. \hide{Also, we interpreted the resulting cost over the fuzzy semiring $F$.}\\ \\
		\noindent\underline{\textbf{Weighted Causality Chain}}: A $wAC(P)$ connector for the Causality Chain, should apply a trigger typing to the sender, i.e., to the
		component that initiates the interaction as well as to any of the involved receivers but the ``last'' one. Then we encode the weight of the resulting synchronization by the $wAC(P)$
		connector
		$$\zeta = [s]'\otimes \big[ [r_{1}]'\otimes [r_{2}]\big].$$
		The connector is shown in Figure \ref{fig:f3d} where triangles are used to denote the trigger for $s$ and $r_{1}$, while the synchron for $r_{2}$ is denoted by a circle. Then we synchronize $[r_{1}]'$ with $[r_{2}]$ and in turn, we synchronize $[s]'$ with $\big[ [r_{1}]'\otimes [r_{2}]\big]$.  The two synchronizations are depicted with a circle in Fig. \ref{fig:f3d}, at lower and higher level, respectively. Hence, the $wAI(P)$ element is computed as follows:
		\begin{align*}
			\left|\zeta\right|=&\bigg| [s]'\otimes \big[ [r_{1}]'\otimes [r_{2}]\big] \bigg|\\
			=&\left|s\right|\otimes \Big(1 \oplus \big| [r_{1}]'\otimes [r_{2}] \big| \Big)\\
			=& \left|s\right|\otimes \big(1 \oplus \left|r_{1}\right|\otimes ( 1\oplus \left| r_{2}\right|) \big)\\
			=& s\otimes \big(1 \oplus r_{1}\otimes ( 1\oplus  r_{2}) \big).
		\end{align*}
		The weight of the above $wAI(P)$ element was computed in Example \ref{eq:4} for $\gamma=\lbrace \lbrace s\rbrace, \lbrace s,r_1\rbrace, \lbrace s,r_1,\\r_2\rbrace\rbrace$, and equals to $k_{s}+ (k_{s}\cdot k_{r_{1}}) +  (k_{s}\cdot k_{r_{1}} \cdot k_{r_{2}})$. Then in $\mathbb{R}_{\mathrm{max}}$ semiring the resulting value corresponds to the maximum sum of the weights associated with the ports occurring in the interactions of $\gamma$.\hide{ Also, we computed the weight of the connector over the $\mathbb{R}_{\mathrm{min}}$ semiring.}
	\end{examp}

	\section{Weighted subalgebras}\label{se6}
	Next we present two subalgebras of $wAC(P)$, namely the weighted Algebra of Synchrons and the weighted Algebra of Triggers over $P$ and $K$, and consider their properties. The former algebra restricts to synchrons, while the latter involves only triggers.\hide{We conclude the section by discussing the relation among the four presented algebras, i.e., among $wAI(P), wAC(P)$, and the two subalgebras.} 
	\subsection{The Weighted Algebra of Synchrons}\label{se6-1}
	Let $P$ be a set of ports. We assign to each port $p \in P$ a unique weight from $K$, denoted by $k_{p}$. Then we consider the subalgebra of $wAC(P)$
	generated by restricting its syntax to synchrons. The resulting algebra over $P$ and $K$ is called weighted Algebra of Synchrons and is denoted by
	$wAS(P)$.
	\begin{defin}
		Given a set of ports $P$, the syntax of the weighted Algebra of Synchrons ($wAS(P)$ for short) over $P$ and $K$ is defined by:
		$$\sigma::= [0] \mid [1]\mid [p] \mid [\zeta]$$
		$$\zeta::=\sigma\mid \zeta \oplus \zeta \mid \zeta \otimes \zeta $$
		where $p\in P$, $\sigma$ denotes a synchron element, and $\zeta\in wAS(P)$.
	\end{defin}
	\noindent The weighted operators ``$\oplus$'' and ``$\otimes$'' are the weighted union and the weighted fusion operator, respectively, from the syntax of $wAC(P)$. Obviously, we get that Proposition \ref{assoc}\ref{synch-assoc} holds for $wAS(P)$, hence the subalgebra satisfies the associativity property of weighted fusion with respect to the synchron typing operator. Furthermore, $wAS(P)$ satisfies Proposition \ref{propos3}, for $\alpha=0$ and $\beta=0$. Intuitively, part a) implies that when we apply the synchron typing operator twice to a $wAS(P)$ connector, then the inner typing can be omitted, part b) expresses that
	the synchron of the weighted union of two $wAS(P)$ connectors coincides with the weighted union of their respective synchrons, and parts c) and d) indicate that the weighted union and weighted fusion operators, respectively, satisfy the commutativity property for the synchron typing operator.

	\subsection{The Weighted Algebra of Triggers}\label{se6-2}
	Let $P$ be a set of ports. We assign to each port $p \in P$ a unique weight from $K$, denoted by $k_{p}$. Then we consider the subalgebra of $wAC(P)$ generated by restricting its syntax to triggers. The resulting algebra over $P$ and $K$ is called weighted Algebra of Triggers and is denoted by $wAT(P)$.
	\begin{defin} Given a set of ports $P$, the syntax of the weighted Algebra of Triggers ($wAT(P)$ for short) over $P$ and $K$ is defined by:
		$$\tau::= [0]' \mid [1]'\mid [p]' \mid [\zeta]'$$
		$$\zeta::=\tau \mid \zeta \oplus \zeta \mid \zeta \otimes \zeta $$
		where $p\in P$, $\tau$ denotes a trigger element, and $\zeta\in wAT(P)$.
	\end{defin}
	\noindent The operators ``$\oplus$'' and ``$\otimes$'' are the weighted union and the weighted fusion operator, respectively, from the syntax of the $wAC(P)$. Moreover, by Proposition \ref{assoc}\ref{trig-assoc} of the previous section we obtain that $wAT(P)$ satisfies the associativity property of weighted fusion with respect to trigger typing operator. Furthermore, $wAT(P)$ satisfies Proposition \ref{propos3}, for $\alpha=1$ and $\beta=1$, where part a) implies that when we doubly type a $wAT(P)$ connector with a trigger, then the inner one can be omitted, part b) expresses that the trigger typing of the weighted union of two $wAT(P)$ connectors coincides with the weighted union of their respective triggers, and parts c) and d) imply that weighted union and fusion operators, respectively, satisfy the commutativity property for the trigger typing operator.

	\par Recall that the $wAC(P)$ connector $[1]$ served as the neutral element of weighted fusion in $wAC(P)$. It should be clear though, that by construction, the weighted Algebra of Triggers does not contain $[1]$, and hence weighted fusion loses its neutral element. 
	To tackle this, we restrict the equivalence relation ``$\equiv$'' defined in $wAC(P)$ (cf. Section \ref{se5}) only to the trigger elements, i.e., to $\zeta\in wAT(P)$. Hence, we define the quotient set $wAT(P)/\equiv$ of ``$\equiv$'' on $wAT(P)$. 
	Then we trivially get that $[0']'$ can serve as an alternative neutral element for the weighted fusion operator in $wAT(P)$, i.e., $\overline{  \left[ \zeta\right]' }\otimes \overline{[0']'}=\overline{\left[ \zeta\right]' }=\overline{ [0']'} \otimes \overline{\left[ \zeta\right] '}$ for $\zeta\in wAT(P)$.
	\hide{\par A similar issue occurred in the work of \cite{Bl:Al}. In turn, the authors introduced a congruence relation for connectors. Congruence relation is important in general because in contrast to equivalence relation, it allows one to replace an existing connector with another one, without causing undesirable alterations. In the next section, we introduce and study a notion of congruence relation for weighted connectors encoded in $wAC(P)$. In turn, we prove two theorems that ensure congruence between $wAC(P)$ connectors.}

	\section{On congruence relation for fusion-$wAC(P)$ connectors}\label{se7}
	
	In Section \ref{se5}, we defined the semantics of a connector $\zeta\in wAC(P)$ as an element $z\in wAI(P)$. Then the semantics of the latter was interpreted as a polynomial $\left\|z\right\|\in K\left\langle \Gamma(P)\right\rangle  $. Moreover, we stated that two $wAC(P)$ connectors $\zeta_1,\zeta_2$, are equivalent, i.e., $\zeta_{1}\equiv \zeta_{2}$ when $\left|\zeta_{1}\right| =\left|\zeta_{2}\right|$. By the semantics of $wAI(P)$ this implied that $\big\|\left|\zeta_{1}\right|\big\|(\gamma)=\big\|\left|\zeta_{2} \right|\big\|(\gamma)$ for every $\gamma\in \Gamma(P)$, i.e., equivalent $wAC(P)$ connectors return the same weight on the same interactions set. 
	\par In this section, we are interested in the congruence problem of $wAC(P)$ connectors. In \cite{Bl:Al}, the authors introduced a congruence relation for the Algebra of Connectors and provided conditions for proving their congruence. Congruence relation is important because in contrast to equivalence relation, it allows to use connectors interchangeably whenever it is required and without causing undesirable alterations in the architecture. 
	
	However, it occurs that providing a congruence relation in the weighted setup, it is not an easy task. Congruence relation implies that given two equivalent elements, if we apply any operator from the algebra on them, then we obtain again two equivalent
	elements \cite{Be:On,Br:St,Dr:Ha}. In our setting, applying the weighted fusion operator to $wAC(P)$ connectors would require specifying a typing. 
	In \cite{Bl:Al}, the authors resolved this issue using the syntactic equality resulting by the several axioms defined
	for the connectors. In the weighted framework we can only derive results by semantic equivalence, and hence we cannot follow a similar method. 
	For this, we restrict the congruence problem on fusion-$wAC(P)$ connectors which are typed by definition. In particular, we show that two equivalent fusion-$wAC(P)$ connectors are not in general interchangeable.
	In turn, we define a concept of congruence relation for fusion-$wAC(P)$ connectors,
	and we provide two theorems for proving such a congruence by extending the respective results from \cite{Bl:Al} in the weighted setup.
	
	Note that our concept of congruence relation is not
	a ``true'' congruence in terms that (i) it does 
	not apply to any $wAC(P)$ connector and (ii) 
	it takes equivalent fusion-$wAC(P)$ connectors and returns equivalent $wAC(P)$ connectors. 
	Extending our results for studying a congruence relation for 
	$wAC(P)$ connectors in general, is an interesting open problem that is left as future work.
	
	\hide{Moreover, applying directly the latter theorem we get an alternative proof that $[0']'$ serves as the neutral element for the weighted fusion operator in the subalgebra of triggers $wAT(P)$.}
	
	\hide{\par By the following example we show that two equivalent connectors cannot always be substituted for one another.}
	\begin{examp} \label{not-wcon}
		Consider the set of ports $P =\left\lbrace p, q\right\rbrace $ and let $k_{p}, k_{q}$, denote the weight of ports $p,q$, respectively. It can be easily verified that the \emph{fusion}-$wAC(P)$ connectors $[p]'$ and $[p]$ are equivalent, i.e., $[p]'\equiv [p]$. In order to prove the equivalence, we compute the $wAI(P)$ elements of the above connectors and we have $\big|[p]'\big|=p$ and $\big|[p]\big|=p$, respectively.
		\par Though, we show that the $wAC(P)$ connectors $\left[p\right]'\otimes \left[ q\right]$ and $\left[p\right]\otimes \left[ q\right]$ are not equivalent. Indeed, the $wAI(P)$ elements for each one of the above $wAC(P)$ connectors is $\big| \left[ p \right]'\otimes \left[ q \right] \big|=\left| p \right|\otimes (1\oplus \left| q \right|)=p\otimes (1\oplus q)=p\oplus (p\otimes q)$ and $\big|[p]\otimes [q]\big|=\left| p \right|\otimes \left| q \right|=p\otimes q$, respectively.
		We let $\gamma = \left\lbrace \left\lbrace p\right\rbrace,\left\lbrace p,q\right\rbrace \right\rbrace\in \Gamma(P)$. Then we compute the weight of the former $wAI(P)$ element on $\gamma$ and we get $\left\|p\otimes (1\oplus q)\right\|(\gamma)=k_{p}+ (k_{p}\cdot k_{q}).$ On the other hand, the weight of the latter $wAI(P)$ element on $\gamma$ is $\left\|p\otimes q\right\|(\gamma)
		=k_{p}\cdot k_{q}.$
		Consequently, we infer that $\left[ p \right]' \otimes \left[ q \right]$ and $\left[ p \right] \otimes \left[ q \right]$  are not equivalent. 
	\end{examp}
	\noindent By the previous example it occurs that when equivalent fusion-$wAC(P)$ connectors are differently typed, then the application of the weighted fusion operator does not preserve the equivalence. For this, next we introduce a concept of congruence relation for fusion-$wAC(P)$ connectors. 
	\begin{defin}\label{de1}
		We denote by `` $\cong$'' the largest congruence relation for \emph{fusion}-$wAC(P)$ connectors contained in $\equiv$ of $wAC(P)$, 
		i.e., the largest relation satisfying the following: For \emph{fusion}-$wAC(P)$ connectors $\zeta_{1}, \zeta_{2}$ and $r\notin P$,
		$$\zeta_{1} \cong\zeta_{2} \Rightarrow \forall E\in wAC\left( P\cup \left\lbrace r\right\rbrace \right), E(\zeta_{1}/r) \equiv E(\zeta_{2}/r)$$
		\noindent where $E(\zeta /r)$ denotes the expression obtained from $E$ by replacing all occurrences of $wAC(P)$ connector $r$ by $\zeta$.
	\end{defin}
	
	\hide{\noindent In the sequel, we refer to weighted congruence relation simply by congruence relation.} 
	
	\hide{\noindent Observe that by definition of congruence relation, the replacement of an arbitrary $\zeta \in wAC(P)$ in expressions with weighted fusion operator, would require specifying the respective typing. This justifies the fact that in this section, we studied the congruence problem for fusion-$wAC(P)$ connectors. In \cite{Bl:Al}, the authors resolved this issue by several axioms defined for their unweighed connectors. As it has been already mentioned, in the weighted framework we can only derive results by semantic equivalence, and hence studying congruence relation for	$wAC(P)$ connectors in general, is an open problem that could be investigated in future work.}
	
	\par Since the weighted fusion operator does not preserve the equivalence of fusion-$wAC(P)$ connectors, we need further conditions for proving their congruence. According to our first result presented below, it suffices to assign identical typing on equivalent fusion-$wAC(P)$ connectors. Hence, we show that two equivalent fusion-$wAC(P)$ connectors which are similarly typed, they are also congruent.
	
	\begin{theo}\label{eq4}
		Let $\zeta_{1}, \zeta_{2}$ be \emph{fusion}-$wAC(P)$ connectors. Then
		$$\zeta_{1}\equiv \zeta_{2} \Leftrightarrow \left[ \zeta_{1}\right] ^{\alpha}\cong \left[ \zeta_{2}\right] ^{\alpha} \
		\text{for any} \ \alpha \in \left\lbrace 0,1\right\rbrace .$$
	\end{theo}
	\begin{prof*}
		Firstly we prove the right-to-left implication. Assume that $\left[ \zeta_{1}\right] ^{\alpha}\cong \left[ \zeta_{2}\right] ^{\alpha}$, i.e., for every $E\in wAC\left( P\cup \left\lbrace r\right\rbrace \right) $ we have $E\left( \left[ \zeta_{1}\right] ^{\alpha}/r\right) \equiv E\left( \left[ \zeta_{2}\right] ^{\alpha}/r\right)$, and hence $\Big| E\left( \left[ \zeta_{1}\right] ^{\alpha}/r\right)\Big| =\Big| E\left( \left[ \zeta_{2}\right] ^{\alpha}/r\right)\Big|.$ For $E=r\in wAC\left( P\cup \left\lbrace r\right\rbrace \right)$ we get that $E\left( \left[ \zeta_{1}\right] ^{\alpha}/r\right) =\left[ \zeta_{1}\right] ^{\alpha}$ and $E\left( \left[ \zeta_{2}\right] ^{\alpha}/r\right) =\left[ \zeta_{2}\right] ^{\alpha}$. By assumption $\Big| E\left( \left[ \zeta_{1}\right] ^{\alpha}/r\right)\Big| =\Big| E\left( \left[ \zeta_{2}\right] ^{\alpha}/r\right)\Big|$, i.e., $\Big| \left[ \zeta_{1}\right] ^{\alpha}\Big| =\Big| \left[ \zeta_{2}\right] ^{\alpha}\Big|.$  We consider the following two cases:
		\begin{itemize}
			\item For $\alpha=0$ we have that $\big| \left[ \zeta_{1}\right] \big| =\big| \left[ \zeta_{2}\right] \big| \Rightarrow \left|\zeta_{1}\right|=\left|\zeta_{2}\right|$.
			\item For $\alpha=1$ we have that $\big| \left[ \zeta_{1}\right]' \big| =\big| \left[ \zeta_{2}\right]' \big| \Rightarrow \left|\zeta_{1}\right|=\left|\zeta_{2}\right|.$
		\end{itemize}
		
		\noindent In any case, we conclude that $\left|\zeta_{1}\right|=\left|\zeta_{2}\right|$. \hide{Therefore, we have proved that if $\left[ \zeta_{1}\right] ^{\alpha}\cong \left[ \zeta_{2}\right] ^{\alpha}$, then $\left|\zeta_{1}\right|=\left|\zeta_{2}\right|$.}
		\hide{which occurs directly, given that the relation ``$\cong$'' is the largest congruence relation contained in ``$\simeq$''.}
		\par Now we prove the left-to-right implication. We assume that $\zeta_{1}\equiv \zeta_{2}$, i.e., for the \emph{fusion}-$wAC(P)$ connectors $\zeta_{1}$ and $\zeta_{2}$ it holds that $\left|\zeta_{1}\right|=\left|\zeta_{2}\right|$. We have to prove that $\left[ \zeta_{1}\right] ^{\alpha}\cong \left[ \zeta_{2}\right] ^{\alpha}$. Therefore, we have to prove that for any expression $E\in wAC\left( P\cup \left\lbrace r\right\rbrace \right) $ it holds that $E(\left[\zeta_{1}\right]^{\alpha}/r) \equiv E(\left[\zeta_{2}\right]^{\alpha}/r)$. We assume that $r$ occurs only once in E. Otherwise, we apply the proof iteratively. By our assumption for $r$ it suffices to consider and prove the following equivalences:  
		\begin{enumerate}
			\item $\left[\zeta_{1}\right]^{\alpha} \equiv \left[\zeta_{2}\right]^{\alpha}$. For $\alpha = 0$, we have that $ \big| \left[\zeta_{1}\right] \big| = \left| \zeta_{1} \right|= \left| \zeta_{2} \right|$ and $\big| \left[\zeta_{2}\right] \big| = \left| \zeta_{2} \right|.$ On the other hand, for $\alpha = 1$, we have that $ \big| \left[\zeta_{1}\right]' \big| = \left| \zeta_{1} \right| = \left| \zeta_{2} \right|$ and $\big| \left[\zeta_{2}\right]' \big| = \left| \zeta_{2} \right|.$ Hence, $\left[\zeta_{1}\right]^{\alpha} \equiv \left[\zeta_{2}\right]^{\alpha}$ for $\alpha\in \lbrace 0,1\rbrace$.
			
			\item $\left[\zeta_{1}\right]^{\alpha} \oplus \zeta \equiv \left[\zeta_{2}\right]^{\alpha} \oplus \zeta $ where $\zeta \in wAC(P)$. For $\alpha = 0$, we have that $ \big| \left[\zeta_{1}\right] \oplus \zeta \big| = \big| \left[\zeta_{1}\right] \big| \oplus \left| \zeta \right| = \left| \zeta_{1} \right| \oplus \left| \zeta \right| = \left| \zeta_{2} \right| \oplus \left| \zeta \right| $ and $ \big| \left[\zeta_{2}\right] \oplus \zeta \big| = \big| \left[\zeta_{2}\right] \big| \oplus \left| \zeta \right| = \left| \zeta_{2} \right| \oplus \left| \zeta \right|. $ On the other hand, for $\alpha = 1$, we have that $ \big| \left[\zeta_{1}\right]' \oplus \zeta \big| = \big| \left[\zeta_{1}\right]' \big| \oplus \left| \zeta \right| = \left| \zeta_{1} \right| \oplus \left| \zeta \right| = \left| \zeta_{2} \right| \oplus \left| \zeta \right| $ and $ \big| \left[\zeta_{2}\right]' \oplus \zeta \big| = \big| \left[\zeta_{2}\right]' \big| \oplus \left| \zeta \right| = \left| \zeta_{2} \right| \oplus \left| \zeta \right|.$ Hence, $\left[\zeta_{1}\right]^{\alpha} \oplus \zeta \equiv \left[\zeta_{2}\right]^{\alpha} \oplus \zeta$ for $\alpha\in \lbrace 0,1\rbrace$.

			\item $\left[\zeta_{1}\right]^{\alpha} \otimes \left[ \xi_{1} \right]^{\alpha_{1}} \otimes \ldots \otimes\left[ \xi_{n} \right]^{\alpha_{n}} \equiv \left[\zeta_{2}\right]^{\alpha} \otimes \left[ \xi_{1} \right]^{\alpha_{1}} \otimes \ldots \otimes\left[ \xi_{n} \right]^{\alpha_{n}}$ where $\xi_{1},\ldots, \xi_{n}\in wAC(P)$ and $\alpha, \alpha_1,\ldots, \alpha_n\in \lbrace 0,1\rbrace$.
			\begin{itemize}
				\item For $\alpha=0$, we consider the following cases:
				\begin{itemize}
					\item[--] \textbf{\underline{Case 1}}: $\#_{T}(\left[ \xi_{1} \right]^{\alpha_{1}} \otimes \ldots \otimes\left[ \xi_{n} \right]^{\alpha_{n}}) = 0$. Then\\
					\begin{align*}
						&\big| \left[\zeta_{1}\right] \otimes \left[ \xi_{1} \right] \otimes \ldots \otimes\left[ \xi_{n} \right] \big| \\
						=&\left|\zeta_{1}\right|\otimes \left|\xi_{1}\right| \otimes \ldots \otimes \left|\xi_{n}\right| \\
						=&\left|\zeta_{2}\right|\otimes \left|\xi_{1}\right| \otimes \ldots \otimes \left|\xi_{n}\right|
					\end{align*}
					and
					\begin{align*}
						&\big| \left[\zeta_{2}\right] \otimes \left[ \xi_{1} \right] \otimes \ldots \otimes\left[ \xi_{n} \right] \big| \\
						=&\left|\zeta_{2}\right|\otimes \left|\xi_{1}\right| \otimes \ldots \otimes \left|\xi_{n}\right|.
					\end{align*}

					\item[--] \textbf{\underline{Case 2}}: $\#_{T}(\left[ \xi_{1} \right]^{\alpha_{1}} \otimes \ldots \otimes\left[ \xi_{n} \right]^{\alpha_{n}}) > 0$. Then\\
					
					\begin{align*}
						&\big| \left[\zeta_{1}\right] \otimes \left[ \xi_{1} \right]^{\alpha_{1}} \otimes \ldots \otimes\left[ \xi_{n} \right]^{\alpha_{n}} \big| \\
						=&\overset{}{\underset{\substack{i\in[n],\\\alpha_{i}=1}}{\bigoplus}} \Big( \left| \xi_{i} \right| \otimes \overset{}{\underset{\substack{k\neq i,\\\alpha_{k}\in \lbrace 0,1 \rbrace}}{\bigotimes}} (1 \oplus \left| \xi_{k}\right|)  \otimes (1\oplus \left|\zeta_{1}\right|) \Big) \\
						=& \overset{}{\underset{\substack{i\in[n],\\\alpha_{i}=1}}{\bigoplus}} \Big( \left| \xi_{i} \right| \otimes \overset{}{\underset{\substack{k\neq i,\\\alpha_{k}\in \lbrace 0,1 \rbrace}}{\bigotimes}} (1 \oplus \left| \xi_{k}\right|)  \otimes (1\oplus \left|\zeta_{2}\right|) \Big) \\
					\end{align*}
					and
					\begin{align*}
						&\big| \left[\zeta_{2}\right] \otimes \left[ \xi_{1} \right]^{\alpha_{1}} \otimes \ldots \otimes\left[ \xi_{n} \right]^{\alpha_{n}} \big| \\
						=& \overset{}{\underset{\substack{i\in[n],\\\alpha_{i}=1}}{\bigoplus}} \Big( \left| \xi_{i} \right| \otimes \overset{}{\underset{\substack{k\neq i,\\\alpha_{k}\in \lbrace 0,1 \rbrace}}{\bigotimes}} (1 \oplus \left| \xi_{k}\right|)  \otimes (1\oplus \left|\zeta_{2}\right|) \Big)  .\\
					\end{align*}
				\end{itemize}

				\item For $\alpha=1$, we consider the respective cases:
				
				\begin{itemize}
					\item[--] \textbf{\underline{Case 1}}: $\#_{T}(\left[ \xi_{1} \right]^{\alpha_{1}} \otimes \ldots \otimes\left[ \xi_{n} \right]^{\alpha_{n}}) = 0$. Then\\
					
					\begin{align*}
						&\big| \left[\zeta_{1}\right]' \otimes \left[ \xi_{1} \right] \otimes \ldots \otimes\left[ \xi_{n} \right] \big|\\
						=& \left|\zeta_{1}\right| \otimes \overset{}{\underset{l\in[n]}{\bigotimes}} (1\oplus \left| \xi_{l} \right|) \\
						=& \left|\zeta_{2}\right| \otimes \overset{}{\underset{l\in[n]}{\bigotimes}} (1\oplus \left| \xi_{l} \right|) \
					\end{align*}
					and
					\begin{align*}
						&\big| \left[\zeta_{2}\right]' \otimes \left[ \xi_{1} \right] \otimes \ldots \otimes\left[ \xi_{n} \right] \big| \\
						=& \left|\zeta_{2}\right| \otimes \overset{}{\underset{l\in[n]}{\bigotimes}} (1\oplus \left| \xi_{l} \right|) .   
					\end{align*}
					
					\item[--] \textbf{\underline{Case 2}}: $\#_{T}(\left[ \xi_{1} \right]^{\alpha_{1}} \otimes \ldots \otimes\left[ \xi_{n} \right]^{\alpha_{n}}) > 0$. Then\\	
					\begin{align*}
						&\big| \left[\zeta_{1}\right]' \otimes \left[ \xi_{1} \right]^{\alpha_{1}} \otimes \ldots \otimes\left[ \xi_{n} \right]^{\alpha_{n}} \big|\\
						=& \Big(\left|\zeta_{1}\right| \otimes \overset{}{\underset{l\in[n]}{\bigotimes}} (1\oplus \left| \xi_{l} \right|) \Big) \oplus \overset{}{\underset{\substack{i\in[n],\\\alpha_{i}=1}}{\bigoplus}} \Big( \left| \xi_{i} \right| \otimes \overset{}{\underset{\substack{k\neq i,\\\alpha_{k}\in \lbrace 0,1 \rbrace}}{\bigotimes}} (1 \oplus \left| \xi_{k}\right|) \otimes (1\oplus \left|\zeta_{1}\right|) \Big) \\
						=& \Big(\left|\zeta_{2}\right| \otimes \overset{}{\underset{l\in[n]}{\bigotimes}} (1\oplus \left| \xi_{l} \right|) \Big) \oplus \overset{}{\underset{\substack{i\in[n],\\\alpha_{i}=1}}{\bigoplus}} \Big( \left| \xi_{i} \right| \otimes \overset{}{\underset{\substack{k\neq i,\\\alpha_{k}\in \lbrace 0,1 \rbrace}}{\bigotimes}} (1 \oplus \left| \xi_{k}\right|) \otimes (1\oplus \left|\zeta_{2}\right|) \Big) \\
					\end{align*}
					and
					\begin{align*}
						&\big| \left[\zeta_{2}\right]' \otimes \left[ \xi_{1} \right]^{\alpha_{1}} \otimes \ldots \otimes\left[ \xi_{n} \right]^{\alpha_{n}} \big| \\
						=& \Big(\left|\zeta_{2}\right| \otimes \overset{}{\underset{l\in[n]}{\bigotimes}} (1\oplus \left| \xi_{l} \right|) \Big) \oplus \overset{}{\underset{\substack{i\in[n],\\\alpha_{i}=1}}{\bigoplus}} \Big( \left| \xi_{i} \right| \otimes \overset{}{\underset{\substack{k\neq i,\\\alpha_{k}\in \lbrace 0,1 \rbrace}}{\bigotimes}} (1 \oplus \left| \xi_{k}\right|) \otimes (1\oplus \left|\zeta_{2}\right|) \Big) . 
					\end{align*}
				\end{itemize}
			\end{itemize}
			\noindent	Hence, $\left[\zeta_{1}\right]^{\alpha} \otimes \left[ \xi_{1} \right]^{\alpha_{1}} \otimes \ldots \otimes \left[ \xi_{n} \right]^{\alpha_{n}} \equiv \left[\zeta_{2}\right]^{\alpha} \otimes \left[ \xi_{1} \right]^{\alpha_{1}} \otimes \ldots \otimes \left[ \xi_{n} \right]^{\alpha_{n}}$, for $\alpha,\alpha_{1},\ldots, \alpha_{n}\in \lbrace 0,1\rbrace$.
			
			\item The symmetric case of 2 and the cases for any other position of $\left[\zeta_{1}\right]^{\alpha}$, $\left[\zeta_{2}\right]^{\alpha}$, where $\alpha \in \lbrace 0,1\rbrace$, in 3. For these cases the proof is analogous. 
		\end{enumerate}
		Finally, for any other form of the expression $E$ we apply the presented cases iteratively, and our proof is completed.\qed
	\end{prof*}

	\noindent Next we prove our second theorem which provides us with a method for checking whether two fusion-$wAC(P)$ connectors are congruent. According to this result, two fusion-$wAC(P)$ connectors can be congruent if they are equivalent, their respective weighted fusion with the $wAC(P)$ connector $[1]'$ preserves their equivalence, 
	and their degree is simultaneously zero or strictly positive. 
	The second condition of the theorem actually serves as the least condition required in order to maintain the equivalence under the weighted fusion operator. 
	Hence, as it is shown by the proof of the theorem, when equivalence is preserved under the weighted fusion for each one of the equivalent connectors with 
	the trivial $wAC(P)$ connector $[1]'$, then it is preserved for its application with any other $wAC(P)$ connector. 
	Regarding the other two conditions of the theorem, they are obviously necessary, since fusion-$wAC(P)$ connectors which are not equivalent cannot be congruent, and similarly for the case that the degree of only one of the two given fusion-$wAC(P)$ connectors is zero or strictly positive. For the proof of the theorem we need the following proposition.
	
	\begin{prop}\label{induct}
		Let $\zeta = \left[ \zeta_{1} \right]^{\alpha_{1}} \otimes \ldots \otimes  \left[ \zeta_{n} \right]^{\alpha_{n}}$, where $\zeta_{i} \in wAC(P)$ and $\alpha_{i} \in \left\lbrace 0,1\right\rbrace $ for $i \in [n]$. Then 
		$$ |\zeta|  \oplus \bigotimes_{i\in [n]} (1\oplus \left| \zeta_{i} \right|) = \bigotimes_{i\in [n]} (1\oplus \left| \zeta_{i} \right|). $$
	\end{prop}
	
	\begin{prof*} We prove the above equality by induction on $n$ and by Corollary \ref{walg_semi}. \qed
	\end{prof*}

	\begin{theo}\label{th1}
		Let $\zeta_{1},\zeta_{2}$ be \emph{fusion}-$wAC(P)$ connectors. Then
		\begin{equation*}
			\zeta_{1}\cong \zeta_{2} \Leftrightarrow \begin{cases*}
				\zeta_{1}\equiv \zeta_{2} \\
				\zeta_{1}\otimes [1]'\equiv \zeta_{2}\otimes [1]'\\
				\#_{Τ}\zeta_{1}>0 \Leftrightarrow \#_{Τ}\zeta_{2}>0.
			\end{cases*}
		\end{equation*}
	\end{theo}
	\begin{prof*}
		We prove the left-to-right implication. We assume that $\zeta_{1}\cong \zeta_{2}$. Hence, for every expression  $E\in wAC(P\cup \left\lbrace r\right\rbrace )$ it holds that $E(\zeta_{1}/r) \equiv E(\zeta_{2}/r)$. Let $E_{1}=r\in wAC\left( P\cup \left\lbrace r\right\rbrace \right) $. Then we get that $E_{1}(\zeta_{1}/r)\equiv E_{1}(\zeta_{2}/r)$, i.e., $\zeta_{1}\equiv \zeta_{2}$. Furthermore, for $E_{2}=r\otimes [1]'\in wAC\left( P\cup \left\lbrace r\right\rbrace \right) $, it holds that $E_{2}(\zeta_{1}/r)\equiv E_{2}(\zeta_{2}/r)$ and hence $\zeta_{1}\otimes [1]'\equiv \zeta_{2}\otimes [1]'$. Now let $\xi=\left[ \xi_{1}\right]^{\alpha_{1}} \otimes \ldots \otimes \left[ \xi_{n}\right]^{\alpha_{n}} \in wAC(P)$ for $\alpha_1, \ldots, \alpha_n\in \lbrace 0,1 \rbrace$\hide{and we consider the $wAC(P)$ connectors $\zeta_{1}\otimes \xi$ and $\zeta_{2}\otimes \xi$}. We claim that $\#_{T}\zeta_{1}>0 \Leftrightarrow \#_{T}\zeta_{2}>0$. On the contrary, we let $\#_{T}\zeta_{1}>0$ and $\#_{T}\zeta_{2}=0$. Specifically, let $\zeta_{1}=[p]', \zeta_{2}=[p]$ and $\xi=[q]$. Then it holds that $\left| \zeta_{1}\otimes \xi \right| = \big| [p]' \otimes [q] \big| = \left|p\right| \otimes (1\oplus \left|q\right|) = p\otimes (1\oplus q)=p \oplus (p\otimes q)$ and $\left| \zeta_{2}\otimes \xi\right| = \big| [p]\otimes [q]\big| = \left|p\right| \otimes \left|q\right|=p\otimes q$. That is, for $E_{3}=r \otimes \xi \in wAC(P\cup \left\lbrace r\right\rbrace) $ it does not hold that $E(\zeta_{1}/r) \equiv E(\zeta_{2}/r)$, which is a contradiction, since $\zeta_{1} \cong \zeta_{2}$. Hence, $\#_{T}\zeta_{1}>0 \Leftrightarrow \#_{T}\zeta_{2} >0$.
		\par Now we prove the right-to-left implication. Since $\zeta_{1}$ and $\zeta_{2}$ are \emph{fusion}-$wAC(P)$ connectors, we let $\zeta_{1} = \left[ \zeta_{1,1} \right]^{\alpha_{1}} \otimes \ldots \otimes  \left[ \zeta_{1,n} \right]^{\alpha_{n}}$ and $\zeta_{2} = \left[ \zeta_{2,1} \right]^{\beta_{1}} \otimes \ldots \otimes  \left[ \zeta_{2,m} \right]^{\beta_{m}}$, where $\zeta_{1,i}, \zeta_{2,j}\in wAC(P)$ and $\alpha_{i}, \beta_{j}\in \left\lbrace 0,1\right\rbrace $, for $i\in [n]$ and $j\in [m]$. We assume that (i) $\zeta_{1}\equiv \zeta_{2}$, i.e., $\left|\zeta_{1}\right|=\left|\zeta_{2}\right|$, (ii) $ \zeta_{1}\otimes [1]' \equiv \zeta_{2} \otimes [1]'$ which implies that $\left| \zeta_{1} \otimes [1]'\right| = \left|\zeta_{2}\otimes [1]'\right|$, and (iii) $\#_{Τ}\zeta_{1} >0 \Leftrightarrow \#_{Τ}\zeta_{2}>0$. We have to prove that $\zeta_{1} \cong \zeta_{2}$, thus we have to prove that for any expression $E\in wAC\left( P\cup \left\lbrace r\right\rbrace \right)$, it holds that $ E(\zeta_{1}/r)\equiv  E(\zeta_{2}/r)$.We assume that $r$ occurs only once in $E$, and otherwise we apply the proof iteratively. By this assumption it suffices to consider and prove the following:
		\begin{enumerate}
			\item $\zeta_{1}\equiv \zeta_{2} $, which holds directly since $\left|\zeta_{1}\right|=\left|\zeta_{2}\right|$.
			
			\item $\zeta_{1} \oplus \zeta \equiv \zeta_{2} \oplus \zeta $ where $\zeta\in wAC(P)$. The equivalence holds since $\left| \zeta_{1}\oplus \zeta \right| = \left|\zeta_{1} \right| \oplus \left|\zeta\right| = \left|\zeta_{2} \right| \oplus \left|\zeta\right| \hide{= \left|\zeta_{2} \right| \oplus \left|\zeta\right|}$ and $\left| \zeta_{2}\oplus \zeta \right| = \left|\zeta_{2} \right| \oplus \left|\zeta\right|.$

			\item  $\zeta_{1}\otimes \left[ \xi_{1} \right]^{\delta_{1}} \otimes \ldots \otimes  \left[ \xi_{r} \right]^{\delta_{r}} \equiv \zeta_{2}\otimes \left[ \xi_{1} \right]^{\delta_{1}} \otimes \ldots \otimes  \left[ \xi_{r} \right]^{\delta_{r}}$ where $ \xi_{1}, \ldots, \xi_{r} \in wAC(P)$ and $\delta_{1},\ldots, \delta_{r} \in \lbrace 0,1\rbrace$. We consider the following cases:
			\begin{itemize}
				\item[--] \textbf{\underline{Case 1}}: ($\#_{T}\zeta_{1} = 0, \#_{T}\zeta_{2} = 0$)\\
				It holds that 
				\begin{align*}\label{s2}
					\left|\zeta_{1}\right| = \left|\zeta_{2}\right| \Rightarrow& \big| \left[ \zeta_{1,1} \right] \otimes \ldots \otimes  \left[ \zeta_{1,n} \right] \big| = \big| \left[ \zeta_{2,1} \right] \otimes \ldots \otimes  \left[ \zeta_{2,m} \right] \big| \\
					\Rightarrow& \left|\zeta_{1,1}\right| \otimes \ldots \otimes \left|\zeta_{1,n}\right| = \left|\zeta_{2,1}\right| \otimes \ldots \otimes \left|\zeta_{2,m}\right|. \tag{$\Sigma_{1}$}
				\end{align*}
				and
				\begin{align*}\label{s4}
					\left|\zeta_{1} \otimes [1]'\right| = \left|\zeta_{2}\otimes [1]'\right| \Rightarrow & \big| \left[ \zeta_{1,1} \right] \otimes \ldots \otimes  \left[ \zeta_{1,n} \right] \otimes [1]' \big| = \big| \left[ \zeta_{2,1} \right] \otimes \ldots \otimes  \left[ \zeta_{2,m} \right] \otimes [1]' \big|\\
					\Rightarrow& \left|1\right| \otimes \overset{}{\underset{i\in [n]}{\bigotimes}} (1\oplus \left|\zeta_{1,i}\right|) = \left|1\right| \otimes \overset{}{\underset{j\in [m]}{\bigotimes}} (1\oplus \left|\zeta_{2,j}\right|)\\
					\Rightarrow& \overset{}{\underset{i\in [n]}{\bigotimes}} (1\oplus \left|\zeta_{1,i}\right|) =  \overset{}{\underset{j\in [m]}{\bigotimes}} (1\oplus \left|\zeta_{2,j}\right|). \tag{$\Sigma_{2}$}
				\end{align*}
				Then we consider the following cases:
				\begin{itemize}
					\item[$\blacktriangleright$] If $\#_{T} \big(\left[ \xi_{1} \right]^{\delta_{1}} \otimes \ldots \otimes  \left[ \xi_{r} \right]^{\delta_{r}}\big) = 0$, then we have that 
					\begin{align*}
						\left|\zeta_{1} \otimes \xi\right|=& \big|\left[ \zeta_{1,1} \right] \otimes \ldots \otimes  \left[ \zeta_{1,n} \right]\otimes \left[ \xi_{1} \right] \otimes \ldots \otimes  \left[ \xi_{r} \right]\big|\\
						=& \left| \zeta_{1,1} \right| \otimes \ldots \otimes  \left| \zeta_{1,n} \right|\otimes \left| \xi_{1} \right| \otimes \ldots \otimes  \left| \xi_{r} \right|\\
						\overset{\eqref{s2}}{=}& \left| \zeta_{2,1} \right| \otimes \ldots \otimes  \left| \zeta_{2,m} \right|\otimes \left| \xi_{1} \right| \otimes \ldots \otimes  \left| \xi_{r} \right|
					\end{align*}
					and 
					\begin{align*}
						\left|\zeta_{2} \otimes \xi\right|=& \big|\left[ \zeta_{2,1} \right] \otimes \ldots \otimes  \left[ \zeta_{2,m} \right]\otimes \left[ \xi_{1} \right] \otimes \ldots \otimes  \left[ \xi_{r} \right]\big|\\
						=& \left| \zeta_{2,1} \right| \otimes \ldots \otimes  \left| \zeta_{2,m} \right|\otimes \left| \xi_{1} \right| \otimes \ldots \otimes  \left| \xi_{r} \right|.
					\end{align*}
					\item[$\blacktriangleright$] If $\#_{T} \big(\left[ \xi_{1} \right]^{\delta_{1}} \otimes \ldots \otimes  \left[ \xi_{r} \right]^{\delta_{r}}\big) > 0$, then we have that
					\begin{align*}
						\left|\zeta_{1} \otimes \xi\right|=& \big|\left[ \zeta_{1,1} \right] \otimes \ldots \otimes  \left[ \zeta_{1,n} \right]\otimes \left[ \xi_{1} \right]^{\delta_{1}} \otimes \ldots \otimes  \left[ \xi_{r} \right]^{\delta_{r}}\big|\\
						=& \overset{}{\underset{\substack{\lambda\in[r],\\\delta_{\lambda}=1}}{\bigoplus}} \Big( \left|\xi_{\lambda}\right| \otimes \overset{}{\underset{\substack{\mu\neq \lambda,\\\delta_{\mu} \in \lbrace 0,1 \rbrace}}{\bigotimes}} (1\oplus \left|\xi_{\mu}\right|) \otimes  \overset{}{\underset{i\in [n]}{\bigotimes}} (1\oplus \left|\zeta_{1,i}\right|)\Big) \\
						=&\Bigg( \overset{}{\underset{\substack{\lambda\in[r],\\\delta_{\lambda}=1}}{\bigoplus}} \Big( \left|\xi_{\lambda}\right| \otimes \overset{}{\underset{\substack{\mu\neq \lambda,\\\delta_{\mu}\in \lbrace 0,1 \rbrace}}{\bigotimes}} (1\oplus \left|\xi_{\mu}\right|)  \Big)\Bigg) \otimes  \overset{}{\underset{i\in [n]}{\bigotimes}} (1\oplus \left|\zeta_{1,i}\right|) \\
						\overset{\eqref{s4}}{=}& \Bigg( \overset{}{\underset{\substack{\lambda\in[r],\\\delta_{\lambda}=1}}{\bigoplus}} \Big( \left|\xi_{\lambda}\right| \otimes \overset{}{\underset{\substack{\mu\neq \lambda,\\\delta_{\mu}\in \lbrace 0,1 \rbrace}}{\bigotimes}} (1\oplus \left|\xi_{\mu}\right|)  \Big)\Bigg) \otimes \overset{}{\underset{j\in [m]}{\bigotimes}} (1\oplus \left|\zeta_{2,j}\right|)
					\end{align*}
					and 
					\begin{align*}
						\left|\zeta_{2} \otimes \xi\right|=& \big|\left[ \zeta_{2,1} \right] \otimes \ldots \otimes  \left[ \zeta_{2,m} \right]\otimes \left[ \xi_{1} \right]^{\delta_{1}} \otimes \ldots \otimes  \left[ \xi_{r} \right]^{\delta_{r}}\big|\\
						=& \overset{}{\underset{\substack{\lambda\in[r],\\\delta_{\lambda}=1}}{\bigoplus}} \Big( \left|\xi_{\lambda}\right| \otimes \overset{}{\underset{\substack{\mu\neq \lambda,\\\delta_{\mu}\in \lbrace 0,1 \rbrace}}{\bigotimes}} (1\oplus \left|\xi_{\mu}\right|) \otimes  \overset{}{\underset{j\in [m]}{\bigotimes}} (1\oplus \left|\zeta_{2,j}\right|)\Big) \\
						=& \Bigg( \overset{}{\underset{\substack{\lambda\in[r],\\\delta_{\lambda}=1}}{\bigoplus}} \Big( \left|\xi_{\lambda}\right| \otimes \overset{}{\underset{\substack{\mu\neq \lambda,\\\delta_{\mu}\in \lbrace 0,1 \rbrace}}{\bigotimes}} (1\oplus \left|\xi_{\mu}\right|)  \Big)\Bigg) \otimes  \overset{}{\underset{j\in [m]}{\bigotimes}} (1\oplus \left|\zeta_{2,j}\right|). 
					\end{align*} Hence, we obtain that $\zeta_{1}\otimes \left[ \xi_{1} \right]^{\delta_{1}} \otimes \ldots \otimes  \left[ \xi_{r} \right]^{\delta_{r}} \equiv \zeta_{2}\otimes \left[ \xi_{1} \right]^{\delta_{1}} \otimes \ldots \otimes  \left[ \xi_{r} \right]^{\delta_{r}}$ for Case 1.\\

					\item[--] \textbf{\underline{Case 2}}: ($\#_{T}\zeta_{1} > 0, \#_{T}\zeta_{2} > 0$)\\
					It holds that 
					\begin{align*}\label{s3}
						\left|\zeta_{1}\right|=\left|\zeta_{2}\right| \Rightarrow& \big| \left[ \zeta_{1,1} \right]^{\alpha_{1}} \otimes \ldots \otimes  \left[ \zeta_{1,n} \right]^{\alpha_{n}} \big| = \big| \left[ \zeta_{2,1} \right]^{\beta_{1}} \otimes \ldots \otimes  \left[ \zeta_{2,m} \right]^{\beta_{m}} \big| \\
						\Rightarrow& \overset{}{\underset{\substack{i_{1}\in[n],\\\alpha_{i_{1}}=1}}{\bigoplus}} \Big( \left|\zeta_{1,i_{1}}\right| \otimes \overset{}{\underset{\substack{k_{1}\neq i_{1},\\\alpha_{k_{1}}\in \lbrace 0,1 \rbrace}}{\bigotimes}} (1\oplus \left|\zeta_{1,k_{1}}\right|)  \Big) \\
						& = \overset{}{\underset{\substack{i_{2}\in[m],\\\beta_{i_{2}}=1}}{\bigoplus}} \Big( \left|\zeta_{2,i_{2}}\right| \otimes \overset{}{\underset{\substack{k_{2}\neq i_{2},\\\alpha_{k_{2}}\in \lbrace 0,1 \rbrace}}{\bigotimes}} (1\oplus \left|\zeta_{2,k_{2}}\right|) \Big). \tag{$\Sigma_{3}$} 
					\end{align*}
					and 
					\begin{align*}\label{s5}
						&\left|\zeta_{1} \otimes [1]'\right| = \left|\zeta_{2}\otimes [1]'\right|  \\
						\Rightarrow& \big| \left[ \zeta_{1,1} \right]^{\alpha_{1}} \otimes \ldots \otimes  \left[ \zeta_{1,n} \right]^{\alpha_{n}} \otimes [1]' \big| = \big| \left[ \zeta_{2,1} \right]^{\beta_{1}} \otimes \ldots \otimes  \left[ \zeta_{2,m} \right]^{\beta_{m}} \otimes [1]' \big|\\
						\Rightarrow& \overset{}{\underset{\substack{i_{1}\in[n],\\\alpha_{i_{1}}=1}}{\bigoplus}} \Big( \left|\zeta_{1,i_{1}}\right| \otimes \overset{}{\underset{\substack{k_{1}\neq i_{1},\\\alpha_{k_{1}}\in \lbrace 0,1\rbrace}}{\bigotimes}} (1\oplus \left|\zeta_{1,k_{1}}\right|) \otimes (1\oplus \left|1\right|) \Big) \oplus \left|1\right| \otimes \overset{}{\underset{i\in [n]}{\bigotimes}} (1\oplus \left|\zeta_{1,i}\right|) \\
						=& \overset{}{\underset{\substack{i_{2}\in[m],\\\beta_{i_{2}}=1}}{\bigoplus}} \Big( \left|\zeta_{2,i_{2}}\right| \otimes \overset{}{\underset{\substack{k_{2}\neq i_{2},\\\beta_{k_{2}}\in \lbrace 0,1\rbrace}}{\bigotimes}} (1\oplus \left|\zeta_{2,k_{2}}\right|) \otimes (1\oplus \left|1\right|) \Big) \oplus \left|1\right| \otimes \overset{}{\underset{j\in [m]}{\bigotimes}} (1\oplus \left|\zeta_{2,j}\right|) \\
						\Rightarrow & \overset{}{\underset{\substack{i_{1}\in[n],\\\alpha_{i_{1}}=1}}{\bigoplus}} \Big( \left|\zeta_{1,i_{1}}\right| \otimes \overset{}{\underset{\substack{k_{1}\neq i_{1},\\\alpha_{k_{1}}\in \lbrace 0,1\rbrace}}{\bigotimes}} (1\oplus \left|\zeta_{1,k_{1}}\right|) \Big) \oplus  \overset{}{\underset{i\in [n]}{\bigotimes}} (1\oplus \left|\zeta_{1,i}\right|) \\ 
						=& \overset{}{\underset{\substack{i_{2}\in[m],\\\beta_{i_{2}}=1}}{\bigoplus}} \Big( \left|\zeta_{2,i_{2}}\right| \otimes \overset{}{\underset{\substack{k_{2}\neq i_{2},\\\beta_{k_{2}}\in \lbrace 0,1\rbrace}}{\bigotimes}} (1\oplus \left|\zeta_{2,k_{2}}\right|) \Big) \oplus  \overset{}{\underset{j\in [m]}{\bigotimes}} (1\oplus \left|\zeta_{2,j}\right|) \\
						\Rightarrow& \overset{}{\underset{i\in [n]}{\bigotimes}} (1\oplus \left|\zeta_{1,i}\right|) = \overset{}{\underset{j\in [m]}{\bigotimes}} (1\oplus \left|\zeta_{2,j}\right|),  \tag{$\Sigma_{4}$}
					\end{align*}
					where the last step holds by Proposition \ref{induct}. Then we consider the following cases:\\
					\begin{itemize}
						\item[$\blacktriangleright$] If $\#_{T} \big(\left[ \xi_{1} \right]^{\delta_{1}} \otimes \ldots \otimes  \left[ \xi_{r} \right]^{\delta_{r}}\big) = 0$, then we have that 
						\begin{align*}
							\left|\zeta_{1} \otimes \xi\right|=& \big|\left[ \zeta_{1,1} \right]^{\alpha_{1}} \otimes \ldots \otimes  \left[ \zeta_{1,n} \right]^{\alpha_{n}}\otimes \left[ \xi_{1} \right] \otimes \ldots \otimes  \left[ \xi_{r} \right]\big|\\
							=& \overset{}{\underset{\substack{i_{1}\in[n],\\\alpha_{i_{1}}=1}}{\bigoplus}} \Big( \left|\zeta_{1,i_{1}}\right| \otimes \overset{}{\underset{\substack{k_{1}\neq i_{1},\\\alpha_{k_{1}}\in \lbrace 0,1\rbrace}}{\bigotimes}} (1\oplus \left|\zeta_{1,k_{1}}\right|) \otimes \overset{}{\underset{l\in [r]}{\bigotimes}} (1\oplus \left|\xi_{l}\right|)\Big)  \\
							=&  \Bigg( \overset{}{\underset{\substack{i_{1}\in[n],\\\alpha_{i_{1}}=1}}{\bigoplus}} \Big( \left|\zeta_{1,i_{1}}\right| \otimes \overset{}{\underset{\substack{k_{1}\neq i_{1},\\\alpha_{k_{1}}\in \lbrace 0,1\rbrace}}{\bigotimes}} (1\oplus \left|\zeta_{1,k_{1}}\right|)\Big) \Bigg)\otimes \overset{}{\underset{l\in [r]}{\bigotimes}} (1\oplus \left|\xi_{l}\right|)  \\
							\overset{\eqref{s3}}{=}& \Bigg( \overset{}{\underset{\substack{i_{2}\in[m],\\\beta_{i_{2}}=1}}{\bigoplus}} \Big( \left|\zeta_{2,i_{2}}\right| \otimes \overset{}{\underset{\substack{k_{2}\neq i_{2},\\\beta_{k_{2}}\in \lbrace 0,1\rbrace}}{\bigotimes}} (1\oplus \left|\zeta_{2,k_{2}}\right|) \Big) \Bigg)\otimes \overset{}{\underset{l\in [r]}{\bigotimes}} (1\oplus \left|\xi_{l}\right|) 
						\end{align*}
						and 
						\begin{align*}
							\left|\zeta_{2} \otimes \xi\right|=& \big|\left[ \zeta_{2,1} \right]^{\beta_{1}} \otimes \ldots \otimes  \left[ \zeta_{2,m} \right]^{\beta_{m}}\otimes \left[ \xi_{1} \right] \otimes \ldots \otimes  \left[ \xi_{r} \right]\big|\\
							=& \overset{}{\underset{\substack{i_{2}\in[m],\\\beta_{i_{2}}=1}}{\bigoplus}} \Big( \left|\zeta_{2,i_{2}}\right| \otimes \overset{}{\underset{\substack{k_{2}\neq i_{2},\\\beta_{k_{2}}\in \lbrace 0,1\rbrace}}{\bigotimes}} (1\oplus \left|\zeta_{2,k_{2}}\right|) \otimes \overset{}{\underset{l\in [r]}{\bigotimes}} (1\oplus \left|\xi_{l}\right|)\Big) \\
							=& \Bigg( \overset{}{\underset{\substack{i_{2}\in[m],\\\beta_{i_{2}}=1}}{\bigoplus}} \Big( \left|\zeta_{2,i_{2}}\right| \otimes \overset{}{\underset{\substack{k_{2}\neq i_{2},\\\beta_{k_{2}}\in \lbrace 0,1\rbrace}}{\bigotimes}} (1\oplus \left|\zeta_{2,k_{2}}\right|)\Big) \Bigg)\otimes \overset{}{\underset{l\in [r]}{\bigotimes}} (1\oplus \left|\xi_{l}\right|). 
						\end{align*}
						
						\item[$\blacktriangleright$] If $\#_{T} \big(\left[ \xi_{1} \right]^{\delta_{1}} \otimes \ldots \otimes  \left[ \xi_{r} \right]^{\delta_{r}}\big) > 0$, then we have that 
						\begin{align*}
							\left|\zeta_{1} \otimes \xi\right|=& \big|\left[ \zeta_{1,1} \right]^{\alpha_{1}} \otimes \ldots \otimes  \left[ \zeta_{1,n} \right]^{\alpha_{n}}\otimes \left[ \xi_{1} \right]^{\delta_{1}} \otimes \ldots \otimes  \left[ \xi_{r} \right]^{\delta_{r}}\big|\\
							=&  \overset{}{\underset{\substack{i_{1}\in[n],\\\alpha_{i_{1}}=1}}{\bigoplus}} \Big( \left|\zeta_{1,i_{1}}\right| \otimes \overset{}{\underset{\substack{k_{1}\neq i_{1},\\\alpha_{k_{1}}\in \lbrace 0,1\rbrace}}{\bigotimes}} (1\oplus \left|\zeta_{1,k_{1}}\right|) \otimes \overset{}{\underset{l\in [r]}{\bigotimes}} (1\oplus \left|\xi_{l}\right|)\Big) \oplus\\
							&\overset{}{\underset{\substack{\lambda\in[r],\\\delta_{\lambda}=1}}{\bigoplus}} \Big( \left|\xi_{\lambda}\right| \otimes \overset{}{\underset{\substack{\mu\neq \lambda,\\\delta_{\mu}\in \lbrace 0,1\rbrace}}{\bigotimes}} (1\oplus \left|\xi_{\mu}\right|) \otimes \overset{}{\underset{i\in [n]}{\bigotimes}} (1\oplus \left|\zeta_{1,i}\right|)\Big) \\
							=& \Bigg(\overset{}{\underset{\substack{i_{1}\in[n],\\\alpha_{i_{1}}=1}}{\bigoplus}} \Big( \left|\zeta_{1,i_{1}}\right| \otimes \overset{}{\underset{\substack{k_{1}\neq i_{1},\\\alpha_{k_{1}}\in \lbrace 0,1\rbrace}}{\bigotimes}} (1\oplus \left|\zeta_{1,k_{1}}\right|) \Big)\Bigg) \otimes \overset{}{\underset{l\in [r]}{\bigotimes}} (1\oplus \left|\xi_{l}\right|) \oplus\\
							&\Bigg( \overset{}{\underset{\substack{\lambda\in[r],\\\delta_{\lambda}=1}}{\bigoplus}} \Big( \left|\xi_{\lambda}\right| \otimes \overset{}{\underset{\substack{\mu\neq \lambda,\\\delta_{\mu}\in \lbrace 0,1\rbrace}}{\bigotimes}} (1\oplus \left|\xi_{\mu}\right|) \Big)\Bigg) \otimes \overset{}{\underset{i\in [n]}{\bigotimes}} (1\oplus \left|\zeta_{1,i}\right|) \\
							\overset{\eqref{s3}}{\underset{\eqref{s5}}{=}}& \Bigg(  \overset{}{\underset{\substack{i_{2}\in[m],\\\beta_{i_{2}}=1}}{\bigoplus}} \Big( \left|\zeta_{2,i_{2}}\right| \otimes \overset{}{\underset{\substack{k_{2}\neq i_{2},\\\beta_{k_{2}}\in \lbrace 0,1\rbrace}}{\bigotimes}} (1\oplus \left|\zeta_{2,k_{2}}\right|) \Big)\Bigg) \otimes \overset{}{\underset{l\in [r]}{\bigotimes}} (1\oplus \left|\xi_{l}\right|) \oplus\\
							&  \Bigg(\overset{}{\underset{\substack{\lambda\in[r],\\\delta_{\lambda}=1}}{\bigoplus}} \Big( \left|\xi_{\lambda}\right| \otimes \overset{}{\underset{\substack{\mu\neq \lambda,\\\delta_{\mu}\in \lbrace 0,1\rbrace}}{\bigotimes}} (1\oplus \left|\xi_{\mu}\right|) \Big)\Bigg) \otimes \overset{}{\underset{j\in [m]}{\bigotimes}} (1\oplus \left|\zeta_{2,j}\right|) 
						\end{align*}
						and 
						\begin{align*}
							\left|\zeta_{2} \otimes \xi\right|=& \big|\left[ \zeta_{2,1} \right]^{\beta_{1}} \otimes \ldots \otimes  \left[ \zeta_{2,m} \right]^{\beta_{m}}\otimes \left[ \xi_{1} \right]^{\delta_{1}} \otimes \ldots \otimes  \left[ \xi_{r} \right]^{\delta_{r}}\big|\\
							=&  \overset{}{\underset{\substack{i_{2}\in[m],\\\beta_{i_{2}}=1}}{\bigoplus}} \Big( \left|\zeta_{2,i_{2}}\right| \otimes \overset{}{\underset{\substack{k_{2}\neq i_{2},\\\beta_{k_{2}}\in \lbrace 0,1\rbrace}}{\bigotimes}} (1\oplus \left|\zeta_{2,k_{2}}\right|) \otimes \overset{}{\underset{l\in [r]}{\bigotimes}} (1\oplus \left|\xi_{l}\right|)\Big) \oplus \\
							&  \overset{}{\underset{\substack{\lambda\in[r],\\\delta_{\lambda}=1}}{\bigoplus}} \Big( \left|\xi_{\lambda}\right| \otimes \overset{}{\underset{\substack{\mu\neq \lambda,\\\delta_{\mu}\in \lbrace 0,1\rbrace}}{\bigotimes}} (1\oplus \left|\xi_{\mu}\right|) \otimes \overset{}{\underset{j\in [m]}{\bigotimes}} (1\oplus \left|\zeta_{2,j}\right|)\Big) \\
							=& \Bigg(  \overset{}{\underset{\substack{i_{2}\in[m],\\\beta_{i_{2}}=1}}{\bigoplus}} \Big( \left|\zeta_{2,i_{2}}\right| \otimes \overset{}{\underset{\substack{k_{2}\neq i_{2},\\\beta_{k_{2}}\in \lbrace 0,1\rbrace}}{\bigotimes}} (1\oplus \left|\zeta_{2,k_{2}}\right|) \Big)\Bigg) \otimes \overset{}{\underset{l\in [r]}{\bigotimes}} (1\oplus \left|\xi_{l}\right|) \oplus \\
							& \Bigg(\overset{}{\underset{\substack{\lambda\in[r],\\\delta_{\lambda}=1}}{\bigoplus}} \Big( \left|\xi_{\lambda}\right| \otimes \overset{}{\underset{\substack{\mu\neq \lambda,\\\delta_{\mu}\in \lbrace 0,1\rbrace}}{\bigotimes}} (1\oplus \left|\xi_{\mu}\right|) \Big)\Bigg) \otimes \overset{}{\underset{j\in [m]}{\bigotimes}} (1\oplus \left|\zeta_{2,j}\right|). 
						\end{align*}
					\end{itemize}
				\end{itemize}
			\end{itemize}
			
			Hence, we get that $\zeta_{1}\otimes \left[ \xi_{1} \right]^{\delta_{1}} \otimes \ldots \otimes  \left[ \xi_{r} \right]^{\delta_{r}} \equiv \zeta_{2}\otimes \left[ \xi_{1} \right]^{\delta_{1}} \otimes \ldots \otimes  \left[ \xi_{r} \right]^{\delta_{r}}$ for Case 2. Therefore, in any case it holds that $ \zeta_{1}\otimes \left[ \xi_{1} \right]^{\delta_{1}} \otimes \ldots \otimes  \left[ \xi_{r} \right]^{\delta_{r}} \equiv \zeta_{2}\otimes \left[ \xi_{1} \right]^{\delta_{1}} \otimes \ldots \otimes  \left[ \xi_{r} \right]^{\delta_{r}}$.

			\item The symmetric case of 2 and the cases for any other position of $\zeta_{1}$, $\zeta_{2}$, in 3. For these cases the proof is analogous.
		\end{enumerate}
		Finally, for any other form of the expression $E$ we apply the presented cases iteratively, and our proof is completed.\qed
	\end{prof*}


	\noindent Next proposition is an application of Theorem \ref{th1} for proving congruence relation of fusion-$wAC(P)$ connectors. 
	\begin{prop}\label{ext}
		Let $\zeta,\zeta_{1}, \zeta_{2}, \zeta_{3}$ be \emph{fusion}-$wAC(P)$ connectors where $\#_{T}\zeta>0$. Then we have 
		\begin{enumerate}[label=\roman*)]
			\item $\zeta\otimes [0']'\cong \zeta$
			\item $\left[ \zeta_{1}\right] '\otimes \left[ \zeta_{2}\right] \otimes \left[ \zeta_{3}\right]  \cong \left[ \zeta_{1}\right] '\otimes\left[  \left[ \zeta_{2}\right]' \otimes \left[ \zeta_{3}\right]'\right] $
			\item $\left[ \zeta_{1} \right] '\otimes \left[ \zeta_{2}\right] ' \cong \left[ \left[ \zeta_{1} \right] '\otimes \left[ \zeta_{2}\right] '\right] '$. \label{as-tr}
		\end{enumerate}
	\end{prop}
	
	\begin{prof*}
		We apply Theorem \ref{th1} and Corollary \ref{walg_semi}, and we are done. \qed
	\end{prof*}

	\section*{Discussion}	
	\par In \cite{Bl:Al}, the authors proved the soundness of their algebras and investigated the conditions under which completeness also holds. Proving such results in the weighted setting, is in general, much harder. In particular, according to \cite{Pa:We},  soundness has been only defined for multi-valued logics, with values in the bounded distributive lattice $[0, 1]$ with the usual $\max$ and $\min$ operations (cf. \cite{Ha:Me}). In turn, in \cite{Pa:We}, the authors introduced a notion of soundness in the context of weighted propositional configuration logic formulas, with weights ranging over a commutative semiring. The formulas of that logic served for encoding the quantitative features of architectures styles. 
	
	Following the work of \cite{Pa:We}, we could provide an analogous definition of soundness for our weighted algebras. In this case, it occurs that proving soundness would require semiring $K$ to be idempotent with respect to its first and second operation. Idempotency for the second operation of $K$ is required by the weighted synchronization and fusion operators in $wAI(P)$ and $wAC(P)$, respectively. However, in this paper, $K$ is idempotent only with respect to its first operation. A further investigation of soundness for our algebras along with the consideration of other algebraic structures is left as future work.
	
	\par On the other hand, the notion of completeness does not comply in general, in the weighted setup. Indeed, due to the presence of weights we cannot ensure that two arbitrary constructs with the same weight have also the same syntax. In our setting, let for instance $z_{1},z_{2}\in wAI(P)$. Then $z_{1}$ and $ z_{2}$ can return the same weight, while they encode different coordination schemes.
	
	\section{Conclusion}\label{se8}
	
	In this paper, we developed an algebraic framework for the formal characterization of the
	quantitative aspects of connectors in architectures of component-based systems. In particular, we firstly
	studied a weighted Algebra of Interactions, $wAI(P)$, over a set of ports $P$ and a commutative and idempotent semiring $K$. Then we interpreted $wAI(P)$ by polynomials in $ K\left\langle 
	\Gamma(P)\right\rangle $ and using the equivalence classes of the algebra we proved several properties for the $wAI(P)$ elements. 
	Specifically, we showed that the structure $(wAI(P)/\equiv,\oplus,\otimes, \bar{0},\bar{1})$ is a commutative and idempotent
	semiring, an important result that was used for the computation of the semantics of the $wAC(P)$ connectors. 
	In turn, we applied the $wAI(P)$ algebra for encoding the weight of well-known coordination schemes.
	
	In the sequel, we studied the weighted Algebra of Connectors over $P$ and $K$, $wAC(P)$, that extended $wAI(P)$ with two typing operators, namely triggers ``$\left[ \cdot \right]' $'' that initiate an interaction and synchrons ``$\left[ \cdot \right] $'' that need synchronization with other ports in order to interact. We expressed the semantics of $wAC(P)$ connectors as $wAI(P)$ elements, and then, applying the semantics of $wAI(P)$, we obtained the corresponding weight of the connectors over a concrete interactions set over $P$. We proved several nice properties for $wAC(P)$ and we showed the expressiveness of our algebra by modeling several connectors in the weighted setup. Moreover, we studied two subalgebras of $wAC(P)$, the weighted Algebra of Synchrons $wAS(P)$ and the weighted Algebra of Triggers $wAT(P)$, over $P$ and $K$, where the former restricted to synchron elements and the latter to trigger elements. Finally, we defined a concept of congruence relation for fusion-$wAC(P)$ connectors and we proved two theorems for checking such a congruence. 
	
	There are several directions for future work. An important open problem is providing a congruence relation for $wAC(P)$ connectors in general, as well as investigating a different modeling methodology for the weighted framework of connectors in order to solve their congruence problem. Future work is also studying our weighted algebras over alternative structures than $K$, in order to prove their soundness. Another work direction includes investigating the concept of $wAC(P)$ connectors over more general structures than semirings, that are used in practical applications,	for instance valuation monoids (cf. \cite{Dr:Re,Ka:We}). 
	
	Moreover, in \cite{Bl:No}, the authors studied the concept of glue operators
	as composition operators, in order to formalize the coordinated behavior in component-based
	systems, while in \cite{Bl:Sy}, they alternatively expressed glue operators as boolean constraints
	between interactions and the state of the coordinated components. 
	Therefore, future research includes developing a framework for modeling the coordination and behavior of component-based systems in the weighted setting \cite{Br:St,Br:Ba,Sa:Us,Su:Re}.
	On the other hand, in several approaches, connectors have been modeled as entities whose interactions may be modified during the operation of the system \cite{Be:On,Pa:Co}. 
	In other words, it would be interesting to extend our results for connectors with dynamic interactions.
	In addition to these theoretical directions, ongoing work includes an implementation of
	the presented formal framework.

	\newpage
	
	\bibliographystyle{alpha}
	\bibliography{Fount_Pit_wCon_archiv_revised}

	\newpage

	\section{Appendix}\label{appen}
	
	\subsection{Weighted Rendezvous}\label{rend-tables}
	Next we present the tables that we used for computing the weight of the $wAI(P)$ element
	$z=s\otimes r_{1}\otimes r_{2}$ on a concrete interactions set. Specifically, for $\gamma=\left\lbrace \left\lbrace s,r_{1},r_{2}\right\rbrace \right\rbrace \in \Gamma(P)$ we obtained the primary 
	Table \ref{tab-re1}, while for
	$\gamma=\left\lbrace \left\lbrace s,r_{1},r_{2}\right\rbrace, \left\lbrace s, r_{2}\right\rbrace \right\rbrace
	\in\Gamma(P)$ we also used the primary Table \ref{tab-re10}. 
	Moreover, we used the primary Table \ref{tab-re10} for computing the weight of $z=s\otimes r_1\otimes r_2$ on $\gamma=\lbrace  \lbrace  s,r_2\rbrace\rbrace$.
	Finally, in all of the three cases we used the auxiliary Tables \ref{tab-re2}-\ref{tab-re9}. 
	\begin{table}[H]
		\centering
		\scalebox{0.85}{
\begin{tabular}{|c!{\vrule width 2pt}c|c!{\vrule width 2pt}c|}
				\hline
				\multicolumn{4}{|c|}{\newline $ \left\|s\otimes r_{1}\otimes  r_{2}\right\|(\left\lbrace a\right\rbrace )$ }
				\\ \hline \hline \hline
				$a=a_{1}\cup a_{2}$ &  $\left\|s\right\|(\left\lbrace a_{1}\right\rbrace ) $ & $ \left\|r_{1}\otimes r_{2}\right\|(\left\lbrace a_{2}\right\rbrace )$ & $\cdot$ 
				\\ \ChangeRT{2pt}
				$a_{1}=\emptyset, a_{2}=\left\lbrace s,r_{1},r_{2}\right\rbrace $ &  $\hat{0}$ & $k_{r_{1}}\cdot k_{r_{2}}$ & $\hat{0}$
				\\ \hline
				$a_{1}=\left\lbrace s,r_{1},r_{2}\right\rbrace, a_{2}=\emptyset $ & $k_{s}$ & $\hat{0}$ & $\hat{0}$ 
				\\ \hline
				$a_{1}=\left\lbrace s\right\rbrace, a_{2}= \left\lbrace r_{1},r_{2} \right\rbrace$ & $k_{s}$ & $k_{r_{1}}\cdot k_{r_{2}}$ & $k_{s}\cdot k_{r_{1}}\cdot k_{r_{2}}$ 
				\\ \hline
				$a_{1}=\left\lbrace r_{1},r_{2} \right\rbrace, a_{2}=\left\lbrace s\right\rbrace $ & $\hat{0}$ & $\hat{0}$ & $\hat{0}$ 
				\\ \hline
				$a_{1}=\left\lbrace s, r_{1} \right\rbrace, a_{2}=\left\lbrace r_{2}\right\rbrace $ & $k_{s}$ & $\hat{0}$ & $\hat{0}$ 
				\\ \hline
				$a_{1}=\left\lbrace r_{2} \right\rbrace, a_{2}=\left\lbrace s, r_{1} \right\rbrace $ & $\hat{0}$ & $\hat{0}$ & $\hat{0}$ 
				\\
\hline
		\end{tabular}}
		\caption{Weighted Rendezvous and $a=\left\lbrace s,r_{1}, r_{2}\right\rbrace $.}
		\label{tab-re1}
	\end{table}	
	
	\begin{table}[H]
		\centering
		\scalebox{0.85}{\begin{tabular}{|c!{\vrule width 2pt}c|c!{\vrule width 2pt}c|}
				\hline
				\multicolumn{4}{|c|}{\newline $ \left\|s\otimes r_{1}\otimes  r_{2}\right\|(\left\lbrace a\right\rbrace )$ }
				\\ \hline \hline \hline
				$a=a_{1}\cup a_{2}$ &  $\left\|s\right\|(\left\lbrace a_{1}\right\rbrace ) $ & $ \left\|r_{1}\otimes r_{2}\right\|(\left\lbrace a_{2}\right\rbrace )$ & $\cdot$ 
				\\ \ChangeRT{2pt}
				$a_{1}=\emptyset, a_{2}=\left\lbrace s,r_{2}\right\rbrace $ &  $\hat{0}$ & $\hat{0}$ & $\hat{0}$
				\\ \hline
				$a_{1}=\left\lbrace s,r_{2}\right\rbrace, a_{2}=\emptyset $ & $k_{s}$ & $\hat{0}$ & $\hat{0}$ 
				\\ \hline
				$a_{1}=\left\lbrace s\right\rbrace, a_{2}= \left\lbrace r_{2} \right\rbrace$ & $k_{s}$ & $\hat{0}$ & $\hat{0}$ 
				\\ \hline
				$a_{1}=\left\lbrace r_{2} \right\rbrace, a_{2}=\left\lbrace s\right\rbrace $ & $\hat{0}$ & $\hat{0}$ & $\hat{0}$ 
				\\ \hline
				$a_{1}=\left\lbrace s \right\rbrace, a_{2}=\left\lbrace s, r_{2}\right\rbrace $ & $k_{s}$ & $\hat{0}$ & $\hat{0}$
				\\ \hline
				$a_{1}=\left\lbrace s, r_{2} \right\rbrace, a_{2}=\left\lbrace s\right\rbrace $ & $k_{s}$ & $\hat{0}$ & $\hat{0}$
				\\ \hline
				$a_{1}=\left\lbrace r_{2} \right\rbrace, a_{2}=\left\lbrace s,r_{2} \right\rbrace $ & $\hat{0}$ & $\hat{0}$ & $\hat{0}$ 
				\\ \hline
				$a_{1}=\left\lbrace s,r_{2} \right\rbrace, a_{2}=\left\lbrace r_{2} \right\rbrace $ & $k_{s}$ & $\hat{0}$ & $\hat{0}$ 
				\\ \hline
				$a_{1}=\left\lbrace s, r_{2} \right\rbrace, a_{2}=\left\lbrace s,r_{2} \right\rbrace $ & $k_{s}$ & $\hat{0}$ & $\hat{0}$ 
				\\ \ChangeRT{2pt}
				$+$ & \multicolumn{2}{c?}{} \vline & \cellcolor{lightgray}$\hat{0}$ 
				\\ \hline
		\end{tabular}}
		\caption{Weighted Rendezvous and $a=\left\lbrace s, r_{2}\right\rbrace$.}
		\label{tab-re10}
	\end{table}

	\begin{table}[H]
		\centering
		\scalebox{0.8}{\begin{tabular}{|c!{\vrule width 2pt}c|c!{\vrule width 2pt}c|}
				\hline
				\multicolumn{4}{|c|}{\newline $ \left\|r_{1}\otimes  r_{2}\right\|(\left\lbrace a_{2}\right\rbrace )$ }
				\\ \hline \hline \hline
				$a_{2}=a_{2,1}\cup a_{2,2}$ &  $\left\|r_{1}\right\|(\left\lbrace a_{2,1}\right\rbrace ) $ & $ \left\|r_{2}\right\|(\left\lbrace a_{2,2}\right\rbrace )$ & $\cdot$ 
				\\ \ChangeRT{2pt}
				$a_{2,1}=\emptyset, a_{2,2}=\emptyset $ &  $\hat{0}$ & $\hat{0}$ & $\hat{0}$
				\\ \ChangeRT{2pt}
				$+$ & \multicolumn{2}{c?}{} \vline & $\hat{0}$
				\\ \hline
		\end{tabular}}
		\caption{$r_{1}\otimes r_{2}$ and $a_{2}=\emptyset$.}
		\label{tab-re2}
	\end{table}
	
	\vspace{5mm}
	
	\begin{table}[H]
		\centering
		\scalebox{0.8}{\begin{tabular}{|c!{\vrule width 2pt}c|c!{\vrule width 2pt}c|}
				\hline
				\multicolumn{4}{|c|}{\newline $ \left\|r_{1}\otimes  r_{2}\right\|(\left\lbrace a_{2}\right\rbrace )$ }
				\\ \hline \hline \hline
				$a_{2}=a_{2,1}\cup a_{2,2}$ &  $\left\|r_{1}\right\|(\left\lbrace a_{2,1}\right\rbrace ) $ & $ \left\|r_{2}\right\|(\left\lbrace a_{2,2}\right\rbrace )$ & $\cdot$ 
				\\ \ChangeRT{2pt}
				$a_{2,1}=\emptyset, a_{2,2}=\left\lbrace s\right\rbrace  $ &  $\hat{0}$ & $\hat{0}$ & $\hat{0}$
				\\ \hline 
				$a_{2,1}=\left\lbrace s\right\rbrace  , a_{2,2}=\emptyset$ &  $\hat{0}$ & $\hat{0}$ & $\hat{0}$
				\\ \hline 
				$a_{2,1}=\left\lbrace s \right\rbrace  , a_{2,2}=\left\lbrace s\right\rbrace $ &  $\hat{0}$ & $\hat{0}$ & $\hat{0}$
				\\ \ChangeRT{2pt}
				$+$ & \multicolumn{2}{c?}{} \vline & $\hat{0}$
				\\ \hline
		\end{tabular}}
		\caption{$r_{1}\otimes r_{2}$ and $a_{2}=\left\lbrace s\right\rbrace $.}
		\label{tab-re3}
	\end{table}
	\vspace{5mm}
	\begin{table}[H]
		\centering
		\scalebox{0.8}{\begin{tabular}{|c!{\vrule width 2pt}c|c!{\vrule width 2pt} c|}
				\hline
				\multicolumn{4}{|c|}{\newline $ \left\|r_{1}\otimes  r_{2}\right\|(\left\lbrace a_{2}\right\rbrace )$ }
				\\ \hline \hline \hline
				$a_{2}=a_{2,1}\cup a_{2,2}$ &  $\left\|r_{1}\right\|(\left\lbrace a_{2,1}\right\rbrace ) $ & $ \left\|r_{2}\right\|(\left\lbrace a_{2,2}\right\rbrace )$ & $\cdot$ 
				\\ \ChangeRT{2pt}
				$a_{2,1}=\emptyset, a_{2,2}=\left\lbrace  r_{1}\right\rbrace  $ &  $\hat{0}$ & $\hat{0}$ & $\hat{0}$
				\\ \hline 
				$a_{2,1}=\left\lbrace r_{1}\right\rbrace  , a_{2,2}=\emptyset$ &  $k_{r_{1}}$ & $\hat{0}$ & $\hat{0}$
				\\ \hline 
				$a_{2,1}=\left\lbrace r_{1} \right\rbrace  , a_{2,2}=\left\lbrace r_{1}\right\rbrace $ &  $k_{r_{1}}$ & $\hat{0}$ & $\hat{0}$
				\\ \ChangeRT{2pt}
				$+$ & \multicolumn{2}{c?}{} \vline & $\hat{0}$
				\\ \hline
		\end{tabular}}
		\caption{$r_{1}\otimes r_{2}$ and $a_{2}=\left\lbrace  r_{1}\right\rbrace $.}
		\label{tab-re4}
	\end{table}
	\vspace{5mm}
	\begin{table}[H]
		\centering
		\scalebox{0.8}{\begin{tabular}{|c!{\vrule width 2pt}c|c!{\vrule width 2pt}c|}
				\hline
				\multicolumn{4}{|c|}{\newline $ \left\|r_{1}\otimes  r_{2}\right\|(\left\lbrace a_{2}\right\rbrace )$ }
				\\ \hline \hline \hline
				$a_{2}=a_{2,1}\cup a_{2,2}$ &  $\left\|r_{1}\right\|(\left\lbrace a_{2,1}\right\rbrace ) $ & $ \left\|r_{2}\right\|(\left\lbrace a_{2,2}\right\rbrace )$ & $\cdot$ 
				\\ \ChangeRT{2pt}
				$a_{2,1}=\emptyset, a_{2,2}=\left\lbrace  r_{2}\right\rbrace  $ &  $\hat{0}$ & $k_{r_{2}}$ & $\hat{0}$
				\\ \hline 
				$a_{2,1}=\left\lbrace r_{2}\right\rbrace  , a_{2,2}=\emptyset$ &  $\hat{0}$ & $\hat{0}$ & $\hat{0}$
				\\ \hline 
				$a_{2,1}=\left\lbrace r_{2} \right\rbrace  , a_{2,2}=\left\lbrace r_{2}\right\rbrace $ &  $\hat{0}$ & $k_{r_{2}}$ & $\hat{0}$
				\\ \ChangeRT{2pt}
				$+$ & \multicolumn{2}{c?}{} \vline & $\hat{0}$
				\\ \hline
		\end{tabular}}
		\caption{$r_{1}\otimes r_{2}$ and $a_{2}=\left\lbrace  r_{2}\right\rbrace $.}
		\label{tab-re5}
	\end{table}
	\vspace{5mm}
	
	\begin{table}[H]
		\centering
		\scalebox{0.8}{\begin{tabular}{|c!{\vrule width 2pt}c|c!{\vrule width 2pt}c|}
				\hline
				\multicolumn{4}{|c|}{\newline $ \left\|r_{1}\otimes r_{2}\right\|(\left\lbrace a_{2}\right\rbrace )$ }
				\\ \hline \hline \hline
				$a_{2}=a_{2,1}\cup a_{2,2}$ &  $\left\|r_{1}\right\|(\left\lbrace a_{2,1}\right\rbrace ) $ & $ \left\|r_{2}\right\|(\left\lbrace a_{2,2}\right\rbrace )$ & $\cdot$ 
				\\ \ChangeRT{2pt}
				$a_{2,1}=\emptyset, a_{2,2}=\left\lbrace s,r_{1}\right\rbrace  $ &  $\hat{0}$ & $\hat{0}$ & $\hat{0}$
				\\ \hline 
				$a_{2,1}=\left\lbrace s,r_{1}\right\rbrace  , a_{2,2}=\emptyset$ &  $k_{r_{1}}$ & $\hat{0}$ & $\hat{0}$
				\\ \hline 
				$a_{2,1}=\left\lbrace s \right\rbrace  , a_{2,2}=\left\lbrace r_{1}\right\rbrace $ &  $\hat{0}$ & $\hat{0}$ & $\hat{0}$
				\\ \hline 
				$a_{2,1}=\left\lbrace r_{1} \right\rbrace  , a_{2,2}=\left\lbrace s\right\rbrace $ &  $k_{r_{1}}$ & $\hat{0}$ & $\hat{0}$
				\\ \hline 
				$a_{2,1}=\left\lbrace s \right\rbrace  , a_{2,2}=\left\lbrace s,r_{1}\right\rbrace $ &  $\hat{0}$ & $\hat{0}$ & $\hat{0}$
				\\ \hline 
				$a_{2,1}=\left\lbrace s,r_{1} \right\rbrace  , a_{2,2}=\left\lbrace s\right\rbrace $ &  $k_{r_{1}}$ & $\hat{0}$ & $\hat{0}$
				\\ \hline 
				$a_{2,1}=\left\lbrace  r_{1} \right\rbrace  , a_{2,2}=\left\lbrace s,r_{1}\right\rbrace $ &  $k_{r_{1}}$ & $\hat{0}$ & $\hat{0}$
				\\ \hline 
				$a_{2,1}=\left\lbrace s,r_{1} \right\rbrace  , a_{2,2}=\left\lbrace r_{1}\right\rbrace $ &  $k_{r_{1}}$ & $\hat{0}$ & $\hat{0}$
				\\ \hline 
				$a_{2,1}=\left\lbrace s,r_{1} \right\rbrace  , a_{2,2}=\left\lbrace s,r_{1}\right\rbrace $ &  $k_{r_{1}}$ & $\hat{0}$ & $\hat{0}$
				\\ \ChangeRT{2pt}
				$+$ & \multicolumn{2}{c?}{} \vline & $\hat{0}$
				\\ \hline
		\end{tabular}}
		\caption{$r_{1}\otimes r_{2}$ and $a_{2}=\left\lbrace s, r_{1}\right\rbrace $.}
		\label{tab-re6}
	\end{table}
	
	\begin{table}[H]
		\centering
		\scalebox{0.8}{\begin{tabular}{|c!{\vrule width 2pt}c|c!{\vrule width 2pt}c|}
				\hline
				\multicolumn{4}{|c|}{\newline $ \left\|r_{1}\otimes  r_{2}\right\|(\left\lbrace a_{2}\right\rbrace )$ }
				\\ \hline \hline \hline
				$a_{2}=a_{2,1}\cup a_{2,2}$ &  $\left\|r_{1}\right\|(\left\lbrace a_{2,1}\right\rbrace ) $ & $ \left\|r_{2}\right\|(\left\lbrace a_{2,2}\right\rbrace )$ & $\cdot$ 
				\\ \ChangeRT{2pt}
				$a_{2,1}=\emptyset, a_{2,2}=\left\lbrace s,r_{2}\right\rbrace  $ &  $\hat{0}$ & $k_{r_{2}}$ & $\hat{0}$
				\\ \hline 
				$a_{2,1}=\left\lbrace s,r_{2}\right\rbrace  , a_{2,2}=\emptyset$ &  $\hat{0}$ & $\hat{0}$ & $\hat{0}$
				\\ \hline 
				$a_{2,1}=\left\lbrace s \right\rbrace  , a_{2,2}=\left\lbrace r_{2}\right\rbrace $ &  $\hat{0}$ & $k_{r_{2}}$ & $\hat{0}$
				\\ \hline 
				$a_{2,1}=\left\lbrace r_{2} \right\rbrace  , a_{2,2}=\left\lbrace s\right\rbrace $ &  $\hat{0}$ & $\hat{0}$ & $\hat{0}$
				\\ \hline 
				$a_{2,1}=\left\lbrace s \right\rbrace  , a_{2,2}=\left\lbrace s,r_{2}\right\rbrace $ &  $\hat{0}$ & $k_{r_{2}}$ & $\hat{0}$
				\\ \hline 
				$a_{2,1}=\left\lbrace s,r_{2} \right\rbrace  , a_{2,2}=\left\lbrace s\right\rbrace $ &  $\hat{0}$ & $\hat{0}$ & $\hat{0}$
				\\ \hline 
				$a_{2,1}=\left\lbrace  r_{2} \right\rbrace  , a_{2,2}=\left\lbrace s,r_{2}\right\rbrace $ &  $\hat{0}$ & $k_{r_{2}}$ & $\hat{0}$
				\\ \hline 
				$a_{2,1}=\left\lbrace s,r_{2} \right\rbrace  , a_{2,2}=\left\lbrace r_{2}\right\rbrace $ &  $\hat{0}$ & $k_{r_{2}}$ & $\hat{0}$
				\\ \hline 
				$a_{2,1}=\left\lbrace s,r_{2} \right\rbrace  , a_{2,2}=\left\lbrace s,r_{2}\right\rbrace $ &  $\hat{0}$ & $k_{r_{2}}$ & $\hat{0}$
				\\ \ChangeRT{2pt}
				$+$ & \multicolumn{2}{c?}{} \vline & $\hat{0}$
				\\ \hline
		\end{tabular}}
		\caption{$r_{1}\otimes r_{2}$ and $a_{2}=\left\lbrace s, r_{2}\right\rbrace $.}
		\label{tab-re7}
	\end{table}
	\begin{table}[H]
		\centering
		\scalebox{0.8}{\begin{tabular}{|c!{\vrule width 2pt}c|c!{\vrule width 2pt}c|}
				\hline
				\multicolumn{4}{|c|}{\newline $ \left\|r_{1}\otimes  r_{2}\right\|(\left\lbrace a_{2}\right\rbrace )$ }
				\\ \hline \hline \hline
				$a_{2}=a_{2,1}\cup a_{2,2}$ &  $\left\|r_{1}\right\|(\left\lbrace a_{2,1}\right\rbrace ) $ & $ \left\|r_{2}\right\|(\left\lbrace a_{2,2}\right\rbrace )$ & $\cdot$ 
				\\ \ChangeRT{2pt}
				$a_{2,1}=\emptyset, a_{2,2}=\left\lbrace r_{1}, r_{2}\right\rbrace  $ &  $\hat{0}$ & $k_{r_{2}}$ & $\hat{0}$
				\\ \hline 
				$a_{2,1}=\left\lbrace r_{1}, r_{2}\right\rbrace  , a_{2,2}=\emptyset$ &  $k_{r_{1}}$ & $\hat{0}$ & $\hat{0}$
				\\ \hline 
				$a_{2,1}=\left\lbrace r_{1} \right\rbrace  , a_{2,2}=\left\lbrace r_{2}\right\rbrace $ &  $k_{r_{1}}$ & $k_{r_{2}}$ & $k_{r_{1}}\cdot k_{r_{2}}$
				\\ \hline 
				$a_{2,1}=\left\lbrace r_{2} \right\rbrace  , a_{2,2}=\left\lbrace r_{1}\right\rbrace $ &  $\hat{0}$ & $\hat{0}$ & $\hat{0}$
				\\ \hline 
				$a_{2,1}=\left\lbrace r_{1} \right\rbrace  , a_{2,2}=\left\lbrace r_{1}, r_{2}\right\rbrace $ &  $k_{r_{1}}$ & $k_{r_{2}}$ & $k_{r_{1}}\cdot k_{r_{2}}$
				\\ \hline 
				$a_{2,1}=\left\lbrace r_{1}, r_{2} \right\rbrace  , a_{2,2}=\left\lbrace r_{1}\right\rbrace $ &  $k_{r_{1}}$ & $\hat{0}$ & $\hat{0}$
				\\ \hline 
				$a_{2,1}=\left\lbrace r_{2} \right\rbrace  , a_{2,2}=\left\lbrace r_{1}, r_{2}\right\rbrace $ &  $\hat{0}$ & $k_{r_{2}}$ & $\hat{0}$
				\\ \hline 
				$a_{2,1}=\left\lbrace r_{1}, r_{2} \right\rbrace  , a_{2,2}=\left\lbrace r_{2}\right\rbrace $ &  $k_{r_{1}}$ & $k_{r_{2}}$ & $k_{r_{1}}\cdot k_{r_{2}}$
				\\ \hline 
				$a_{2,1}=\left\lbrace r_{1}, r_{2} \right\rbrace  , a_{2,2}=\left\lbrace r_{1},r_{2}\right\rbrace $ &  $k_{r_{1}}$ & $k_{r_{2}}$ & $k_{r_{1}}\cdot k_{r_{2}}$
				\\ \ChangeRT{2pt}
				$+$ & \multicolumn{2}{c?}{} \vline & $k_{r_{1}}\cdot k_{r_{2}}$
				\\ \hline
		\end{tabular}}
		\caption{$r_{1}\otimes r_{2}$ and $a_{2}=\left\lbrace r_{1}, r_{2}\right\rbrace $.}
		\label{tab-re8}
	\end{table}
	
	\begin{table}[H]
		\centering
		\scalebox{0.8}{\begin{tabular}{|c!{\vrule width 2pt}c|c!{\vrule width 2pt}c|}
				\hline
				\multicolumn{4}{|c|}{\newline $ \left\|r_{1}\otimes  r_{2}\right\|(\left\lbrace a_{2}\right\rbrace )$ }
				\\ \hline \hline \hline
				$a_{2}=a_{2,1}\cup a_{2,2}$ &  $\left\|r_{1}\right\|(\left\lbrace a_{2,1}\right\rbrace ) $ & $ \left\|r_{2}\right\|(\left\lbrace a_{2,2}\right\rbrace )$ & $\cdot$ 
				\\ \ChangeRT{2pt}
				$a_{2,1}=\emptyset, a_{2,2}=\left\lbrace s,r_{1},r_{2}\right\rbrace $ &  $\hat{0}$ & $k_{r_{2}}$ & $\hat{0}$
				\\ \hline
				$a_{2,1}=\left\lbrace s,r_{1},r_{2}\right\rbrace, a_{2,2}=\emptyset $ & $k_{r_{1}}$ & $\hat{0}$ & $\hat{0}$ 
				\\ \hline
				$a_{2,1}=\left\lbrace s\right\rbrace, a_{2,2}= \left\lbrace r_{1},r_{2} \right\rbrace$ & $\hat{0}$ & $k_{r_{2}}$ & $\hat{0}$ 
				\\ \hline
				$a_{2,1}=\left\lbrace r_{1},r_{2} \right\rbrace, a_{2,2}=\left\lbrace s\right\rbrace $ & $k_{r_{1}}$ & $\hat{0}$ & $\hat{0}$ 
				\\ \hline
				$a_{2,1}=\left\lbrace s, r_{1} \right\rbrace, a_{2,2}=\left\lbrace r_{2}\right\rbrace $ & $k_{r_{1}}$ & $k_{r_{2}}$ & $k_{r_{1}}\cdot k_{r_{2}}$ 
				\\ \hline
				$a_{2,1}=\left\lbrace r_{2} \right\rbrace, a_{2,2}=\left\lbrace s, r_{1} \right\rbrace $ & $\hat{0}$ & $\hat{0}$ & $\hat{0}$ 
				\\
\hline
		\end{tabular}}
		\caption{$r_{1}\otimes r_{2}$ and $a_{2}=\left\lbrace s,r_{1},r_{2}\right\rbrace$.}
		\label{tab-re9}
	\end{table}
	
	\subsection{Weighted Broadcast}	
	The following tables were used for computing the weight of the $wAI(P)$ element $z=s\otimes (1\oplus r_{1})\otimes (1\oplus r_{2})$ for the Broadcast scheme on the interactions set $\gamma=\left\lbrace  \left\lbrace  s\right\rbrace , \left\lbrace  s,r_{1}\right\rbrace , \left\lbrace  s,r_{2}\right\rbrace , \left\lbrace  s,r_{1},r_{2}\right\rbrace  \right\rbrace \in \Gamma(P)$. The primary tables are Tables \ref{tab-bra}-\ref{tab-brd}, while Tables \ref{tab-brd1}-\ref{tab-brd8} are the auxiliary ones. 
	\begin{table}[H]
		\centering
		\scalebox{0.85}{
\begin{tabular}{|c!{\vrule width 2pt}c|c!{\vrule width 2pt}c|}
				\hline
				\multicolumn{4}{|c|}{\newline $ \left\|s\otimes (1\oplus r_{1})\otimes  (1\oplus r_{2})\right\|(\left\lbrace a\right\rbrace )$ }
				\\ \hline \hline \hline
				$a=a_{1}\cup a_{2}$ &  $\left\|s\right\|(\left\lbrace a_{1}\right\rbrace ) $ & $ \left\|(1\oplus r_{1})\otimes (1\oplus r_{2})\right\|(\left\lbrace a_{2}\right\rbrace )$ & $\cdot$ 
				\\ \ChangeRT{2pt}
				$a_{1}=\emptyset, a_{2}=\left\lbrace s\right\rbrace  $ &  $\hat{0}$ & $\hat{0}$ & $\hat{0}$
				\\ \hline 
				$a_{1}=\left\lbrace s\right\rbrace  , a_{2}=\emptyset$ &  $k_{s}$ & $\hat{1}$ & $k_{s}$
				\\ \hline 
				$a_{1}=\left\lbrace s \right\rbrace  , a_{2}=\left\lbrace s\right\rbrace $ &  $k_{s}$ & $\hat{0}$ & $\hat{0}$
				\\ \ChangeRT{2pt}
				$+$ & \multicolumn{2}{c?}{} \vline & \cellcolor{lightgray}$k_{s}$
				\\ \hline
		\end{tabular}}
		\caption{Weighted Broadcast and $a=\left\lbrace s\right\rbrace $.}
		\label{tab-bra}
	\end{table}
	
	\begin{table}[H]
		\centering
		\scalebox{0.85}{\begin{tabular}{|c!{\vrule width 2pt}c|c!{\vrule width 2pt}c|}
				\hline
				\multicolumn{4}{|c|}{\newline $ \left\|s\otimes (1\oplus r_{1})\otimes  (1\oplus r_{2})\right\|(\left\lbrace a\right\rbrace )$ }
				\\ \hline \hline \hline
				$a=a_{1}\cup a_{2}$ &  $\left\|s\right\|(\left\lbrace a_{1}\right\rbrace ) $ & $ \left\|(1\oplus r_{1})\otimes (1\oplus r_{2})\right\|(\left\lbrace a_{2}\right\rbrace )$ & $\cdot$ 
				\\ \ChangeRT{2pt}
				$a_{1}=\emptyset, a_{2}=\left\lbrace s,r_{1}\right\rbrace  $ &  $\hat{0}$ & $k_{r_{1}}$ & $\hat{0}$
				\\ \hline 
				$a_{1}=\left\lbrace s,r_{1}\right\rbrace  , a_{2}=\emptyset$ &  $k_{s}$ & $\hat{1}$ & $k_{s}$
				\\ \hline 
				$a_{1}=\left\lbrace s \right\rbrace  , a_{2}=\left\lbrace r_{1}\right\rbrace $ &  $k_{s}$ & $k_{r_{1}}$ & $k_{s}\cdot k_{r_{1}}$
				\\ \hline 
				$a_{1}=\left\lbrace r_{1} \right\rbrace  , a_{2}=\left\lbrace s\right\rbrace $ &  $\hat{0}$ & $\hat{0}$ & $\hat{0}$
				\\ \hline 
				$a_{1}=\left\lbrace s \right\rbrace  , a_{2}=\left\lbrace s,r_{1}\right\rbrace $ &  $k_{s}$ & $k_{r_{1}}$ & $k_{s}\cdot k_{r_{1}}$
				\\ \hline 
				$a_{1}=\left\lbrace s,r_{1} \right\rbrace  , a_{2}=\left\lbrace s\right\rbrace $ &  $k_{s}$ & $\hat{0}$ & $\hat{0}$
				\\ \hline 
				$a_{1}=\left\lbrace  r_{1} \right\rbrace  , a_{2}=\left\lbrace s,r_{1}\right\rbrace $ &  $\hat{0}$ & $k_{r_{1}}$ & $\hat{0}$
				\\ \hline 
				$a_{1}=\left\lbrace s,r_{1} \right\rbrace  , a_{2}=\left\lbrace r_{1}\right\rbrace $ &  $k_{s}$ & $k_{r_{1}}$ & $k_{s}\cdot k_{r_{1}}$
				\\ \hline 
				$a_{1}=\left\lbrace s,r_{1} \right\rbrace  , a_{2}=\left\lbrace s,r_{1}\right\rbrace $ &  $k_{s}$ & $k_{r_{1}}$ & $k_{s} \cdot k_{r_{1}}$
				\\ \ChangeRT{2pt}
				$+$ & \multicolumn{2}{c?}{} \vline & \cellcolor{lightgray}$k_{s}+ (k_{s}\cdot k_{r_{1}})$
				\\ \hline
		\end{tabular}}
		\caption{Weighted Broadcast and $a=\left\lbrace s, r_{1}\right\rbrace $.}
		\label{tab-brb}
	\end{table}
	
	\begin{table}[H]
		\centering
		\scalebox{0.85}{\begin{tabular}{|c!{\vrule width 2pt}c|c!{\vrule width 2pt}c|}
				\hline
				\multicolumn{4}{|c|}{\newline $ \left\|s\otimes (1\oplus r_{1})\otimes  (1\oplus r_{2})\right\|(\left\lbrace a\right\rbrace )$ }
				\\ \hline \hline \hline
				$a=a_{1}\cup a_{2}$ &  $\left\|s\right\|(\left\lbrace a_{1}\right\rbrace ) $ & $ \left\|(1\oplus r_{1})\otimes (1\oplus r_{2})\right\|(\left\lbrace a_{2}\right\rbrace )$ & $\cdot$ 
				\\ \ChangeRT{2pt}
				$a_{1}=\emptyset, a_{2}=\left\lbrace s,r_{2}\right\rbrace  $ &  $\hat{0}$ & $k_{r_{2}}$ & $\hat{0}$
				\\ \hline 
				$a_{1}=\left\lbrace s,r_{2}\right\rbrace  , a_{2}=\emptyset$ &  $k_{s}$ & $\hat{1}$ & $k_{s}$
				\\ \hline 
				$a_{1}=\left\lbrace s \right\rbrace  , a_{2}=\left\lbrace r_{2}\right\rbrace $ &  $k_{s}$ & $k_{r_{2}}$ & $k_{s}\cdot k_{r_{2}}$
				\\ \hline 
				$a_{1}=\left\lbrace r_{2} \right\rbrace  , a_{2}=\left\lbrace s\right\rbrace $ &  $\hat{0}$ & $\hat{0}$ & $\hat{0}$
				\\ \hline 
				$a_{1}=\left\lbrace s \right\rbrace  , a_{2}=\left\lbrace s,r_{2}\right\rbrace $ &  $k_{s}$ & $k_{r_{2}}$ & $k_{s}\cdot k_{r_{2}}$
				\\ \hline 
				$a_{1}=\left\lbrace s,r_{2} \right\rbrace  , a_{2}=\left\lbrace s\right\rbrace $ &  $k_{s}$ & $\hat{0}$ & $\hat{0}$
				\\ \hline 
				$a_{1}=\left\lbrace  r_{2} \right\rbrace  , a_{2}=\left\lbrace s,r_{2}\right\rbrace $ &  $\hat{0}$ & $k_{r_{2}}$ & $\hat{0}$
				\\ \hline 
				$a_{1}=\left\lbrace s,r_{2} \right\rbrace  , a_{2}=\left\lbrace r_{2}\right\rbrace $ &  $k_{s}$ & $k_{r_{2}}$ & $k_{s}\cdot k_{r_{2}}$
				\\ \hline 
				$a_{1}=\left\lbrace s,r_{2} \right\rbrace  , a_{2}=\left\lbrace s,r_{2}\right\rbrace $ &  $k_{s}$ & $k_{r_{2}}$ & $k_{s}\cdot k_{r_{2}}$
				\\ \ChangeRT{2pt}
				$+$ & \multicolumn{2}{c?}{} \vline & \cellcolor{lightgray}$k_{s} + (k_{s}\cdot k_{r_{2}})$
				\\ \hline
		\end{tabular}}
		\caption{Weighted Broadcast and $a=\left\lbrace s, r_{2}\right\rbrace $.}
		\label{tab-brc}
	\end{table}
	
	\begin{table}[H]
		\centering
		\scalebox{0.8}{\begin{tabular}{|c!{\vrule width 2pt}c|c!{\vrule width 2pt}c|}
				\hline
				\multicolumn{4}{|c|}{\newline $ \left\|s\otimes (1\oplus r_{1})\otimes  (1\oplus r_{2})\right\|(\left\lbrace a\right\rbrace )$ }
				\\ \hline \hline \hline
				$a=a_{1}\cup a_{2}$ &  $\left\|s\right\|(\left\lbrace a_{1}\right\rbrace ) $ & $ \left\|(1\oplus r_{1})\otimes (1\oplus r_{2})\right\|(\left\lbrace a_{2}\right\rbrace )$ & $\cdot$ 
				\\ \ChangeRT{2pt}
				$a_{1}=\emptyset, a_{2}=\left\lbrace s,r_{1},r_{2}\right\rbrace $ &  $\hat{0}$ & $k_{r_{1}}+ k_{r_{2}}+ (k_{r_{1}}\cdot k_{r_{2}})$ & $\hat{0}$
				\\ \hline
				$a_{1}=\left\lbrace s,r_{1},r_{2}\right\rbrace, a_{2}=\emptyset $ & $k_{s}$ & $\hat{1}$ & $k_{s}$ 
				\\ \hline
				\multirow{2}{*}{$a_{1}=\left\lbrace s\right\rbrace, a_{2}= \left\lbrace r_{1},r_{2} \right\rbrace$} & \multirow{2}{*}{$k_{s}$} & \multirow{2}{*}{$k_{r_{1}}+ k_{r_{2}}+ (k_{r_{1}}\cdot k_{r_{2}})$}  & $(k_{s}\cdot k_{r_{1}})+ (k_{s}\cdot k_{r_{2}})+ $
				\\ 
				& & & $(k_{s}\cdot k_{r_{1}}\cdot k_{r_{2}})$ 
				\\ \hline
				$a_{1}=\left\lbrace r_{1},r_{2} \right\rbrace, a_{2}=\left\lbrace s\right\rbrace $ & $\hat{0}$ & $\hat{0}$ & $\hat{0}$ 
				\\ \hline
				$a_{1}=\left\lbrace s, r_{1} \right\rbrace, a_{2}=\left\lbrace r_{2}\right\rbrace $ & $k_{s}$ & $k_{r_{2}}$ & $k_{s}\cdot k_{r_{2}}$ 
				\\
\hline
		\end{tabular}}
		\caption{Weighted Broadcast and $a=\left\lbrace s,r_{1},r_{2}\right\rbrace$.}
		\label{tab-brd}
	\end{table}	
	
	\vspace{3mm}

	\begin{table}[H]
		\centering
		\scalebox{0.8}{\begin{tabular}{|c!{\vrule width 2pt}c|c!{\vrule width 2pt}c|}
				\hline
				\multicolumn{4}{|c|}{\newline $ \left\|(1\oplus r_{1})\otimes (1\oplus r_{2})\right\|(\left\lbrace a_{2}\right\rbrace )$ }
				\\ \hline \hline \hline
				$a_{2}=a_{2,1}\cup a_{2,2}$ &  $\left\|1\oplus r_{1}\right\|(\left\lbrace a_{2,1}\right\rbrace ) $ & $ \left\|1\oplus r_{2}\right\|(\left\lbrace a_{2,2}\right\rbrace )$ & $\cdot$ 
				\\ \ChangeRT{2pt}
				$a_{2,1}=\emptyset, a_{2,2}=\emptyset $ &  $\hat{1}+ \hat{0}$ & $\hat{1}+ \hat{0}$ & $\hat{1}$
				\\ \ChangeRT{2pt}
				$+$ & \multicolumn{2}{c?}{} \vline & $\hat{1}$
				\\ \hline
		\end{tabular}}
		\caption{$(1\oplus r_{1})\otimes (1\oplus r_{2})$ and $a_{2}=\emptyset$.}
		\label{tab-brd1}
	\end{table}
	\begin{table}[H]
		\centering
		\scalebox{0.8}{\begin{tabular}{|c!{\vrule width 2pt}c|c!{\vrule width 2pt}c|}
				\hline
				\multicolumn{4}{|c|}{\newline $ \left\|(1\oplus r_{1})\otimes (1\oplus r_{2})\right\|(\left\lbrace a_{2}\right\rbrace )$ }
				\\ \hline \hline \hline
				$a_{2}=a_{2,1}\cup a_{2,2}$ &  $\left\|1\oplus r_{1}\right\|(\left\lbrace a_{2,1}\right\rbrace ) $ & $ \left\|1\oplus r_{2}\right\|(\left\lbrace a_{2,2}\right\rbrace )$ & $\cdot$ 
				\\ \ChangeRT{2pt}
				$a_{2,1}=\emptyset, a_{2,2}=\left\lbrace s\right\rbrace  $ &  $\hat{1} + \hat{0}$ & $\hat{0}+ \hat{0}$ & $\hat{0}$
				\\ \hline 
				$a_{2,1}=\left\lbrace s\right\rbrace  , a_{2,2}=\emptyset$ &  $\hat{0}+ \hat{0}$ & $\hat{1}+ \hat{0}$ & $\hat{0}$
				\\ \hline 
				$a_{2,1}=\left\lbrace s \right\rbrace  , a_{2,2}=\left\lbrace s\right\rbrace $ &  $\hat{0}+ \hat{0}$ & $\hat{0}+ \hat{0}$ & $\hat{0}$
				\\ \ChangeRT{2pt}
				$+$ & \multicolumn{2}{c?}{} \vline & $\hat{0}$
				\\ \hline
		\end{tabular}}
		\caption{$(1\oplus r_{1})\otimes (1\oplus r_{2})$ and $a_{2}=\left\lbrace s\right\rbrace $.}
		\label{tab-brd2}
	\end{table}
	\vspace{3mm}
	\begin{table}[H]
		\centering
		\scalebox{0.8}{\begin{tabular}{|c!{\vrule width 2pt}c|c!{\vrule width 2pt} c|}
				\hline
				\multicolumn{4}{|c|}{\newline $ \left\|(1\oplus r_{1})\otimes (1\oplus r_{2})\right\|(\left\lbrace a_{2}\right\rbrace )$ }
				\\ \hline \hline \hline
				$a_{2}=a_{2,1}\cup a_{2,2}$ &  $\left\|1\oplus r_{1}\right\|(\left\lbrace a_{2,1}\right\rbrace ) $ & $ \left\|1\oplus r_{2}\right\|(\left\lbrace a_{2,2}\right\rbrace )$ & $\cdot$ 
				\\ \ChangeRT{2pt}
				$a_{2,1}=\emptyset, a_{2,2}=\left\lbrace  r_{1}\right\rbrace  $ &  $\hat{1}+ \hat{0}$ & $\hat{0}+ \hat{0}$ & $\hat{0}$
				\\ \hline 
				$a_{2,1}=\left\lbrace r_{1}\right\rbrace  , a_{2,2}=\emptyset$ &  $\hat{0}+ k_{r_{1}}$ & $\hat{1}+ \hat{0}$ & $k_{r_{1}}$
				\\ \hline 
				$a_{2,1}=\left\lbrace r_{1} \right\rbrace  , a_{2,2}=\left\lbrace r_{1}\right\rbrace $ &  $\hat{0}+ k_{r_{1}}$ & $\hat{0}+ \hat{0}$ & $\hat{0}$
				\\ \ChangeRT{2pt}
				$+$ & \multicolumn{2}{c?}{} \vline & $k_{r_{1}}$
				\\ \hline
		\end{tabular}}
		\caption{$(1\oplus r_{1})\otimes (1\oplus r_{2})$ and $a_{2}=\left\lbrace  r_{1}\right\rbrace $.}
		\label{tab-brd3}
	\end{table}
	\vspace{3mm}
	\begin{table}[H]
		\centering
		\scalebox{0.8}{\begin{tabular}{|c!{\vrule width 2pt}c|c!{\vrule width 2pt}c|}
				\hline
				\multicolumn{4}{|c|}{\newline $ \left\|(1\oplus r_{1})\otimes (1\oplus r_{2})\right\|(\left\lbrace a_{2}\right\rbrace )$ }
				\\ \hline \hline \hline
				$a_{2}=a_{2,1}\cup a_{2,2}$ &  $\left\|1\oplus r_{1}\right\|(\left\lbrace a_{2,1}\right\rbrace ) $ & $ \left\|1\oplus r_{2}\right\|(\left\lbrace a_{2,2}\right\rbrace )$ & $\cdot$ 
				\\ \ChangeRT{2pt}
				$a_{2,1}=\emptyset, a_{2,2}=\left\lbrace  r_{2}\right\rbrace  $ &  $\hat{1}+ \hat{0}$ & $\hat{0}+ k_{r_{2}}$ & $k_{r_{2}}$
				\\ \hline 
				$a_{2,1}=\left\lbrace r_{2}\right\rbrace  , a_{2,2}=\emptyset$ &  $\hat{0}+ \hat{0}$ & $\hat{1}+\hat{0}$ & $\hat{0}$
				\\ \hline 
				$a_{2,1}=\left\lbrace r_{2} \right\rbrace  , a_{2,2}=\left\lbrace r_{2}\right\rbrace $ &  $\hat{0}+ \hat{0}$ & $\hat{0}+ k_{r_{2}}$ & $\hat{0}$
				\\ \ChangeRT{2pt}
				$+$ & \multicolumn{2}{c?}{} \vline & $k_{r_{2}}$
				\\ \hline
		\end{tabular}}
		\caption{$(1\oplus r_{1})\otimes (1\oplus r_{2})$ and $a_{2}=\left\lbrace  r_{2}\right\rbrace $.}
		\label{tab-brd4}
	\end{table}
	\vspace{3mm}
	\begin{table}[H]
		\centering
		\scalebox{0.8}{\begin{tabular}{|c!{\vrule width 2pt}c|c!{\vrule width 2pt}c|}
				\hline
				\multicolumn{4}{|c|}{\newline $ \left\|(1\oplus r_{1})\otimes (1\oplus r_{2})\right\|(\left\lbrace a_{2}\right\rbrace )$ }
				\\ \hline \hline \hline
				$a_{2}=a_{2,1}\cup a_{2,2}$ &  $\left\|1\oplus r_{1}\right\|(\left\lbrace a_{2,1}\right\rbrace ) $ & $ \left\|1\oplus r_{2}\right\|(\left\lbrace a_{2,2}\right\rbrace )$ & $\cdot$ 
				\\ \ChangeRT{2pt}
				$a_{2,1}=\emptyset, a_{2,2}=\left\lbrace s,r_{1}\right\rbrace  $ &  $\hat{1}+ \hat{0}$ & $\hat{0}+ \hat{0}$ & $\hat{0}$
				\\ \hline 
				$a_{2,1}=\left\lbrace s,r_{1}\right\rbrace  , a_{2,2}=\emptyset$ &  $\hat{0}+ k_{r_{1}}$ & $\hat{1}+ \hat{0}$ & $k_{r_{1}}$
				\\ \hline 
				$a_{2,1}=\left\lbrace s \right\rbrace  , a_{2,2}=\left\lbrace r_{1}\right\rbrace $ &  $\hat{0}+ \hat{0}$ & $\hat{0}+ \hat{0}$ & $\hat{0}$
				\\ \hline 
				$a_{2,1}=\left\lbrace r_{1} \right\rbrace  , a_{2,2}=\left\lbrace s\right\rbrace $ &  $\hat{0}+ k_{r_{1}}$ & $\hat{0}+ \hat{0}$ & $\hat{0}$
				\\ \hline 
				$a_{2,1}=\left\lbrace s \right\rbrace  , a_{2,2}=\left\lbrace s,r_{1}\right\rbrace $ &  $\hat{0}+ \hat{0}$ & $\hat{0}+ \hat{0}$ & $\hat{0}$
				\\ \hline 
				$a_{2,1}=\left\lbrace s,r_{1} \right\rbrace  , a_{2,2}=\left\lbrace s\right\rbrace $ &  $\hat{0}+ k_{r_{1}}$ & $\hat{0}+ \hat{0}$ & $\hat{0}$
				\\ \hline 
				$a_{2,1}=\left\lbrace  r_{1} \right\rbrace  , a_{2,2}=\left\lbrace s,r_{1}\right\rbrace $ &  $\hat{0}+ k_{r_{1}}$ & $\hat{0}+ \hat{0}$ & $\hat{0}$
				\\ \hline 
				$a_{2,1}=\left\lbrace s,r_{1} \right\rbrace  , a_{2,2}=\left\lbrace r_{1}\right\rbrace $ &  $\hat{0}+ k_{r_{1}}$ & $\hat{0}+ \hat{0}$ & $\hat{0}$
				\\ \hline 
				$a_{2,1}=\left\lbrace s,r_{1} \right\rbrace  , a_{2,2}=\left\lbrace s,r_{1}\right\rbrace $ &  $\hat{0}+ k_{r_{1}}$ & $\hat{0}+ \hat{0}$ & $\hat{0}$
				\\ \ChangeRT{2pt}
				$+$ & \multicolumn{2}{c?}{} \vline & $k_{r_{1}}$
				\\ \hline
		\end{tabular}}
		\caption{$(1\oplus r_{1})\otimes (1\oplus r_{2})$ and $a_{2}=\left\lbrace s, r_{1}\right\rbrace $.}
		\label{tab-brd5}
	\end{table}
	
	\vspace{-5mm}
	
	\begin{table}[H]
		\centering
		\scalebox{0.8}{\begin{tabular}{|c!{\vrule width 2pt}c|c!{\vrule width 2pt}c|}
				\hline
				\multicolumn{4}{|c|}{\newline $ \left\|(1\oplus r_{1})\otimes (1\oplus r_{2})\right\|(\left\lbrace a_{2}\right\rbrace )$ }
				\\ \hline \hline \hline
				$a_{2}=a_{2,1}\cup a_{2,2}$ &  $\left\|1\oplus r_{1}\right\|(\left\lbrace a_{2,1}\right\rbrace ) $ & $ \left\|1\oplus r_{2}\right\|(\left\lbrace a_{2,2}\right\rbrace )$ & $\cdot$ 
				\\ \ChangeRT{2pt}
				$a_{2,1}=\emptyset, a_{2,2}=\left\lbrace s,r_{2}\right\rbrace  $ &  $\hat{1}+ \hat{0}$ & $\hat{0}+ k_{r_{2}}$ & $k_{r_{2}}$
				\\ \hline 
				$a_{2,1}=\left\lbrace s,r_{2}\right\rbrace  , a_{2,2}=\emptyset$ &  $\hat{0}+ \hat{0}$ & $\hat{1}+ \hat{0}$ & $\hat{0}$
				\\ \hline 
				$a_{2,1}=\left\lbrace s \right\rbrace  , a_{2,2}=\left\lbrace r_{2}\right\rbrace $ &  $\hat{0}+ \hat{0}$ & $\hat{0}+ k_{r_{2}}$ & $\hat{0}$
				\\ \hline 
				$a_{2,1}=\left\lbrace r_{2} \right\rbrace  , a_{2,2}=\left\lbrace s\right\rbrace $ &  $\hat{0}+ \hat{0}$ & $\hat{0}+ \hat{0}$ & $\hat{0}$
				\\ \hline 
				$a_{2,1}=\left\lbrace s \right\rbrace  , a_{2,2}=\left\lbrace s,r_{2}\right\rbrace $ &  $\hat{0}+ \hat{0}$ & $\hat{0}+ k_{r_{2}}$ & $\hat{0}$
				\\ \hline 
				$a_{2,1}=\left\lbrace s,r_{2} \right\rbrace  , a_{2,2}=\left\lbrace s\right\rbrace $ &  $\hat{0}+ \hat{0}$ & $\hat{0}+ \hat{0}$ & $\hat{0}$
				\\ \hline 
				$a_{2,1}=\left\lbrace  r_{2} \right\rbrace  , a_{2,2}=\left\lbrace s,r_{2}\right\rbrace $ &  $\hat{0}+ \hat{0}$ & $\hat{0}+ k_{r_{2}}$ & $\hat{0}$
				\\ \hline 
				$a_{2,1}=\left\lbrace s,r_{2} \right\rbrace  , a_{2,2}=\left\lbrace r_{2}\right\rbrace $ &  $\hat{0}+ \hat{0}$ & $\hat{0}+ k_{r_{2}}$ & $\hat{0}$
				\\ \hline 
				$a_{2,1}=\left\lbrace s,r_{2} \right\rbrace  , a_{2,2}=\left\lbrace s,r_{2}\right\rbrace $ &  $\hat{0}+ \hat{0}$ & $\hat{0}+ k_{r_{2}}$ & $\hat{0}$
				\\ \ChangeRT{2pt}
				$+$ & \multicolumn{2}{c?}{} \vline & $k_{r_{2}}$
				\\ \hline
		\end{tabular}}
		\caption{$(1\oplus r_{1})\otimes (1\oplus r_{2})$ and $a_{2}=\left\lbrace s, r_{2}\right\rbrace $.}
		\label{tab-brd6}
	\end{table}
	\vspace{10mm}
	\begin{table}[H]
		\centering
		\scalebox{0.8}{\begin{tabular}{|c!{\vrule width 2pt}c|c!{\vrule width 2pt}c|}
				\hline
				\multicolumn{4}{|c|}{\newline $ \left\|(1\oplus r_{1})\otimes (1\oplus r_{2})\right\|(\left\lbrace a_{2}\right\rbrace )$ }
				\\ \hline \hline \hline
				$a_{2}=a_{2,1}\cup a_{2,2}$ &  $\left\|1\oplus r_{1}\right\|(\left\lbrace a_{2,1}\right\rbrace ) $ & $ \left\|1\oplus r_{2}\right\|(\left\lbrace a_{2,2}\right\rbrace )$ & $\cdot$ 
				\\ \ChangeRT{2pt}
				$a_{2,1}=\emptyset, a_{2,2}=\left\lbrace r_{1}, r_{2}\right\rbrace  $ &  $\hat{1}+ \hat{0}$ & $\hat{0}+ k_{r_{2}}$ & $k_{r_{2}}$
				\\ \hline 
				$a_{2,1}=\left\lbrace r_{1}, r_{2}\right\rbrace  , a_{2,2}=\emptyset$ &  $\hat{0}+k_{r_{1}}$ & $\hat{1}+ \hat{0}$ & $k_{r_{1}}$
				\\ \hline 
				$a_{2,1}=\left\lbrace r_{1} \right\rbrace  , a_{2,2}=\left\lbrace r_{2}\right\rbrace $ &  $\hat{0}+ k_{r_{1}}$ & $\hat{0}+ k_{r_{2}}$ & $k_{r_{1}}\cdot k_{r_{2}}$
				\\ \hline 
				$a_{2,1}=\left\lbrace r_{2} \right\rbrace  , a_{2,2}=\left\lbrace r_{1}\right\rbrace $ &  $\hat{0}+ \hat{0}$ & $\hat{0}+ \hat{0}$ & $\hat{0}$
				\\ \hline 
				$a_{2,1}=\left\lbrace r_{1} \right\rbrace  , a_{2,2}=\left\lbrace r_{1}, r_{2}\right\rbrace $ &  $\hat{0}+ k_{r_{1}}$ & $\hat{0}+ k_{r_{2}}$ & $k_{r_{1}}\cdot k_{r_{2}}$
				\\ \hline 
				$a_{2,1}=\left\lbrace r_{1}, r_{2} \right\rbrace  , a_{2,2}=\left\lbrace r_{1}\right\rbrace $ &  $\hat{0}+ k_{r_{1}}$ & $\hat{0}+ \hat{0}$ & $\hat{0}$
				\\
\hline
		\end{tabular}}
		\caption{$(1\oplus r_{1})\otimes (1\oplus r_{2})$ and $a_{2}=\left\lbrace r_{1}, r_{2}\right\rbrace $.}
		\label{tab-brd7}
	\end{table}

	\begin{table}[H]
		\centering
		\scalebox{0.8}{\begin{tabular}{|c!{\vrule width 2pt}c|c!{\vrule width 2pt}c|}
				\hline
				\multicolumn{4}{|c|}{\newline $ \left\|(1\oplus r_{1})\otimes (1\oplus r_{2})\right\|(\left\lbrace a_{2}\right\rbrace )$ }
				\\ \hline \hline \hline
				$a_{2}=a_{2,1}\cup a_{2,2}$ &  $\left\|1\oplus r_{1}\right\|(\left\lbrace a_{2,1}\right\rbrace ) $ & $ \left\|1\oplus r_{2}\right\|(\left\lbrace a_{2,2}\right\rbrace )$ & $\cdot$ 
				\\ \ChangeRT{2pt}
				$a_{2,1}=\emptyset, a_{2,2}=\left\lbrace s,r_{1},r_{2}\right\rbrace $ &  $\hat{1}+ \hat{0}$ & $\hat{0}+ k_{r_{2}}$ & $k_{r_{2}}$
				\\ \hline
				$a_{2,1}=\left\lbrace s,r_{1},r_{2}\right\rbrace, a_{2,2}=\emptyset $ & $\hat{0}+ k_{r_{1}}$ & $\hat{1}+ \hat{0}$ & $k_{r_{1}}$ 
				\\ \hline
				$a_{2,1}=\left\lbrace s\right\rbrace, a_{2,2}= \left\lbrace r_{1},r_{2} \right\rbrace$ & $\hat{0}+ \hat{0}$ & $\hat{0}+ k_{r_{2}}$ & $\hat{0}$ 
				\\ \hline
				$a_{2,1}=\left\lbrace r_{1},r_{2} \right\rbrace, a_{2,2}=\left\lbrace s\right\rbrace $ & $\hat{0}+ k_{r_{1}}$ & $\hat{0}+ \hat{0}$ & $\hat{0}$ 
				\\ \hline
				$a_{2,1}=\left\lbrace s, r_{1} \right\rbrace, a_{2,2}=\left\lbrace r_{2}\right\rbrace $ & $\hat{0}+ k_{r_{1}}$ & $\hat{0}+ k_{r_{2}}$ & $k_{r_{1}}\cdot k_{r_{2}}$ 
				\\ \hline
				$a_{2,1}=\left\lbrace r_{2} \right\rbrace, a_{2,2}=\left\lbrace s, r_{1} \right\rbrace $ & $\hat{0}+ \hat{0}$ & $\hat{0}+ \hat{0}$ & $\hat{0}$ 
				\\
\hline
		\end{tabular}}
		\caption{$(1\oplus r_{1})\otimes (1\oplus r_{2})$ and $a_{2}=\left\lbrace s,r_{1},r_{2}\right\rbrace$.}
		\label{tab-brd8}
	\end{table}
	
	\subsection{Weighted Atomic Broadcast}	
	The weight of the $wAI(P)$ element $z=s\otimes (1\oplus r_{1}\otimes r_{2})$ on the interactions set $\gamma=\lbrace \lbrace s\rbrace, \lbrace s, r_{1},r_{2}\rbrace\rbrace \in$ $\Gamma(P)$ was computed using the tables presented below. Specifically, we used the primary Tables \ref{tab-ata}-\ref{tab-atb} as well as the auxiliary Tables \ref{tab-re2}-\ref{tab-re9} presented in Subsection \ref{rend-tables} for the weighted Rendezvous scheme.

	\begin{table}[H]
		\centering
		\scalebox{0.85}{\begin{tabular}{|c!{\vrule width 2pt}c|c!{\vrule width 2pt}c|}
				\hline
				\multicolumn{4}{|c|}{\newline $ \left\|s\otimes (1\oplus r_{1}\otimes  r_{2})\right\|(\left\lbrace a\right\rbrace )$ }
				\\ \hline \hline \hline
				$a=a_{1}\cup a_{2}$ &  $\left\|s\right\|(\left\lbrace a_{1}\right\rbrace ) $ & $ \left\|1\oplus r_{1}\otimes  r_{2}\right\|(\left\lbrace a_{2}\right\rbrace )$ & $\cdot$ 
				\\ \ChangeRT{2pt}
				$a_{1}=\emptyset, a_{2}=\left\lbrace s\right\rbrace  $ &  $\hat{0}$ & $\hat{0}+ \hat{0}$ & $\hat{0}$
				\\ \hline 
				$a_{1}=\left\lbrace s\right\rbrace  , a_{2}=\emptyset$ &  $k_{s}$ & $\hat{1}+ \hat{0}$ & $k_{s}$
				\\ \hline 
				$a_{1}=\left\lbrace s \right\rbrace  , a_{2}=\left\lbrace s\right\rbrace $ &  $k_{s}$ & $\hat{0}+ \hat{0}$ & $\hat{0}$
				\\ \ChangeRT{2pt}
				$+$ & \multicolumn{2}{c?}{} \vline & \cellcolor{lightgray}$k_{s}$
				\\ \hline
		\end{tabular}}
		\caption{Weighted Atomic Broadcast and $a=\left\lbrace s\right\rbrace $.}
		\label{tab-ata}
	\end{table}
	
	\begin{table}[H]
		\centering
		\scalebox{0.85}{\begin{tabular}{|c!{\vrule width 2pt}c|c!{\vrule width 2pt}c|}
				\hline
				\multicolumn{4}{|c|}{\newline $ \left\|s\otimes (1\oplus r_{1}\otimes  r_{2})\right\|(\left\lbrace a\right\rbrace )$ }
				\\ \hline \hline \hline
				$a=a_{1}\cup a_{2}$ &  $\left\|s\right\|(\left\lbrace a_{1}\right\rbrace ) $ & $ \left\|1\oplus r_{1}\otimes  r_{2}\right\|(\left\lbrace a_{2}\right\rbrace )$ & $\cdot$ 
				\\ \ChangeRT{2pt}
				$a_{1}=\emptyset, a_{2}=\left\lbrace s,r_{1},r_{2}\right\rbrace $ &  $\hat{0}$ & $\hat{0}+ (k_{r_{1}}\cdot k_{r_{2}})$ & $\hat{0}$
				\\ \hline
				$a_{1}=\left\lbrace s,r_{1},r_{2}\right\rbrace, a_{2}=\emptyset $ & $k_{s}$ & $\hat{1}+ \hat{0}$ & $k_{s}$ 
				\\ \hline
				$a_{1}=\left\lbrace s\right\rbrace, a_{2}= \left\lbrace r_{1},r_{2} \right\rbrace$ & $k_{s}$ & $\hat{0}+ (k_{r_{1}}\cdot k_{r_{2}})$ & $k_{s}\cdot k_{r_{1}}\cdot k_{r_{2}}$ 
				\\ \hline
				$a_{1}=\left\lbrace r_{1},r_{2} \right\rbrace, a_{2}=\left\lbrace s\right\rbrace $ & $\hat{0}$ & $\hat{0}+ \hat{0}$ & $\hat{0}$ 
				\\ \hline
				$a_{1}=\left\lbrace s, r_{1} \right\rbrace, a_{2}=\left\lbrace r_{2}\right\rbrace $ & $k_{s}$ & $\hat{0}+ \hat{0}$ & $\hat{0}$ 
				\\ \hline
				$a_{1}=\left\lbrace r_{2} \right\rbrace, a_{2}=\left\lbrace s, r_{1} \right\rbrace $ & $\hat{0}$ & $\hat{0}+ \hat{0}$ & $\hat{0}$ 
				\\ \hline
				$a_{1}=\left\lbrace s,r_{2} \right\rbrace, a_{2}=\left\lbrace r_{1} \right\rbrace $ & $k_{s}$ & $\hat{0}+ \hat{0}$ & $\hat{0}$ 
				\\ \hline
				$a_{1}=\left\lbrace r_{1} \right\rbrace, a_{2}=\left\lbrace s,r_{2} \right\rbrace $ & $\hat{0}$ & $\hat{0}+ \hat{0}$ & $\hat{0}$ 
				\\ \hline
				$a_{1}=\left\lbrace s \right\rbrace, a_{2}=\left\lbrace s,r_{1},r_{2} \right\rbrace $ & $k_{s}$ & $\hat{0}+ (k_{r_{1}}\cdot k_{r_{2}})$ & $k_{s}\cdot k_{r_{1}}\cdot k_{r_{2}}$ 
				\\ \hline
				$a_{1}=\left\lbrace s, r_{1}, r_{2} \right\rbrace, a_{2}=\left\lbrace s \right\rbrace $ & $k_{s}$ & $\hat{0}+ \hat{0}$ & $\hat{0}$ 
				\\ \hline
				$a_{1}=\left\lbrace r_{1} \right\rbrace, a_{2}=\left\lbrace s,r_{1},r_{2} \right\rbrace $ & $\hat{0}$ & $\hat{0}+ (k_{r_{1}}\cdot k_{r_{2}})$ & $\hat{0}$ 
				\\ \hline
				$a_{1}=\left\lbrace s,r_{1},r_{2} \right\rbrace, a_{2}=\left\lbrace r_{1} \right\rbrace $ & $k_{s}$ & $\hat{0}+ \hat{0}$ & $\hat{0}$ 
				\\ \hline
				$a_{1}=\left\lbrace r_{2} \right\rbrace, a_{2}=\left\lbrace s,r_{1},r_{2} \right\rbrace $ & $\hat{0}$ & $\hat{0}+ (k_{r_{1}}\cdot k_{r_{2}})$ & $\hat{0}$ 
				\\ \hline
				$a_{1}=\left\lbrace s,r_{1},r_{2} \right\rbrace, a_{2}=\left\lbrace r_{2} \right\rbrace $ & $k_{s}$ & $\hat{0}+ \hat{0}$ & $\hat{0}$ 
				\\ \hline
				$a_{1}=\left\lbrace s,r_{1} \right\rbrace, a_{2}=\left\lbrace r_{1},r_{2} \right\rbrace $ & $k_{s}$ & $\hat{0}+ (k_{r_{1}}\cdot k_{r_{2}})$ & $k_{s}\cdot k_{r_{1}}\cdot k_{r_{2}}$ 
				\\ \hline
				$a_{1}=\left\lbrace r_{1},r_{2} \right\rbrace, a_{2}=\left\lbrace s,r_{1} \right\rbrace $ & $\hat{0}$ & $\hat{0}+ \hat{0}$ & $\hat{0}$
				\\ \hline
				$a_{1}=\left\lbrace s, r_{1} \right\rbrace, a_{2}=\left\lbrace s,r_{2} \right\rbrace $ & $k_{s}$ & $\hat{0}+ \hat{0}$ & $\hat{0}$
				\\ \hline
				$a_{1}=\left\lbrace s,r_{2} \right\rbrace, a_{2}=\left\lbrace s,r_{1} \right\rbrace $ & $k_{s}$ & $\hat{0}+\hat{0}$ & $\hat{0}$
				\\ \hline
				$a_{1}=\left\lbrace s,r_{1} \right\rbrace, a_{2}=\left\lbrace s,r_{1},r_{2} \right\rbrace $ & $k_{s}$ & $\hat{0}+ (k_{r_{1}}\cdot k_{r_{2}})$ & $k_{s}\cdot k_{r_{1}}\cdot k_{r_{2}}$ 
				\\ \hline
				$a_{1}=\left\lbrace s,r_{1},r_{2}  \right\rbrace, a_{2}=\left\lbrace s,r_{1}\right\rbrace $ & $k_{s}$ & $\hat{0}+ \hat{0}$ & $\hat{0}$
				\\ \hline
				$a_{1}=\left\lbrace s,r_{2} \right\rbrace, a_{2}=\left\lbrace r_{1},r_{2} \right\rbrace $ & $k_{s}$ & $\hat{0}+ (k_{r_{1}}\cdot k_{r_{2}})$ & $k_{s}\cdot k_{r_{1}}\cdot k_{r_{2}}$ 
				\\ \hline
				$a_{1}=\left\lbrace r_{1}, r_{2} \right\rbrace, a_{2}=\left\lbrace s,r_{2} \right\rbrace $ & $\hat{0}$ & $\hat{0}+ \hat{0}$ & $\hat{0}$  
				\\ \hline
				$a_{1}=\left\lbrace s,r_{2} \right\rbrace, a_{2}=\left\lbrace s,r_{1},r_{2} \right\rbrace $ & $k_{s}$ & $\hat{0}+ (k_{r_{1}}\cdot k_{r_{2}})$ & $k_{s}\cdot k_{r_{1}}\cdot k_{r_{2}}$ 
				\\ \hline
				$a_{1}=\left\lbrace s,r_{1},r_{2} \right\rbrace, a_{2}=\left\lbrace s,r_{2} \right\rbrace $ & $k_{s}$ & $\hat{0}+ \hat{0}$ & $\hat{0}$ 
				\\ \hline
				$a_{1}=\left\lbrace r_{1},r_{2} \right\rbrace, a_{2}=\left\lbrace s,r_{1},r_{2} \right\rbrace $ & $\hat{0}$ & $\hat{0}+ (k_{r_{1}}\cdot k_{r_{2}})$ & $\hat{0}$ 
				\\ \hline
				$a_{1}=\left\lbrace s,r_{1},r_{2} \right\rbrace, a_{2}=\left\lbrace r_{1},r_{2} \right\rbrace $ & $k_{s}$ & $\hat{0}+ (k_{r_{1}}\cdot k_{r_{2}})$ & $k_{s}\cdot k_{r_{1}}\cdot k_{r_{2}}$
				\\ \hline
				$a_{1}=\left\lbrace s,r_{1},r_{2} \right\rbrace, a_{2}=\left\lbrace s,r_{1},r_{2} \right\rbrace $ & $k_{s}$ & $\hat{0}+ (k_{r_{1}}\cdot k_{r_{2}})$ & $k_{s}\cdot k_{r_{1}}\cdot k_{r_{2}}$ 
				\\ \ChangeRT{2pt}
				$+$ & \multicolumn{2}{c?}{} \vline & \cellcolor{lightgray}$k_{s}+ (k_{s}\cdot k_{r_{1}}\cdot k_{r_{2}})$ 
				\\ \hline
		\end{tabular}}
		\caption{Weighted Atomic Broadcast and $a=\left\lbrace s,r_{1},r_{2}\right\rbrace$.}
		\label{tab-atb}
	\end{table}

	\subsection{Weighted Causality Chain}
	The weight of the $wAI(P)$ element $z=s\otimes (1\oplus r_{1} \otimes (1\oplus r_{2}))$ on the interactions set $\gamma=\left\lbrace \left\lbrace s \right\rbrace ,\left\lbrace s,r_{1}\right\rbrace , \left\lbrace s, r_{1}, r_{2}\right\rbrace \right\rbrace \in \Gamma(P)$ was computed by using the following tables. The primary tables are Tables \ref{tab-caua}-\ref{tab-cauc}, while Tables \ref{tab-cauc1}-\ref{tab-cauc8} are the auxiliary ones.
	
	\begin{table}[H]
		\centering
		\scalebox{0.85}{
\begin{tabular}{|c!{\vrule width 2pt}c|c!{\vrule width 2pt}c|}
				\hline
				\multicolumn{4}{|c|}{\newline $ \left\|s\otimes \big(1\oplus r_{1}\otimes (1\oplus  r_{2})\big)\right\|(\left\lbrace a\right\rbrace )$ }
				\\ \hline \hline \hline
				$a=a_{1}\cup a_{2}$ &  $\left\|s\right\|(\left\lbrace a_{1}\right\rbrace ) $ & $ \left\|1\oplus r_{1}\otimes  (1\oplus r_{2})\right\|(\left\lbrace a_{2}\right\rbrace )$ & $\cdot$ 
				\\ \ChangeRT{2pt}
				$a_{1}=\emptyset, a_{2}=\left\lbrace s\right\rbrace  $ &  $\hat{0}$ & $\hat{0}+ \hat{0}$ & $\hat{0}$
				\\ \hline 
				$a_{1}=\left\lbrace s\right\rbrace  , a_{2}=\emptyset$ &  $k_{s}$ & $\hat{1}+ \hat{0}$ & $k_{s}$
				\\ \hline 
				$a_{1}=\left\lbrace s \right\rbrace  , a_{2}=\left\lbrace s\right\rbrace $ &  $k_{s}$ & $\hat{0}+ \hat{0}$ & $\hat{0}$
				\\ \ChangeRT{2pt}
				$+$ & \multicolumn{2}{c?}{} \vline & \cellcolor{lightgray}$k_{s}$
				\\ \hline
		\end{tabular}}
		\caption{Weighted Causality Chain and $a=\left\lbrace s\right\rbrace $.}
		\label{tab-caua}
	\end{table}

	\begin{table}[H]
		\centering
		\scalebox{0.85}{\begin{tabular}{|c!{\vrule width 2pt}c|c!{\vrule width 2pt}c|}
				\hline
				\multicolumn{4}{|c|}{\newline $ \left\|s\otimes \big(1\oplus r_{1}\otimes (1\oplus  r_{2})\big)\right\|(\left\lbrace a\right\rbrace )$ }
				\\ \hline \hline \hline
				$a=a_{1}\cup a_{2}$ &  $\left\|s\right\|(\left\lbrace a_{1}\right\rbrace ) $ & $ \left\|1\oplus r_{1}\otimes (1\oplus r_{2})\right\|(\left\lbrace a_{2}\right\rbrace )$ & $\cdot$ 
				\\ \ChangeRT{2pt}
				$a_{1}=\emptyset, a_{2}=\left\lbrace s,r_{1}\right\rbrace  $ &  $\hat{0}$ & $\hat{0}+ k_{r_{1}}$ & $\hat{0}$
				\\ \hline 
				$a_{1}=\left\lbrace s,r_{1}\right\rbrace  , a_{2}=\emptyset$ &  $k_{s}$ & $\hat{1}+ \hat{0}$ & $k_{s}$
				\\ \hline 
				$a_{1}=\left\lbrace s \right\rbrace  , a_{2}=\left\lbrace r_{1}\right\rbrace $ &  $k_{s}$ & $\hat{0}+ k_{r_{1}}$ & $k_{s}\cdot k_{r_{1}}$
				\\ \hline 
				$a_{1}=\left\lbrace r_{1} \right\rbrace  , a_{2}=\left\lbrace s\right\rbrace $ &  $\hat{0}$ & $\hat{0}+ \hat{0}$ & $\hat{0}$
				\\ \hline 
				$a_{1}=\left\lbrace s \right\rbrace  , a_{2}=\left\lbrace s,r_{1}\right\rbrace $ &  $k_{s}$ & $\hat{0}+ k_{r_{1}}$ & $k_{s}\cdot k_{r_{1}}$
				\\ \hline 
				$a_{1}=\left\lbrace s,r_{1} \right\rbrace  , a_{2}=\left\lbrace s\right\rbrace $ &  $k_{s}$ & $\hat{0}+ \hat{0}$ & $\hat{0}$
				\\ \hline 
				$a_{1}=\left\lbrace  r_{1} \right\rbrace  , a_{2}=\left\lbrace s,r_{1}\right\rbrace $ &  $\hat{0}$ & $\hat{0}+ k_{r_{1}}$ & $\hat{0}$
				\\ \hline 
				$a_{1}=\left\lbrace s,r_{1} \right\rbrace  , a_{2}=\left\lbrace r_{1}\right\rbrace $ &  $k_{s}$ & $\hat{0} + k_{r_{1}}$ & $k_{s}\cdot k_{r_{1}}$
				\\ \hline 
				$a_{1}=\left\lbrace s,r_{1} \right\rbrace  , a_{2}=\left\lbrace s,r_{1}\right\rbrace $ &  $k_{s}$ & $\hat{0}+ k_{r_{1}}$ & $k_{s}\cdot k_{r_{1}}$
				\\ \ChangeRT{2pt}
				$+$ & \multicolumn{2}{c?}{} \vline & \cellcolor{lightgray}$k_{s}+ (k_{s}\cdot k_{r_{1}})$
				\\ \hline
		\end{tabular}}
		\caption{Weighted Causality Chain and $a=\left\lbrace s, r_{1}\right\rbrace $.}
		\label{tab-caub}
	\end{table}
	
	\begin{table}[H]
		\centering
		\scalebox{0.85}{\begin{tabular}{|c!{\vrule width 2pt}c|c!{\vrule width 2pt}c|}
				\hline
				\multicolumn{4}{|c|}{\newline $ \left\|s\otimes \big(1\oplus r_{1}\otimes (1\oplus  r_{2})\big)\right\|(\left\lbrace a\right\rbrace )$ }
				\\ \hline \hline \hline
				$a=a_{1}\cup a_{2}$ &  $\left\|s\right\|(\left\lbrace a_{1}\right\rbrace ) $ & $ \left\|1\oplus (r_{1}\otimes  (1\oplus r_{2})\right\|(\left\lbrace a_{2}\right\rbrace )$ & $\cdot$ 
				\\ \ChangeRT{2pt}
				$a_{1}=\emptyset, a_{2}=\left\lbrace s,r_{1},r_{2}\right\rbrace $ &  $\hat{0}$ & $\hat{0}+ \big(k_{r_{1}}+ (k_{r_{1}}\cdot k_{r_{2}})\big)$ & $\hat{0}$
				\\ \hline
				$a_{1}=\left\lbrace s,r_{1},r_{2}\right\rbrace, a_{2}=\emptyset $ & $k_{s}$ & $\hat{1}+ \hat{0}$ & $k_{s}$ 
				\\ \hline
				$a_{1}=\left\lbrace s\right\rbrace, a_{2}= \left\lbrace r_{1},r_{2} \right\rbrace$ & $k_{s}$ & $\hat{0}+ \big(k_{r_{1}}+ (k_{r_{1}}\cdot k_{r_{2}})\big)$ & $(k_{s}\cdot k_{r_{1}})+ (k_{s}+ k_{r_{1}}\cdot k_{r_{2}})$ 
				\\ \hline
				$a_{1}=\left\lbrace r_{1},r_{2} \right\rbrace, a_{2}=\left\lbrace s\right\rbrace $ & $\hat{0}$ & $\hat{0}+ \hat{0}$ & $\hat{0}$ 
				\\ \hline
				$a_{1}=\left\lbrace s, r_{1} \right\rbrace, a_{2}=\left\lbrace r_{2}\right\rbrace $ & $k_{s}$ & $\hat{0}+ \hat{0}$ & $\hat{0}$ 
				\\ \hline
				$a_{1}=\left\lbrace r_{2} \right\rbrace, a_{2}=\left\lbrace s, r_{1} \right\rbrace $ & $\hat{0}$ & $\hat{0}+ k_{r_{1}}$ & $\hat{0}$ 
				\\
\hline
		\end{tabular}}
		\caption{Weighted Causality Chain and $a=\left\lbrace s,r_{1},r_{2}\right\rbrace$.}
		\label{tab-cauc}
	\end{table}

	\begin{table}[H]
		\centering
		\scalebox{0.8}{\begin{tabular}{|c!{\vrule width 2pt}c|c!{\vrule width 2pt}c|}
				\hline
				\multicolumn{4}{|c|}{\newline $ \left\|r_{1}\otimes (1\oplus r_{2})\right\|(\left\lbrace a_{2}\right\rbrace )$ }
				\\ \hline \hline \hline
				$a_{2}=a_{2,1}\cup a_{2,2}$ &  $\left\| r_{1}\right\|(\left\lbrace a_{2,1}\right\rbrace ) $ & $ \left\|1\oplus r_{2}\right\|(\left\lbrace a_{2,2}\right\rbrace )$ & $\cdot$ 
				\\ \ChangeRT{2pt}
				$a_{2,1}=\emptyset, a_{2,2}=\emptyset $ &  $\hat{0}$ & $\hat{1}+ \hat{0}$ & $\hat{0}$
				\\ \ChangeRT{2pt}
				$+$ & \multicolumn{2}{c?}{} \vline & $\hat{0}$
				\\ \hline
		\end{tabular}}
		\caption{$r_{1}\otimes (1\oplus r_{2})$ and $a_{2}=\emptyset$.}
		\label{tab-cauc1}
	\end{table}
	\vspace{-3mm}
	\begin{table}[H]
		\centering
		\scalebox{0.8}{\begin{tabular}{|c!{\vrule width 2pt}c|c!{\vrule width 2pt}c|}
				\hline
				\multicolumn{4}{|c|}{\newline $ \left\|r_{1}\otimes (1\oplus r_{2})\right\|(\left\lbrace a_{2}\right\rbrace )$ }
				\\ \hline \hline \hline
				$a_{2}=a_{2,1}\cup a_{2,2}$ &  $\left\| r_{1}\right\|(\left\lbrace a_{2,1}\right\rbrace ) $ & $ \left\|1\oplus r_{2}\right\|(\left\lbrace a_{2,2}\right\rbrace )$ & $\cdot$ 
				\\ \ChangeRT{2pt}
				$a_{2,1}=\emptyset, a_{2,2}=\left\lbrace s\right\rbrace  $ &  $\hat{0}$ & $\hat{0}+ \hat{0}$ & $\hat{0}$
				\\ \hline 
				$a_{2,1}=\left\lbrace s\right\rbrace  , a_{2,2}=\emptyset$ &  $\hat{0}$ & $\hat{1}+ \hat{0}$ & $\hat{0}$
				\\ \hline 
				$a_{2,1}=\left\lbrace s \right\rbrace  , a_{2,2}=\left\lbrace s\right\rbrace $ &  $\hat{0}$ & $\hat{0}+ \hat{0}$ & $\hat{0}$
				\\ \ChangeRT{2pt}
				$+$ & \multicolumn{2}{c?}{} \vline & $\hat{0}$
				\\ \hline
		\end{tabular}}
		\caption{$r_{1}\otimes (1\oplus r_{2})$ and $a_{2}=\left\lbrace s\right\rbrace $.}
		\label{tab-cauc2}
	\end{table}
	\vspace{-3mm}
	\begin{table}[H]
		\centering
		\scalebox{0.8}{\begin{tabular}{|c!{\vrule width 2pt}c|c!{\vrule width 2pt} c|}
				\hline
				\multicolumn{4}{|c|}{\newline $ \left\|r_{1}\otimes (1\oplus r_{2})\right\|(\left\lbrace a_{2}\right\rbrace )$ }
				\\ \hline \hline \hline
				$a_{2}=a_{2,1}\cup a_{2,2}$ &  $\left\| r_{1}\right\|(\left\lbrace a_{2,1}\right\rbrace ) $ & $ \left\|1\oplus r_{2}\right\|(\left\lbrace a_{2,2}\right\rbrace )$ & $\cdot$ 
				\\ \ChangeRT{2pt}
				$a_{2,1}=\emptyset, a_{2,2}=\left\lbrace  r_{1}\right\rbrace  $ &  $\hat{0}$ & $\hat{0}+ \hat{0}$ & $\hat{0}$
				\\ \hline 
				$a_{2,1}=\left\lbrace r_{1}\right\rbrace  , a_{2,2}=\emptyset$ &  $k_{r_{1}}$ & $\hat{1}+ \hat{0}$ & $k_{r_{1}}$
				\\ \hline 
				$a_{2,1}=\left\lbrace r_{1} \right\rbrace  , a_{2,2}=\left\lbrace r_{1}\right\rbrace $ &  $k_{r_{1}}$ & $\hat{0}+ \hat{0}$ & $\hat{0}$
				\\ \ChangeRT{2pt}
				$+$ & \multicolumn{2}{c?}{} \vline & $k_{r_{1}}$
				\\ \hline
		\end{tabular}}
		\caption{$r_{1}\otimes (1\oplus r_{2})$ and $a_{2}=\left\lbrace  r_{1}\right\rbrace $.}
		\label{tab-cauc3}
	\end{table}
	\vspace{-3mm}
	\begin{table}[H]
		\centering
		\scalebox{0.8}{\begin{tabular}{|c!{\vrule width 2pt}c|c!{\vrule width 2pt}c|}
				\hline
				\multicolumn{4}{|c|}{\newline $ \left\|r_{1}\otimes (1\oplus r_{2})\right\|(\left\lbrace a_{2}\right\rbrace )$ }
				\\ \hline \hline \hline
				$a_{2}=a_{2,1}\cup a_{2,2}$ &  $\left\| r_{1}\right\|(\left\lbrace a_{2,1}\right\rbrace ) $ & $ \left\|1\oplus r_{2}\right\|(\left\lbrace a_{2,2}\right\rbrace )$ & $\cdot$ 
				\\ \ChangeRT{2pt}
				$a_{2,1}=\emptyset, a_{2,2}=\left\lbrace  r_{2}\right\rbrace  $ &  $\hat{0}$ & $\hat{0}+ k_{r_{2}}$ & $\hat{0}$
				\\ \hline 
				$a_{2,1}=\left\lbrace r_{2}\right\rbrace  , a_{2,2}=\emptyset$ &  $\hat{0}$ & $\hat{1}+ \hat{0}$ & $\hat{0}$
				\\ \hline 
				$a_{2,1}=\left\lbrace r_{2} \right\rbrace  , a_{2,2}=\left\lbrace r_{2}\right\rbrace $ &  $\hat{0}$ & $\hat{0}+ k_{r_{2}}$ & $\hat{0}$
				\\ \ChangeRT{2pt}
				$+$ & \multicolumn{2}{c?}{} \vline & $\hat{0}$
				\\ \hline
		\end{tabular}}
		\caption{$r_{1}\otimes (1\oplus r_{2})$ and $a_{2}=\left\lbrace  r_{2}\right\rbrace $.}
		\label{tab-cauc4}
	\end{table}
	\vspace{-2mm}
	\begin{table}[H]
		\centering
		\scalebox{0.8}{\begin{tabular}{|c!{\vrule width 2pt}c|c!{\vrule width 2pt}c|}
				\hline
				\multicolumn{4}{|c|}{\newline $ \left\|r_{1}\otimes (1\oplus r_{2})\right\|(\left\lbrace a_{2}\right\rbrace )$ }
				\\ \hline \hline \hline
				$a_{2}=a_{2,1}\cup a_{2,2}$ &  $\left\| r_{1}\right\|(\left\lbrace a_{2,1}\right\rbrace ) $ & $ \left\|1\oplus r_{2}\right\|(\left\lbrace a_{2,2}\right\rbrace )$ & $\cdot$ 
				\\ \ChangeRT{2pt}
				$a_{2,1}=\emptyset, a_{2,2}=\left\lbrace s,r_{1}\right\rbrace  $ &  $\hat{0}$ & $\hat{0}+ \hat{0}$ & $\hat{0}$
				\\ \hline 
				$a_{2,1}=\left\lbrace s,r_{1}\right\rbrace  , a_{2,2}=\emptyset$ &  $k_{r_{1}}$ & $\hat{1}+ \hat{0}$ & $k_{r_{1}}$
				\\ \hline 
				$a_{2,1}=\left\lbrace s \right\rbrace  , a_{2,2}=\left\lbrace r_{1}\right\rbrace $ &  $\hat{0}$ & $\hat{0}+ \hat{0}$ & $\hat{0}$
				\\ \hline 
				$a_{2,1}=\left\lbrace r_{1} \right\rbrace  , a_{2,2}=\left\lbrace s\right\rbrace $ &  $k_{r_{1}}$ & $\hat{0}+ \hat{0}$ & $\hat{0}$
				\\ \hline 
				$a_{2,1}=\left\lbrace s \right\rbrace  , a_{2,2}=\left\lbrace s,r_{1}\right\rbrace $ &  $\hat{0}$ & $\hat{0}+ \hat{0}$ & $\hat{0}$
				\\ \hline 
				$a_{2,1}=\left\lbrace s,r_{1} \right\rbrace  , a_{2,2}=\left\lbrace s\right\rbrace $ &  $k_{r_{1}}$ & $\hat{0}+ \hat{0}$ & $\hat{0}$
				\\ \hline 
				$a_{2,1}=\left\lbrace  r_{1} \right\rbrace  , a_{2,2}=\left\lbrace s,r_{1}\right\rbrace $ &  $k_{r_{1}}$ & $\hat{0}+ \hat{0}$ & $\hat{0}$
				\\ \hline 
				$a_{2,1}=\left\lbrace s,r_{1} \right\rbrace  , a_{2,2}=\left\lbrace r_{1}\right\rbrace $ &  $k_{r_{1}}$ & $\hat{0}+ \hat{0}$ & $\hat{0}$
				\\ \hline 
				$a_{2,1}=\left\lbrace s,r_{1} \right\rbrace  , a_{2,2}=\left\lbrace s,r_{1}\right\rbrace $ &  $k_{r_{1}}$ & $\hat{0}+ \hat{0}$ & $\hat{0}$
				\\ \ChangeRT{2pt}
				$+$ & \multicolumn{2}{c?}{} \vline & $k_{r_{1}}$
				\\ \hline
		\end{tabular}}
		\caption{$r_{1}\otimes (1\oplus r_{2})$ and $a_{2}=\left\lbrace s, r_{1}\right\rbrace $.}
		\label{tab-cauc5}
	\end{table}
	
	\begin{table}[H]
		\centering
		\scalebox{0.8}{\begin{tabular}{|c!{\vrule width 2pt}c|c!{\vrule width 2pt}c|}
				\hline
				\multicolumn{4}{|c|}{\newline $ \left\|r_{1}\otimes (1\oplus r_{2})\right\|(\left\lbrace a_{2}\right\rbrace )$ }
				\\ \hline \hline \hline
				$a_{2}=a_{2,1}\cup a_{2,2}$ &  $\left\| r_{1}\right\|(\left\lbrace a_{2,1}\right\rbrace ) $ & $ \left\|1\oplus r_{2}\right\|(\left\lbrace a_{2,2}\right\rbrace )$ & $\cdot$ 
				\\ \ChangeRT{2pt}
				$a_{2,1}=\emptyset, a_{2,2}=\left\lbrace s,r_{2}\right\rbrace  $ &  $\hat{0}$ & $\hat{0}+ k_{r_{2}}$ & $\hat{0}$
				\\ \hline 
				$a_{2,1}=\left\lbrace s,r_{2}\right\rbrace  , a_{2,2}=\emptyset$ &  $\hat{0}$ & $\hat{1}+ \hat{0}$ & $\hat{0}$
				\\ \hline 
				$a_{2,1}=\left\lbrace s \right\rbrace  , a_{2,2}=\left\lbrace r_{2}\right\rbrace $ &  $\hat{0}$ & $\hat{0}+ k_{r_{2}}$ & $\hat{0}$
				\\ \hline 
				$a_{2,1}=\left\lbrace r_{2} \right\rbrace  , a_{2,2}=\left\lbrace s\right\rbrace $ &  $\hat{0}$ & $\hat{0}+ \hat{0}$ & $\hat{0}$
				\\ \hline 
				$a_{2,1}=\left\lbrace s \right\rbrace  , a_{2,2}=\left\lbrace s,r_{2}\right\rbrace $ &  $\hat{0}$ & $\hat{0}+ k_{r_{2}}$ & $\hat{0}$
				\\ \hline 
				$a_{2,1}=\left\lbrace s,r_{2} \right\rbrace  , a_{2,2}=\left\lbrace s\right\rbrace $ &  $\hat{0}$ & $\hat{0}+ \hat{0}$ & $\hat{0}$
				\\ \hline 
				$a_{2,1}=\left\lbrace  r_{2} \right\rbrace  , a_{2,2}=\left\lbrace s,r_{2}\right\rbrace $ &  $\hat{0}$ & $\hat{0}+ k_{r_{2}}$ & $\hat{0}$
				\\ \hline 
				$a_{2,1}=\left\lbrace s,r_{2} \right\rbrace  , a_{2,2}=\left\lbrace r_{2}\right\rbrace $ &  $\hat{0}$ & $\hat{0}+ k_{r_{2}}$ & $\hat{0}$
				\\ \hline 
				$a_{2,1}=\left\lbrace s,r_{2} \right\rbrace  , a_{2,2}=\left\lbrace s,r_{2}\right\rbrace $ &  $\hat{0}$ & $\hat{0}+ k_{r_{2}}$ & $\hat{0}$
				\\ \ChangeRT{2pt}
				$+$ & \multicolumn{2}{c?}{} \vline & $\hat{0}$
				\\ \hline
		\end{tabular}}
		\caption{$r_{1}\otimes (1\oplus r_{2})$ and $a_{2}=\left\lbrace s, r_{2}\right\rbrace $.}
		\label{tab-cauc6}
	\end{table}
	
	\vspace{10mm}
	
	\begin{table}[H]
		\centering
		\scalebox{0.8}{\begin{tabular}{|c!{\vrule width 2pt}c|c!{\vrule width 2pt}c|}
				\hline
				\multicolumn{4}{|c|}{\newline $ \left\|r_{1}\otimes (1\oplus r_{2})\right\|(\left\lbrace a_{2}\right\rbrace )$ }
				\\ \hline \hline \hline
				$a_{2}=a_{2,1}\cup a_{2,2}$ &  $\left\| r_{1}\right\|(\left\lbrace a_{2,1}\right\rbrace ) $ & $ \left\|1\oplus r_{2}\right\|(\left\lbrace a_{2,2}\right\rbrace )$ & $\cdot$ 
				\\ \ChangeRT{2pt}
				$a_{2,1}=\emptyset, a_{2,2}=\left\lbrace r_{1}, r_{2}\right\rbrace  $ &  $\hat{0}$ & $\hat{0}+ k_{r_{2}}$ & $\hat{0}$
				\\ \hline 
				$a_{2,1}=\left\lbrace r_{1}, r_{2}\right\rbrace  , a_{2,2}=\emptyset$ &  $k_{r_{1}}$ & $\hat{1}+ \hat{0}$ & $k_{r_{1}}$
				\\ \hline 
				$a_{2,1}=\left\lbrace r_{1} \right\rbrace  , a_{2,2}=\left\lbrace r_{2}\right\rbrace $ &  $k_{r_{1}}$ & $\hat{0}+ k_{r_{2}}$ & $k_{r_{1}}\cdot k_{r_{2}}$
				\\ \hline 
				$a_{2,1}=\left\lbrace r_{2} \right\rbrace  , a_{2,2}=\left\lbrace r_{1}\right\rbrace $ &  $\hat{0}$ & $\hat{0}+ \hat{0}$ & $\hat{0}$
				\\ \hline 
				$a_{2,1}=\left\lbrace r_{1} \right\rbrace  , a_{2,2}=\left\lbrace r_{1}, r_{2}\right\rbrace $ &  $k_{r_{1}}$ & $\hat{0}+ k_{r_{2}}$ & $k_{r_{1}}\cdot k_{r_{2}}$
				\\ \hline 
				$a_{2,1}=\left\lbrace r_{1}, r_{2} \right\rbrace  , a_{2,2}=\left\lbrace r_{1}\right\rbrace $ &  $k_{r_{1}}$ & $\hat{0}+ \hat{0}$ & $\hat{0}$
				\\
\hline
		\end{tabular}}
		\caption{$r_{1}\otimes (1\oplus r_{2})$ and $a_{2}=\left\lbrace r_{1}, r_{2}\right\rbrace $.}
		\label{tab-cauc7}
	\end{table}

	\begin{table}[H]
		\centering
		\scalebox{0.8}{\begin{tabular}{|c!{\vrule width 2pt}c|c!{\vrule width 2pt}c|}
				\hline
				\multicolumn{4}{|c|}{\newline $ \left\|r_{1}\otimes (1\oplus r_{2})\right\|(\left\lbrace a_{2}\right\rbrace )$ }
				\\ \hline \hline \hline
				$a_{2}=a_{2,1}\cup a_{2,2}$ &  $\left\| r_{1}\right\|(\left\lbrace a_{2,1}\right\rbrace ) $ & $ \left\|1\oplus r_{2}\right\|(\left\lbrace a_{2,2}\right\rbrace )$ & $\cdot$ 
				\\ \ChangeRT{2pt}
				$a_{2,1}=\emptyset, a_{2,2}=\left\lbrace s,r_{1},r_{2}\right\rbrace $ &  $\hat{0}$ & $\hat{0}+ k_{r_{2}}$ & $\hat{0}$
				\\ \hline
				$a_{2,1}=\left\lbrace s,r_{1},r_{2}\right\rbrace, a_{2,2}=\emptyset $ & $k_{r_{1}}$ & $\hat{1}+ \hat{0}$ & $k_{r_{1}}$ 
				\\ \hline
				$a_{2,1}=\left\lbrace s\right\rbrace, a_{2,2}= \left\lbrace r_{1},r_{2} \right\rbrace$ & $\hat{0}$ & $\hat{0}+ k_{r_{2}}$ & $\hat{0}$ 
				\\ \hline
				$a_{2,1}=\left\lbrace r_{1},r_{2} \right\rbrace, a_{2,2}=\left\lbrace s\right\rbrace $ & $k_{r_{1}}$ & $\hat{0}+ \hat{0}$ & $\hat{0}$ 
				\\ \hline
				$a_{2,1}=\left\lbrace s, r_{1} \right\rbrace, a_{2,2}=\left\lbrace r_{2}\right\rbrace $ & $k_{r_{1}}$ & $\hat{0}+ k_{r_{2}}$ & $k_{r_{1}}\cdot k_{r_{2}}$ 
				\\ \hline
				$a_{2,1}=\left\lbrace r_{2} \right\rbrace, a_{2,2}=\left\lbrace s, r_{1} \right\rbrace $ & $\hat{0}$ & $\hat{0}+ \hat{0}$ & $\hat{0}$ 
				\\
\hline
		\end{tabular}}
		\caption{$r_{1}\otimes (1\oplus r_{2})$ and $a_{2}=\left\lbrace s,r_{1},r_{2}\right\rbrace$.}
		\label{tab-cauc8}
	\end{table}

\end{document}